\documentclass[journal]{IEEEtran}
\ifCLASSINFOpdf
\else
\fi
\usepackage{graphicx}
\usepackage{color}
\usepackage{placeins}
\usepackage{float}
\usepackage{tabularx,colortbl}
\usepackage{times,amsmath}
\usepackage{amsmath,graphicx}
\usepackage{amssymb}

\usepackage{subfig}
\usepackage[labelfont=bf, justification=justified]{caption}
\usepackage{hhline}

\usepackage{psfrag}
\usepackage[table,xcdraw,dvipsnames]{xcolor}
\usepackage{setspace}
\usepackage{multirow}
\usepackage{multicol}
\usepackage{arydshln}

\usepackage{rotating}
\newcommand\tabrotate[1]{\begin{turn}{90}\rlap{#1}\end{turn}}

\usepackage{placeins}

\usepackage{pgfplots, pgfplotstable}
\pgfplotsset{compat=newest}
\usetikzlibrary{shapes, arrows, calc, patterns, matrix, positioning}

\usepgfplotslibrary{groupplots}
\usepackage{tikz}
\usetikzlibrary{decorations.pathmorphing}
\usepackage{xifthen}
\newlength\figureheight
\newlength\figurewidth

\begin{document}
%
\title{Graph-Based Compensated Wavelet Lifting for Scalable Lossless Coding of Dynamic Medical Data}
%
%

\author{Daniela Lanz,~\IEEEmembership{Member,~IEEE}, and Andr\'{e} Kaup,~\IEEEmembership{Fellow,~IEEE}
\thanks{The authors are with the Chair of Multimedia Communications and Signal Processing, Friedrich-Alexander-University Erlangen-N\"{u}rnberg (FAU), Cauerstr. 7, 91058 Erlangen, Germany (e-mail: \mbox{daniela.lanz@fau.de}, \mbox{andre.kaup@fau.de}).}
}

\maketitle

\begin{abstract}
Lossless compression of dynamic \mbox{2-D+t} and \mbox{3-D+t} medical data is challenging regarding the huge amount of data, the characteristics of the inherent noise, and the high bit depth. Beyond that, a scalable representation is often required in telemedicine applications.
Motion Compensated Temporal Filtering works well for lossless compression of medical volume data and additionally provides temporal, spatial, and quality scalability features. 
To achieve a high quality lowpass subband, which shall be used as a downscaled representative of the original data, graph-based motion compensation was recently introduced to this framework. However, encoding the motion information, which is stored in adjacency matrices, is not well investigated so far. 
This work focuses on coding these adjacency matrices to make the graph-based motion compensation feasible for data compression.
We propose a novel coding scheme based on constructing so-called motion maps. This allows for the first time to compare the performance of graph-based motion compensation to traditional block- and mesh-based approaches. For high quality lowpass subbands our method is able to outperform the block- and mesh-based approaches by increasing the visual quality in terms of PSNR by 0.53\,dB and 0.28\,dB for CT data, as well as 1.04\,dB and 1.90\,dB for MR data, respectively, while the bit rate is reduced at the same time. 

\end{abstract}


%
\IEEEpeerreviewmaketitle

\section{Introduction}
\label{sec:introduction}
%
%
%
%

\begin{figure}[t]
	\begin{center}
		\psfragscanon
		\psfrag{x}{$x$}
		\psfrag{y}{$y$}
		\psfrag{z}{$z$}
		\psfrag{t}{$t$}
		\psfrag{X}{$X$}
		\psfrag{Y}{$Y$}
		\psfrag{Z}{$Z$}
		\psfrag{T}{$T$}
		\psfrag{1}{$1$}
		\psfrag{2}{$2$}
		\includegraphics[width=0.48\textwidth]{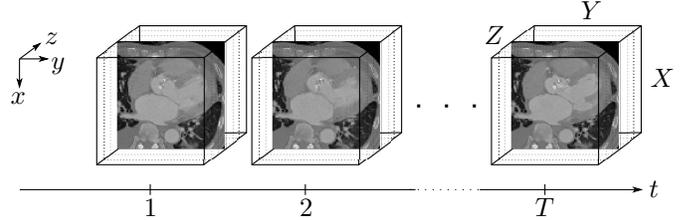}
		\psfragscanoff
		\caption{Example of a dynamic medical data set: The sketch shows a 3-D+t CT volume of a thorax which consists of temporally succeeding 3-D volumes.}
		\label{fig:data}
		\vspace{-2pt}
	\end{center}
\end{figure}

\IEEEPARstart{I}{n} the daily medical routine, dynamic volume data from Computed Tomography (CT) and Magnetic Resonance Imaging (MR) provide a good basis for analyses and predictions of spatio-temporal movements of particular parts of the human body. 
According to~\cite{Ohnesorge2007}, modern multi-slice CT scanners allow for recording volumes of the human body over a specific time. For example, when examining cardiac diseases, this offers the possibility to record the human thorax over multiple cardiac cycles. By simultaneously recording an electrocardiography trace (ECG), all recorded frames can be sorted retrospectively resulting in a \mbox{3-D+t} data set covering an entire cardiac cycle. Fig.\,\ref{fig:data} shows the resulting representation of such a \mbox{3-D+t} volume consisting of $T$ temporally equidistant spatial \mbox{3-D} volumes of size \mbox{$X{\times}Y{\times}Z$}.

Data rates of such systems depend on the number of detector rows and gantry rotation times. According to~\cite{Ulzheimer2009}, a $4$-slice CT system with $0.5$ seconds rotation time roughly generates $5.6$\,MB of data per rotation. This corresponds to $11.2$\,MB/s, while a 16-slice CT scanner with the same rotation time generates already 45\,MB/s, and a 64-slice CT system produces up to $200$\,MB/s.

This huge amount of data is challenging for data transmission off the gantry and for archiving the \mbox{3-D+t} volumes. According to~\cite{Doukas:2008:ATM:1556147.1362837}, various telemedicine applications exist, including remote surgery systems, patient remote telemonitoring facilities, and transmission of medical content for remote assessment. The exchange of medical content between medical experts for educational purposes is also of high relevance. Since lossless reconstruction is a crucial condition in medical environments, a lossless scalable representation is advantageous which allows for tasks like browsing and fast previewing. 
Two additional properties need to be taken into account for lossless compression. Firstly, medical volume data contain a lot of sensor noise. For CT, this is caused by the radiation, which has to be kept low to reduce the risks for the patients, as well as the short acquisition time which is kept as low as possible to avoid motion artifacts. According to~\cite{McVeigh1985}, the noise in an MR imaging system arises from the resistance of the coil, dielectric and inductive losses in the sample, and the preamplifier.
Secondly, medical image data have typically a bit depth of 12 bits per pixel whereas natural video sequences originating from the entertainment industry are commonly stored in 8 bit format. 

All these conditions require an appropriate coding scheme for dynamic volume data. One way is to apply common video codecs to medical dynamic volumes by generating a single sequence over $t$ for every slice position $z$, resulting in $Z$ temporal sequences. In Fig.\,\ref{fig:data}, the resulting sequence for slice position $z{=}3$ is highlighted. Every single temporal sequence can then be compressed using common video coding schemes under the condition of lossless compression and adaption of the range of the bit depth.

\begin{figure}[t]
	\centering
	\begin{tikzpicture}[scale=0.69, >=latex'] 
	\input{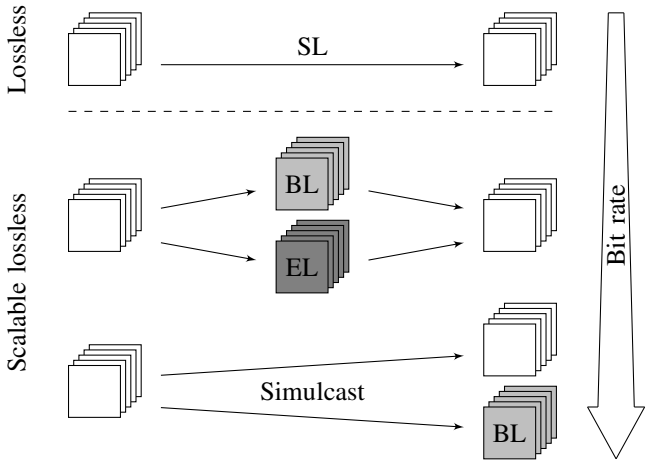}
	\tikzstyle{box} = [draw]
	\draw[fill=white] (0.4,2.6) rectangle (1.4,3.6);
	\draw[fill=white] (0.3,2.5) rectangle (1.3,3.5);
	\draw[fill=white] (0.2,2.4) rectangle (1.2,3.4);
	\draw[fill=white] (0.1,2.3) rectangle (1.1,3.3);
	\draw[fill=white] (0,2.2) rectangle (1,3.2);
	\draw[fill=white] (8.4,2.6) rectangle (9.4,3.6);
	\draw[fill=white] (8.3,2.5) rectangle (9.3,3.5);
	\draw[fill=white] (8.2,2.4) rectangle (9.2,3.4);
	\draw[fill=white] (8.1,2.3) rectangle (9.1,3.3);
	\draw[fill=white] (8,2.2) rectangle (9,3.2);
	\node (n_sl_orig) at (1.6,2.6) {};
	\node (n_sl_rec) at (7.8,2.6) {};
	\draw[->] (n_sl_orig) -- node[above]{SL}(n_sl_rec);
	\draw[fill=white] (0.4,0.4) rectangle (1.4,-0.6);
	\draw[fill=white] (0.3,0.3) rectangle (1.3,-0.7);
	\draw[fill=white] (0.2,0.2) rectangle (1.2,-0.8);
	\draw[fill=white] (0.1,0.1) rectangle (1.1,-0.9);
	\draw[fill=white] (0,0) rectangle (1,-1);	
	\draw[fill=lightgray] (4.4,0.2) rectangle (5.4,1.2);
	\draw[fill=lightgray] (4.3,0.1) rectangle (5.3,1.1);
	\draw[fill=lightgray] (4.2,0) rectangle (5.2,1);
	\draw[fill=lightgray] (4.1,-0.1) rectangle (5.1,0.9);
	\draw[fill=lightgray] (4,-0.2) rectangle (5,0.8);
	\draw[fill=gray] (4.4,-1.4) rectangle (5.4,-0.4);
	\draw[fill=gray] (4.3,-1.5) rectangle (5.3,-0.5);
	\draw[fill=gray] (4.2,-1.6) rectangle (5.2,-0.6);
	\draw[fill=gray] (4.1,-1.7) rectangle (5.1,-0.7);
	\draw[fill=gray] (4,-1.8) rectangle (5,-0.8);
	\draw[fill=white] (8.4,0.4) rectangle (9.4,-0.6);
	\draw[fill=white] (8.3,0.3) rectangle (9.3,-0.7);
	\draw[fill=white] (8.2,0.2) rectangle (9.2,-0.8);
	\draw[fill=white] (8.1,0.1) rectangle (9.1,-0.9);
	\draw[fill=white] (8,0) rectangle (9,-1);
	\node (bl) at (4.5,0.3) {BL};
	\node (bl) at (4.5,-1.3) {EL};
	\node (ml_orig) at (1.6,-0.2) {};
	\node (ml_orig2) at (1.6,-0.8) {};
	\node (ml_bl) at (3.8,0.2) {};
	\node (ml_bl_end) at (5.6,0.2) {};
	\node (ml_el) at (3.8,-1.2) {};
	\node (ml_el_end) at (5.6,-1.2) {};
	\node (ml_rec) at (7.8,-0.2) {};
	\node (ml_rec2) at (7.8,-0.8) {};
	\draw[->] (ml_orig) -- (ml_bl);
	\draw[->] (ml_bl_end) -- (ml_rec);
	\draw[->] (ml_orig2) -- (ml_el);
	\draw[->] (ml_el_end) -- (ml_rec2);
	\draw[fill=white] (0.4,-3.8) rectangle (1.4,-2.8);
	\draw[fill=white] (0.3,-3.9) rectangle (1.3,-2.9);
	\draw[fill=white] (0.2,-4) rectangle (1.2,-3);
	\draw[fill=white] (0.1,-4.1) rectangle (1.1,-3.1);
	\draw[fill=white] (0,-4.2) rectangle (1,-3.2);	
	\draw[fill=lightgray] (8.4,-4.6) rectangle (9.4,-3.6);
	\draw[fill=lightgray] (8.3,-4.7) rectangle (9.3,-3.7);
	\draw[fill=lightgray] (8.2,-4.8) rectangle (9.2,-3.8);
	\draw[fill=lightgray] (8.1,-4.9) rectangle (9.1,-3.9);
	\draw[fill=lightgray] (8,-5) rectangle (9,-4);
	\draw[fill=white] (8.4,-3.0) rectangle (9.4,-2.0);
	\draw[fill=white] (8.3,-3.1) rectangle (9.3,-2.1);
	\draw[fill=white] (8.2,-3.2) rectangle (9.2,-2.2);
	\draw[fill=white] (8.1,-3.3) rectangle (9.1,-2.3);
	\draw[fill=white] (8,-3.4) rectangle (9,-2.4);
	\node (bl) at (8.5,-4.5) {BL};
	\node (slc_orig) at (1.6,-3.4) {};
	\node (slc_orig2) at (1.6,-4.0) {};
	\node (slc_b1) at (7.8,-3.0) {};
	\node (slc_b2) at (7.8,-4.4) {};
	\node (slc) at (4.7,-3.7) {Simulcast};
	\draw[->] (slc_orig) -- (slc_b1);
	\draw[->] (slc_orig2) -- (slc_b2);
	\draw[dashed] (0,1.7) -- (9.4,1.7);
	\node[rotate=90] (lossless) at (-1,2.9) {Lossless};
	\node[rotate=90] (scal) at (-1,-1.6) {Scalable lossless};
	\coordinate (ul) at (10.4,3.6);
	\coordinate (ur) at (10.7,3.6);
	\coordinate (bl) at (10.2,-4);
	\coordinate (br) at (10.9,-4);
	\coordinate (l) at (10,-4);
	\coordinate (r) at (11.1,-4);
	\coordinate (b) at (10.55,-5);
	\draw[-] (ul) -- (ur);
	\draw[-] (ur) -- (br);
	\draw[-] (br) -- (r);
	\draw[-] (r) -- (b);
	\draw[-] (b) -- (l);
	\draw[-] (l) -- (bl);
	\draw[-] (bl) -- (ul);
	\node[rotate=90] (scal) at (10.55,-0.5) {Bit rate};
	\end{tikzpicture}
	\caption{Different approaches for lossless coding of video data. While SL coding typically costs less bits than scalable approaches, EL coding and simulcast provide scalability features.}
	\label{fig:model}
	\vspace{-2pt}
\end{figure}

Fig.\,\ref{fig:model} describes the basic concepts for lossless compression of video data. Lossless single-layer (SL) coding represents the most direct way. Beside conventional still image compression schemes allowing lossless compression, like JPEG-LS~\cite{855427}, also motion compensated predictive coders, like the HEVC codec~\cite{2016} in lossless mode, can be applied. 
Although SL video coding typically costs less bits compared to scalable lossless video coding~\cite{7172510}, the time needed for transmission of an entire medical volume, considering for example any wireless network, may still be high. Therefore, scalable lossless video coding would be preferable, where videos are coded in multiple layers and each layer represents a different quality representation of the same video scene.

Basically, three different types of video scalability can be distinguished. Temporal scalability affects the frame rate of the video, spatial scalability controls the spatial resolution of the video, and quality scalability manipulates the fidelity of the video. 

In this context, two different approaches of scalable lossless video coding can be differentiated: Encoding a video separately at different bit rates and transmitting all layers simultaneously is called simulcast. In Fig.\,\ref{fig:model}, one base layer (BL), which is of lower quality, is transmitted in addition to the original video sequence. This BL is encoded with less bits and can be sent, if the physical channel is limited. Obviously, the required bit rate for transmitting both layers is always higher than SL coding.

The second approach is characterized by transmitting in addition to the BL, one or more enhancement layers (EL), which provide additional data necessary for lossless reconstruction of the coded video sequence. In Fig.\,\ref{fig:model}, the residual data is transmitted by one EL. Typical approaches based on this concept are Scalable High Efficiency Video Coding (SHVC)~\cite{Chen2014} and Sample-Based Weighted Prediction for Enhancement Layer Coding (SELC)~\cite{lnt2017-42,lnt2014-20}. Both approaches employ conventional DCT-based hybrid coding schemes for encoding  the BL and differ regarding the encoding of the EL. An alternative coding scheme is represented by 3-D subband coding (SBC)~\cite{Karlsson1988}. A wavelet-based approach naturally achieves scalability features without additional overhead~\cite{lnt2011-23}. In every transformation step, the signal is decomposed into a lowpass (LP) and a highpass (HP) subband with the energy concentrated in the LP subband. The HP subband contains the structural information of the video sequence and corresponds therefore to the EL. Accordingly, the LP subband serves as the BL. Both layers offer only half the frame rate compared to the original sequence. If both layers are coded for example by the wavelet-based volume coder JPEG\,2000~\cite{ITU-T-800}, all of the above mentioned scalability types are supported by SBC. 

Since data fidelity is a very important aspect in the medical environment, a high visual quality of the BL is of huge relevance. However, SBC causes ghosting artifacts in the LP subband due to temporal displacements in the sequence. These can be reduced by incorporating motion compensation (MC) methods into the subband coder. This is called Motion Compensated Temporal Filtering (MCTF) and was introduced by Ohm in~\cite{334985}. Beside block-based MC, as investigated in~\cite{lnt2012-40}, also mesh-based MC can successfully be applied to medical volume data. The latter one is optimized for medical volume data in~\cite{8066388}, by performing one compensated Haar wavelet transform (WT) in temporal direction. The results are promising regarding the visual quality of the LP subband and the compression ratio of the entire coding scheme. However, a further improvement of the data fidelity without a major increase of the required bit rate is desirable. By applying graph-based MC as introduced in~\cite{Lanz2016} instead of mesh-based MC in the lifting structure of the MCTF coder, the geometric structure of the data can directly be incorporated into the WT~\cite{shuman2013emerging} and allows for a superior visual quality of the LP subband.
The motion information used for prediction in graph-based MC is stored in adjacency matrices. However, appropriate ways to encode the resulting prediction matrices are not investigated so far.

In this article, we introduce a novel method for encoding the adjacency matrices based on the construction of motion maps. To make this approach feasible for data compression, we apply sparse sampling to these motion maps. After a scanning operation, the resulting symbol streams are entropy coded using multiple-context adaptive arithmetic coding~\cite{Witten:1987}. 
An important step within the lifting structure is the update step, where the MC has to be inverted. So far, inversion has been done by recalculating the required adjacency matrix. However, this leads to a rising bit rate, since this update matrix also has to be transmitted. To omit this additional overhead, we suggest to estimate the update matrix from the already encoded prediction matrix. This can be done by applying the specification of the optimum update for motion-compensated lifting as introduced in~\cite{1381473}. 

This article is outlined as follows. After a short review on compensated wavelet lifting in Section~\ref{subsec:compLifting}, Section~\ref{subsec:graph} provides an overview of graph-based MC, followed by the description of the inversion of graph-based MC in Section~\ref{subsec:inversion}. The proposed coding scheme is presented in Section~\ref{sec:prop} including the construction of motion maps and the encoding process. Simulation results are presented in Section~\ref{sec:results}. In Section~\ref{conclusion}, a conclusion and outlook is given.

\section{Graph-Based Compensated Wavelet Lifting}
\label{sec:lifting}

\begin{figure}[t]
	\centering
	\begin{tikzpicture}[scale=1, >=latex'] 
	\input{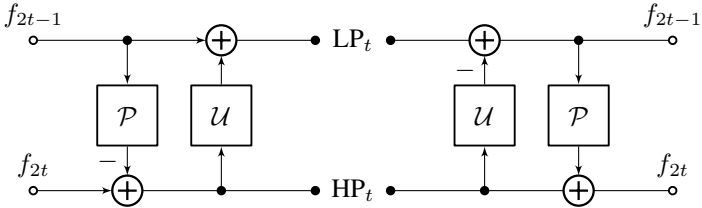}
	\tikzstyle{box} = [draw]
	\coordinate (c1) at (0,0);
	\coordinate (c2) at (0,-2);	
	\coordinate (c3) at (8.5,0);
	\coordinate (c4) at (8.5,-2);
	\node[dspnodeopen,dsp/label=above] (n1) at (c1) {$f_{2t-1}$}; 
	\node[dspnodeopen,dsp/label=above] (n2) at (c2) {$f_{2t}$}; 
	\node[dspnodeopen,dsp/label=above] (n3) at (c3) {$f_{2t-1}$}; 
	\node[dspnodeopen,dsp/label=above] (n4) at (c4) {$f_{2t}$};
	\node[dspadder] (add1) at (2.5,0) {};
	\node[dspadder] (add2) at (6,0) {};
	\node[dspadder,dsp/label=below] (add3) at (1.25,-2) {};
	\node[dspadder] (add4) at (7.25,-2) {};
	\node[dspsquare](b1) at (1.25,-1) {$\mathcal{P}$};
	\node[dspsquare](b2) at (2.5,-1) {$\mathcal{U}$};
	\node[dspsquare](b3) at (7.25,-1) {$\mathcal{P}$};
	\node[dspsquare](b4) at (6,-1) {$\mathcal{U}$};
	\node[dspnodefull](n5) at (1.25,0) {};
	\node[dspnodefull](n6) at (2.5,-2) {};
	\node[dspnodefull](n7) at (6,-2) {};
	\node[dspnodefull](n8) at (7.25,0) {};
	\node (n9) at (4.25,0) {$\text{LP}_t$};
	\node (n10) at (4.25,-2) {$\text{HP}_t$};
	\node[dspnodefull] (n11) at (3.75,0) {};
	\node[dspnodefull] (n12) at (3.75,-2) {};
	\node[dspnodefull] (n13) at (4.75,0) {};
	\node[dspnodefull] (n14) at (4.75,-2) {};
	\draw[-] (n1) -- (n5);
	\draw[->] (n5) -- (b1);
	\draw[->] (n2) -- (add3);
	\draw[->] (b1) -- node[left]{$-$}(add3);
	\draw[->] (n5) -- (add1);
	\draw[->] (b2) -- (add1);
	\draw[->] (n6) -- (b2);
	\draw[-] (add3) -- (n6);
	\draw[-] (add1) -- (n11);
	\draw[-] (n6) -- (n12);
	\draw[-] (n13) -- (add2);
	\draw[-] (add2) -- (n8);
	\draw[-] (n8) -- (n3);
	\draw[-] (n14) -- (n7);
	\draw[-] (n7) -- (add4);
	\draw[-] (add4) -- (n4);
	\draw[->] (n7) -- (b4);
	\draw[->] (b4) -- node[left]{$-$}(add2);
	\draw[->] (n8) -- (b3);
	\draw[->] (b3) -- (add4);
	\end{tikzpicture}
	\caption{Block diagram of the lifting structure.}
	\label{fig:lifting}
	\vspace{-5pt}
\end{figure}

An efficient implementation of the discrete WT was proposed by Swelden~\cite{Sweldens1995}. The so-called lifting structure consists of three steps: split, predict, and update. The principal scheme is to decorrelate data quickly by using polynomial factorizations instead of employing Fourier Transforms. Fig.\,\ref{fig:lifting} shows a block diagram of the lifting structure. In the first step, splitting is performed by decomposing the input video signal into even- and odd-indexed frames $f_{2t}$ and $f_{2t-1}$. In a second step, the even frames are predicted from the odd frames by a prediction operator $\mathcal{P}$ which is independent of the data to get a more compact representation of the input signal. Subtracting the predicted value $\mathcal{P}(f_{2t-1})$ from the even frames, results in the HP frames. If the prediction works well, the HP frames will contain less information than the original signal. To maintain global properties of the original video signal, in an update step, the HP frames are filtered by an update operator $\mathcal{U}$ and are added back to the odd frames, resulting in the LP frames. Thus, the LP subband can be used as a downscaled representative for the original video sequence, corresponding to a BL with lower temporal resolution and different visual quality.
Accordingly, the HP and LP coefficients are generated by
\begin{equation}
	\begin{aligned}
	\text{HP}_t &= f_{2t}-\mathcal{P}(f_{2t-1}), \\
	\text{LP}_t &= f_{2t-1}+\mathcal{U}(\text{HP}_t).
	\end{aligned}
\end{equation}
The inversion of the lifting structure works straightforward. After reversing the prediction and update operations and toggle $+$ and $-$ signs, the original video signal can easily be regenerated by
\begin{equation}
	\begin{aligned}
	f_{2t-1} &= \text{LP}_t - \mathcal{U}(\text{HP}_t), \\
	f_{2t} &= \text{HP}_t +\mathcal{P}(f_{2t-1}).
	\end{aligned}
\end{equation}
Hereby, the lifting structure offers a flexible framework which can be modified in multiple ways. Common wavelet filters often have floating point coefficients. By introducing rounding operators as introduced in~\cite{647983}, integer to integer transforms can be achieved. Hence, perfect reconstruction can be guaranteed which makes the WT highly attractive for telemedicine applications by offering a scalable representation and lossless reconstruction at the same time.

In case of the Haar wavelet, the HP coefficients are computed by taking the difference of $f_{2t}$ and $f_{2t-1}$. The LP coefficients are achieved by calculating the average of these two frames $\frac{f_{2t-1}+f_{2t}}{2} = f_{2t-1} + \frac{1}{2}\text{HP}_t$. Applying rounding operators to ensure lossless reconstruction, leads to
\begin{equation}
	\begin{aligned}
	\text{HP}_t &= f_{2t}-f_{2t-1},\\
	\text{LP}_t &= f_{2t-1}+\left \lfloor \frac{1}{2}\text{HP}_t \right\rfloor.
	\end{aligned}
\end{equation}

However, blurriness and ghosting artifacts will appear in the LP frames due to displacements over time. This can be alleviated by incorporating compensation methods directly into the lifting structure without losing the perfect reconstruction property.

\subsection{Compensated Wavelet Lifting}
\label{subsec:compLifting} 

A compensated wavelet transform in temporal direction is called Motion Compensated Temporal Filtering (MCTF)~\cite{334985}. 

To achieve a compensated transform, the prediction \mbox{operator $\mathcal{P}$} is realized by the warping operator $\mathcal{W}_{2t-1\rightarrow2t}$. Instead of the original odd frames $f_{2t-1}$, a compensated version is subtracted from the even frames $f_{2t}$. Then, the prediction step of the Haar transform is given by
\begin{align}
\text{HP}_t &= f_{2t}-\lfloor\mathcal{W}_{2t-1\rightarrow2t}(f_{2t-1})\rfloor. \label{eq:HP_mc}
\end{align}  
To preserve perfect reconstruction, rounding operators are set. However, to achieve an equivalent wavelet transform, the compensation has to be inverted in the update step~\cite{bozinovic2005}. By reversing the index of $\mathcal{W}$, the LP coefficients of the Haar transform can be calculated by
\begin{equation}
\text{LP}_t = f_{2t-1}+\left\lfloor\frac{1}{2}\mathcal{W}_{2t\rightarrow2t-1}(\text{HP}_t)\right\rfloor. \label{eq:LP_mc}
\end{equation} 
To guarantee lossless reconstruction, rounding operators are required. If $\mathcal{W}$ is realized by a graph-based MC, the geometric structure of the data is directly incorporated, providing a high-performing motion compensation.

\subsection{Graph-Based Motion Compensation}
\label{subsec:graph}

\begin{figure}[t]
	\begin{tikzpicture}[scale=1, >=latex'] 
		\input{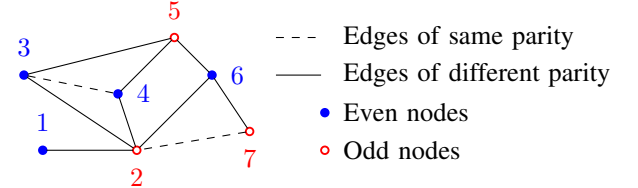}
		\tikzstyle{box} = [draw]		
		\node[dspnodefull, white] (n0) at (0,1) {};
		\node[dspnodefull, blue] (n1) at (1.25,1.5) {$1$};
		\node[dspnodeopen, red, dsp/label=below] (n2) at (2.5,1.5) {$2$};
		\node[dspnodeopen, red] (n3) at (3,3) {$5$};
		\node[dspnodeopen, red, dsp/label=below] (n4) at (4,1.75) {$7$};
		\node[dspnodefull, blue, dsp/label=right] (n5) at (3.5,2.5) {$6$};
		\node[dspnodefull, blue] (n6) at (1,2.5) {$3$};
		\node[dspnodefull, blue, dsp/label=right] (n7) at (2.25,2.25) {$4$};
		\node[dspnodefull, blue, dsp/label=right] (n8) at (5,2) {\textcolor{black}{Even nodes}};
		\node[dspnodeopen, red, dsp/label=right] (n9) at (5,1.5) {\textcolor{black}{Odd nodes}};
		\node[dspnodefull, white, dsp/label=right] (n10) at (4.3,3) {};
		\node[dspnodefull, white, dsp/label=right] (n11) at (5,3) {\textcolor{black}{Edges of same parity}};
		\node[dspnodefull, white, dsp/label=right] (n12) at (4.3,2.5) {};
		\node[dspnodefull, white, dsp/label=right] (n13) at (5,2.5) {\textcolor{black}{Edges of different parity}};
		\draw[-] (n1) -- (n2);
		\draw[-] (n2) -- (n6);
		\draw[-] (n2) -- (n7);
		\draw[-] (n7) -- (n3);
		\draw[-] (n3) -- (n6);
		\draw[-,dashed] (n6) -- (n7);
		\draw[-] (n3) -- (n5);
		\draw[-] (n5) -- (n4);
		\draw[-,dashed] (n4) -- (n2);
		\draw[-] (n2) -- (n5);
		\draw[-] (n12) -- (n13);
		\draw[-,dashed] (n10) -- (n11);
	\end{tikzpicture}	
	\caption{Arbitrary graph showing a possible splitting into even and odd subsets is shown.}
	\label{fig:graphs}
	\vspace{-5pt}
\end{figure}

\begin{figure*}[tb]
	\centering
	\psfragscanon	
	\psfrag{nz}{$n_z {=} 169$}
	\psfrag{n}{$n_z {=} 25$}
	\psfrag{0}{$0$}
	\psfrag{5}{$5$}
	\psfrag{10}{$10$}
	\psfrag{15}{$15$}
	\psfrag{20}{$20$}
	\psfrag{25}{$25$}
	\psfrag{r}{$r{=}1$}
	\psfrag{i}{Start node $i$}
	\psfrag{I}{\rotatebox{180}{Start nodes $i$}}
	\psfrag{d1}{\textcolor{black}{$1$}}
	\psfrag{d2}{\textcolor{black}{$2$}}
	\psfrag{d3}{\textcolor{black}{$3$}}
	\psfrag{d4}{\textcolor{black}{$4$}}
	\psfrag{d5}{\textcolor{black}{$5$}}
	\psfrag{d6}{\textcolor{black}{$6$}}
	\psfrag{d7}{\textcolor{black}{$7$}}
	\psfrag{d8}{\textcolor{black}{$8$}}
	\psfrag{d9}{\textcolor{black}{$9$}}
	\psfrag{j}{Possible end nodes}
	\psfrag{J}{End nodes $j$}	
	\subfloat[Node assignment for graph signals on videos with a predefined neighborhood with radius $r {=} 1$. One single node $i$ of the even frame can get linked to $9$ different end nodes $j$ of the odd frame.]{	
		\begin{tikzpicture}[scale=1, >=latex'] 
		\input{tikzlibrarydsp}
		\tikzstyle{box} = [draw]		
		\node[dspnodefull, white] (n0) at (1.5,-2.75) {};
		\node[dspnodefull, white] (n_even) at (5.5,1) {\textcolor{black}{$f_{2t}$}};
		\node[dspnodefull, white] (n_odd) at (2.5,1) {\textcolor{black}{$f_{2t-1}$}};	
		\node[dspnodefull, white] (n_even) at (5.5,1.5) {\textcolor{black}{Even frame}};
		\node[dspnodefull, white] (n_odd) at (2.5,1.5) {\textcolor{black}{Odd frame}};
		\pgftransformcm{1}{0.2}{0}{1}{\pgfpoint{0cm}{0cm}}{}
		\draw [decorate,decoration={brace,amplitude=2pt}] (1.5,0.125) -- (2,0.125) node [black,above,xshift=-0.75em,yshift=0em] {$r{=}1$};
		\draw[step=0.5,gray,thin,dotted] (1.49,-2) grid (3.5,0);	
		\foreach \x in {1.5,2,...,3.5} {
			\foreach \y in {-2,-1.5,...,0} {
				\draw[red,fill = white,thick] (\x,\y) circle (0.05);
			}
		}	
		\draw[step=0.5,gray,thin,dotted] (4.49,-3) grid (6.5,-1);
		\foreach \x in {4.5,5,...,6.5} {
			\foreach \y in {-3,-2.5,...,-1} {
				\fill[color=blue] (\x,\y) circle (0.05);
			}
		}		
		\draw[-,very thin] (5,-1.5) -- (2,0);
		\draw[-,very thin] (5,-1.5) -- (2,-1);
		\draw[-,very thin] (5,-1.5) -- (2,-0.5);
		\draw[-,very thin] (5,-1.5) -- (1.5,-1);
		\draw[-,very thin] (5,-1.5) -- (1.5,0);
		\draw[-,very thin] (5,-1.5) -- (1.5,-0.5);
		\draw[-,very thin] (5,-1.5) -- (2.5,-1);
		\draw[-,very thin] (5,-1.5) -- (2.5,0);
		\draw[-,very thin] (5,-1.5) -- (2.5,-0.5);
		\node at (1.3,-0.2) {$1$};
		\node at (1.3,-0.7) {$2$};
		\node at (1.3,-1.2) {$3$};
		\node at (1.8,-0.2) {$4$};
		\node at (1.8,-0.7) {$5$};
		\node at (1.8,-1.2) {$6$};
		\node at (2.3,-0.2) {$7$};
		\node at (2.3,-0.7) {$8$};
		\node at (2.3,-1.2) {$9$};
		\end{tikzpicture}	
		\label{fig:posA}
	}\hfill
	\subfloat[Adjacency matrix for a fully connected neighborhood with radius $r {=} 1$ for a frame of size $5\times 5$ pixels. It consists of $n_z{=}169$ nonzero entries distributed over 9 possible diagonals.]{\includegraphics[width=0.3\textwidth]{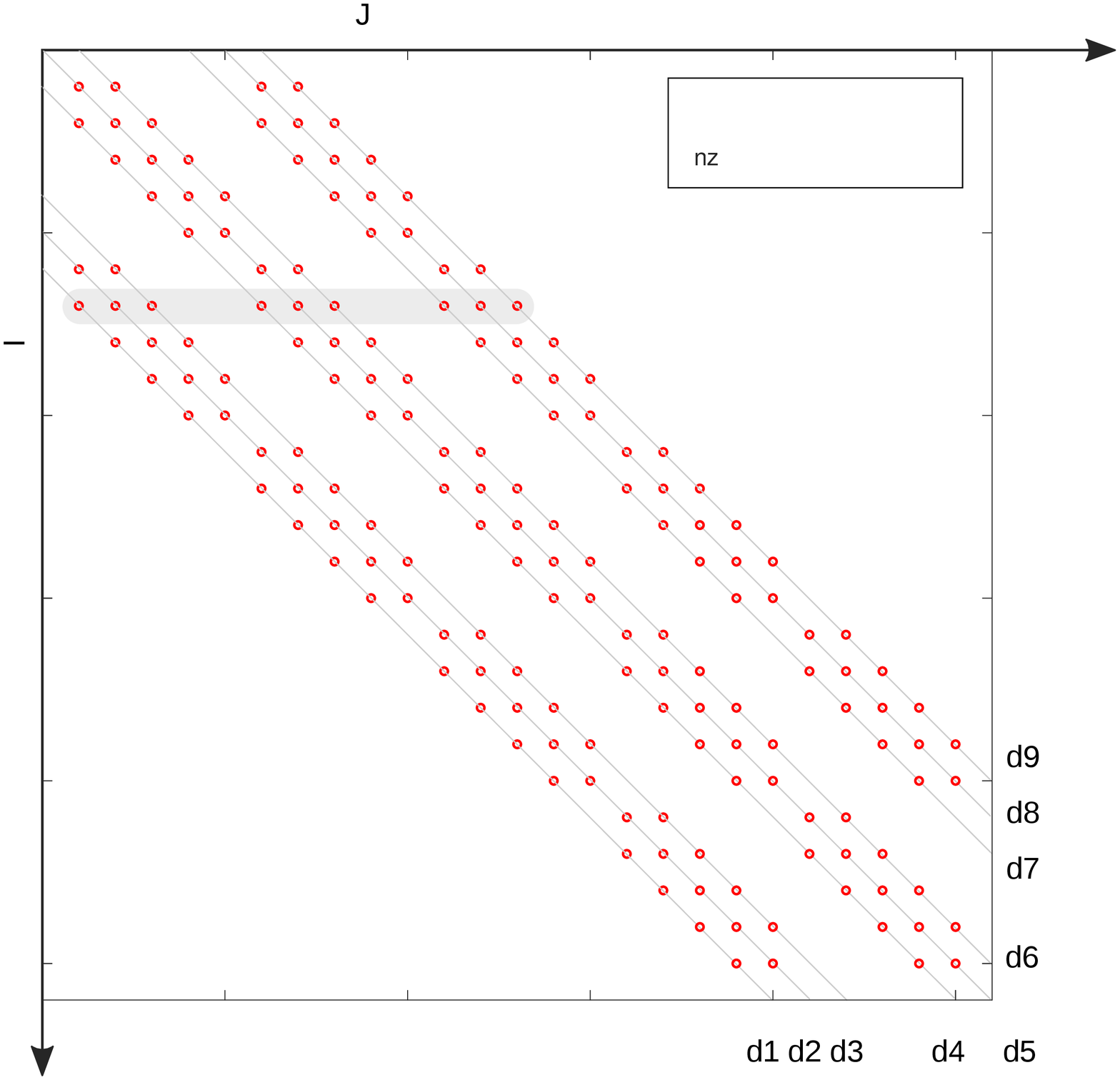}\label{fig:posB}}\hfill
	\subfloat[Adjacency matrix for a reduced neighborhood with radius $r{=}1$ for a frame size of $5\times 5$ pixels. It consists of $n_z{=}25$ nonzero entries distributed over 9 possible diagonals.]{\includegraphics[width=0.3\textwidth]{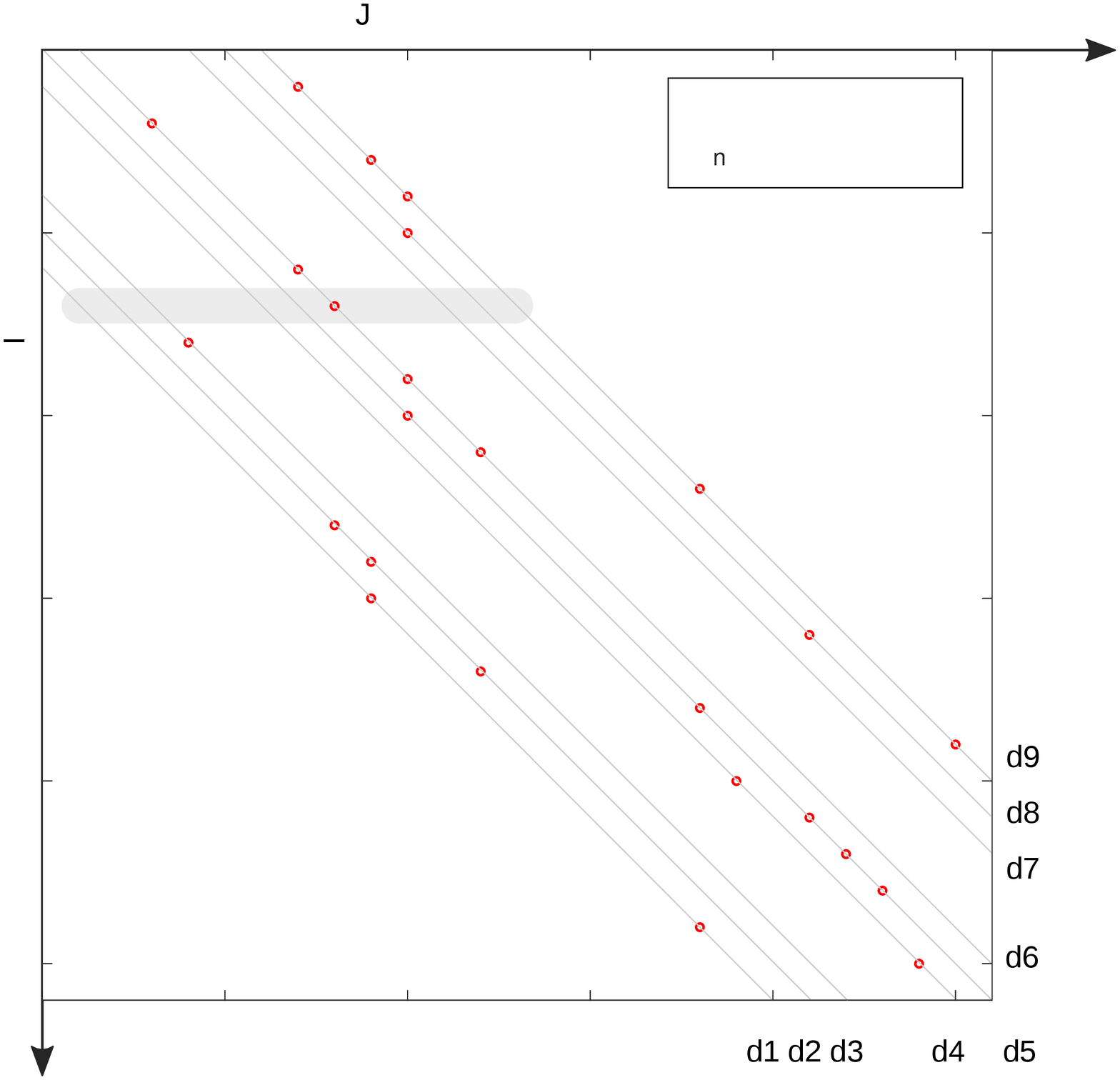}\label{fig:posC}}
	\psfragscanoff
	\caption{Correspondences of positions in a predefined neighborhood and the diagonals in the related adjacency matrices are shown. If only one edge is set for every start node $i$, the number of nonzero entries can be reduced significantly compared to a fully connected neighborhood.}
	\label{fig:pos}
\end{figure*}

In Fig.\,\ref{fig:graphs}, an arbitrary graph $G(\mathcal V,\mathcal{E})$ is shown, where $\mathcal{V}$ is the set of nodes, indexed as $1,2,3,...,N$ and $\mathcal{E}$ is the set of links between the nodes. Every link is defined by a triplet $(i,j,w_{ij}) $, where $i$ and $j$ are the start and end nodes, respectively, and $w_{ij}$ is the weight which has a \mbox{value {$\neq 0$}}, if $i$ and $j$ are linked to each other. These relations are stored in an adjacency matrix $\boldsymbol{A}$. Further, a vector $\boldsymbol{x}$ is introduced which contains the amplitude values of every single node. 
In~\cite{narang2009lifting}, a lifting-based WT on arbitrary graphs is introduced, where the first step comprises the splitting of the nodes into even and odd subsets. As a consequence, the corresponding adjacency matrix $\boldsymbol{A}$ has to be rearranged accordingly:
\begin{equation}
\boldsymbol{x} = \begin{pmatrix}
\boldsymbol{x}_\text{even} \\ 
\boldsymbol{x}_\text{odd}
\end{pmatrix}, \quad 
\boldsymbol{A} = \begin{pmatrix}
\boldsymbol{F} & \boldsymbol{J} \\ \boldsymbol{K} & \boldsymbol{L}
\end{pmatrix}.
\end{equation}
The submatrices $\boldsymbol{F}$ and $\boldsymbol{L}$ contain edges which connect nodes of same parity and the submatrices $\boldsymbol{J}$ and $\boldsymbol{K}$ contain all edges which connect nodes of different parity. Now, the HP and LP coefficients can be calculated by
\begin{equation}
\begin{aligned}
\boldsymbol{H} &= \boldsymbol{x}_\text{even}-\boldsymbol{J_P}\times \boldsymbol{x}_\text{odd},\\
\boldsymbol{L} &= \boldsymbol{x}_\text{odd}+\boldsymbol{K_U}\times \boldsymbol{H},
\end{aligned}
\label{eq:WT}
\end{equation}
where $\boldsymbol{H}$ and $\boldsymbol{L}$ correspond to the vector notation of the HP and LP coefficients.
The matrices $\boldsymbol{J_P}$ and $\boldsymbol{K_U}$ are computed from $\boldsymbol{J}$ and $\boldsymbol{K}$ by assigning prediction and update weights, respectively. Since the matrices $\boldsymbol{F}$ and $\boldsymbol{L}$ are not used in~(\ref{eq:WT}) anymore, a perfect splitting of nodes should be intended~\cite{hidane2013lifting}.

Considering videos as graph signals, as introduced in~\cite{Lanz2016} and shown in Fig.\,\ref{fig:pos} on the left side, and applying the graph-based wavelet transform, every single pixel of a frame corresponds to a node and its associated intensity value is stored in vector $\boldsymbol{x}$. To fulfill the constraint of perfect splitting, every node of an even indexed frame $f_{2t}$ gets solely connected to nodes of the odd indexed frame $f_{2t-1}$. In Fig.\,\ref{fig:posA}, this procedure is shown for one single node using a \mbox{9-grid} neighborhood which corresponds to a radius $r {=} 1$.
The prediction and update weights of $\boldsymbol{J_P}$ and $\boldsymbol{K_U}$ are calculated by measuring the similarity of two connected nodes. The higher the weight, the more similar two linked nodes are. Hence, every node of the even frame $f_{2t}$ is predicted by a weighted average of its assigned neighboring nodes in frame $f_{2t-1}$. 
According to~\cite{Lanz2016}, an increasing radius for the considered neighborhood of a single node results in an increasing visual quality of the LP subband and decreasing mean energy in the HP subband. However, to assure perfect reconstruction, the chosen neighborhood as well as the prediction and update weights have to be known at the decoder side. While the transmission of the corresponding radius of the chosen neighborhood is trivial, the transmission of the prediction and update weights is challenging. 

\subsection{Inversion of the Graph-Based Motion Compensation}
\label{subsec:inversion}
So far, the update step is performed by predicting from $f_{2t-1}$ to $f_{2t}$~\cite{Lanz2016}. The resulting edge weights are stored in $\boldsymbol{K_U}$ and have to be known at decoder side, too. However, concerning the coding costs to transmit all these update weights in addition to the prediction weights, we force a solution which reuses the prediction weights to generate the update weights. The formulation of an optimum update $\mathcal{U_\text{opt}}$ based on a given general linear predictor $\mathcal{P}$ was already provided in~\cite{1381473}. They formulated the closed-form expression
\begin{equation}
\mathcal{U}_\text{opt} = ( \boldsymbol{I}+ \mathcal{P}^T\mathcal{P})^{-1}\mathcal{P}^T
\end{equation}
for motion compensated wavelet lifting, where $\boldsymbol{I}$ describes the identity matrix. 
This formula requires the prediction operator $\mathcal{P}$ to be given in matrix notation. Fortunately, the calculation rules for graph-based MC are already given in matrix and vector notation by equation~(\ref{eq:WT}). Accordingly, $\mathcal{P}$ can directly be substituted by the prediction matrix $\boldsymbol{J_P}$. In addition, they provide a sparse matrix technique that enables the practical implementation of the optimal update step for MCTF, which is also used in this work. This allows us to omit the calculation of $\boldsymbol{K_U}$ and thus also the transmission of the update weights. 

Hence, the remaining task is to find a proper way to encode the graph description by considering the chosen radius and the corresponding prediction weights.

\section{Proposed Coding Scheme}
\label{sec:prop}

A larger radius for the chosen neighborhood results in a better visual quality of the LP frames and a lower mean energy of the HP frames, which can be used as an indicator for the corresponding bit rate needed to encode the HP subband. However, a larger radius comprises also more prediction weights which have to be transmitted. 
Since the prediction weights are highly uncorrelated, common coding schemes like transformation and quantization are not suitable for this application. Therefore, we suggest to reduce the graph description and convert the remaining adjacency matrices to motion maps.

\begin{figure}[t]	
	\captionsetup[subfigure]{justification=centering}
	\begin{center}
		\subfloat[Reference frame $f_{2t-1}$]{\includegraphics[trim = 14cm 2cm 14cm 1cm,clip,width = 0.25\textwidth]{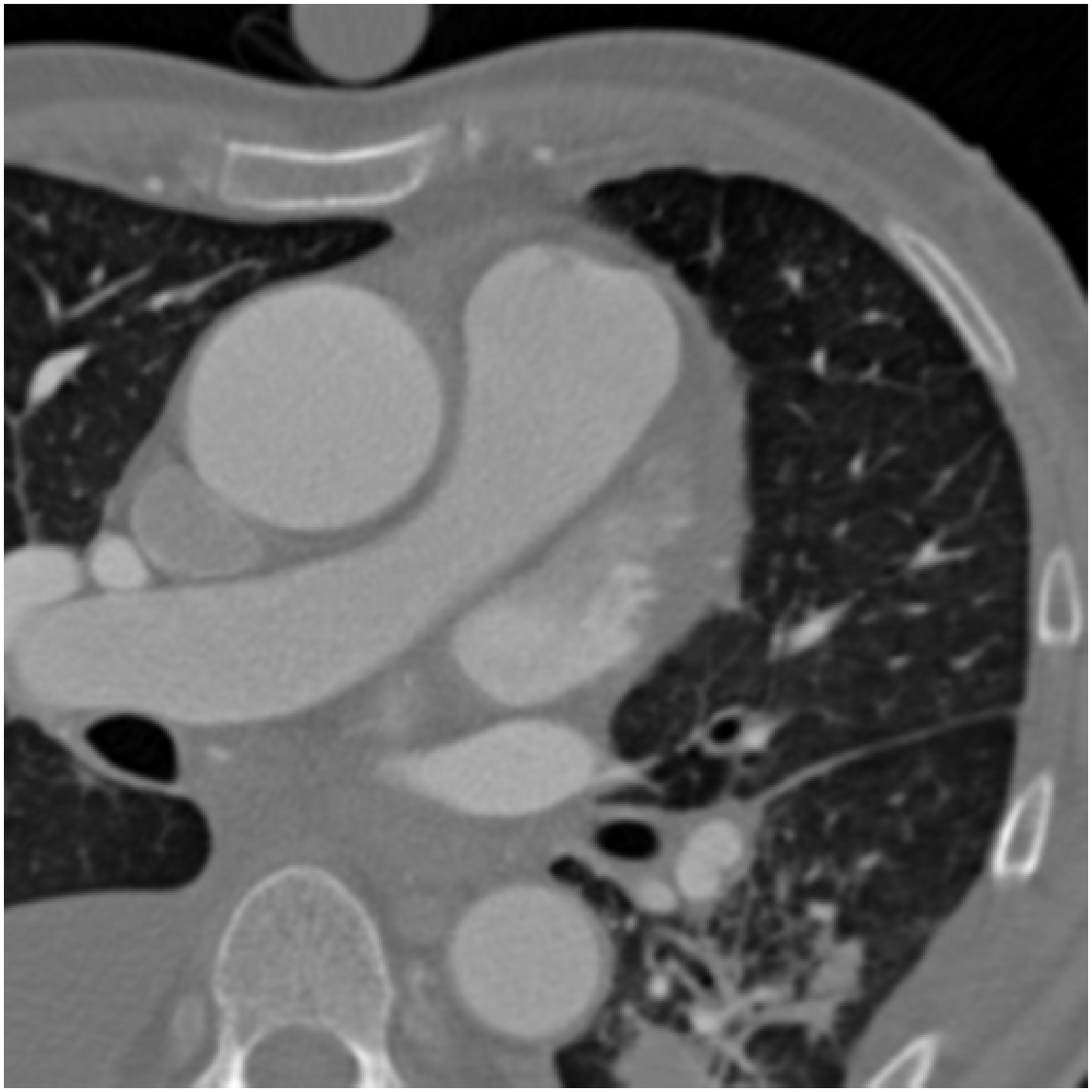}\label{fig:ref}}
		\subfloat[Current frame $f_{2t}$]{\includegraphics[trim = 14cm 2cm 14cm 1cm,clip,width = 0.25\textwidth]{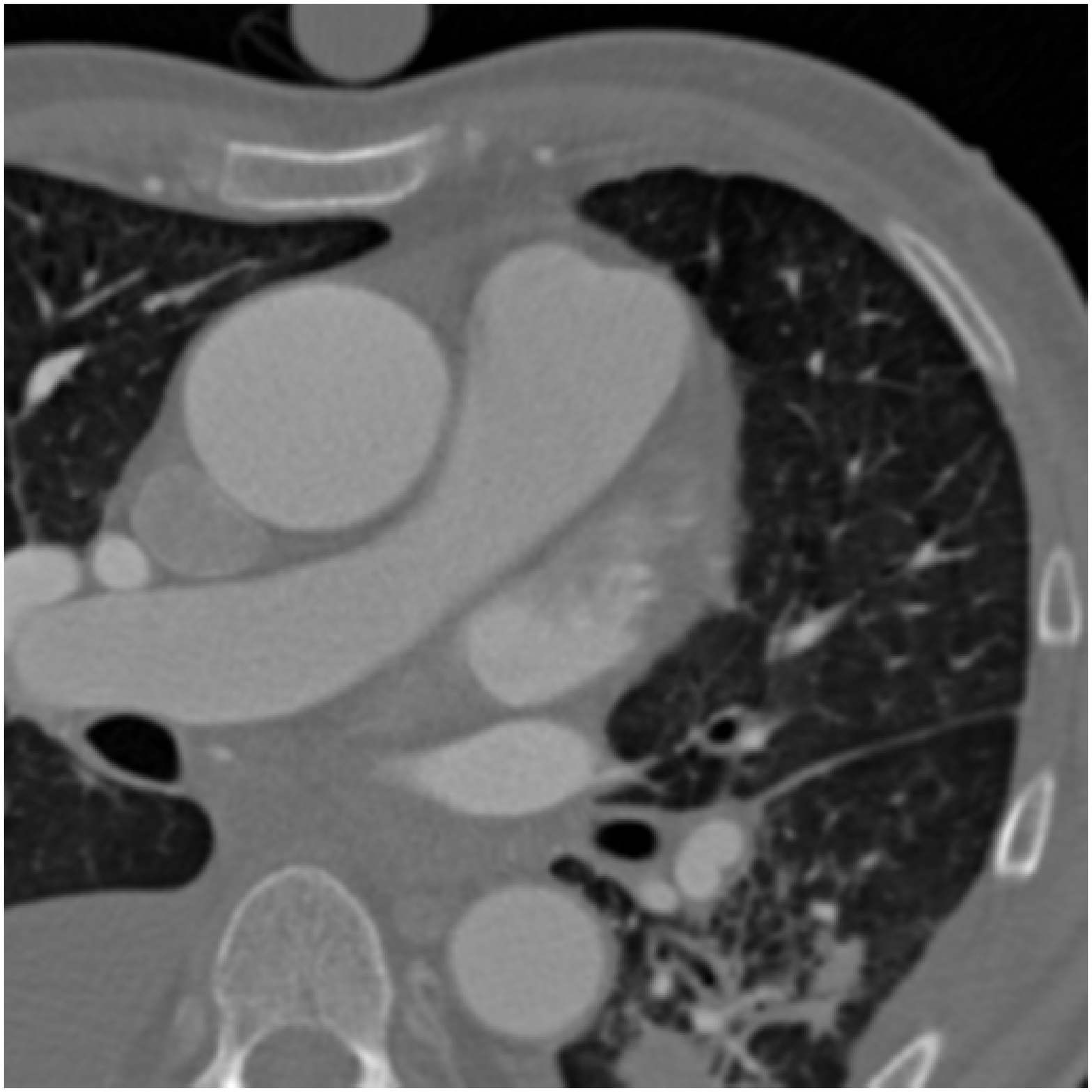}\label{fig:cur}}\par
		\subfloat[$\boldsymbol{M}$ for $r{=}1$ ]{\includegraphics[trim = 14cm 2cm 14cm 1cm,clip,width = 0.25\textwidth]{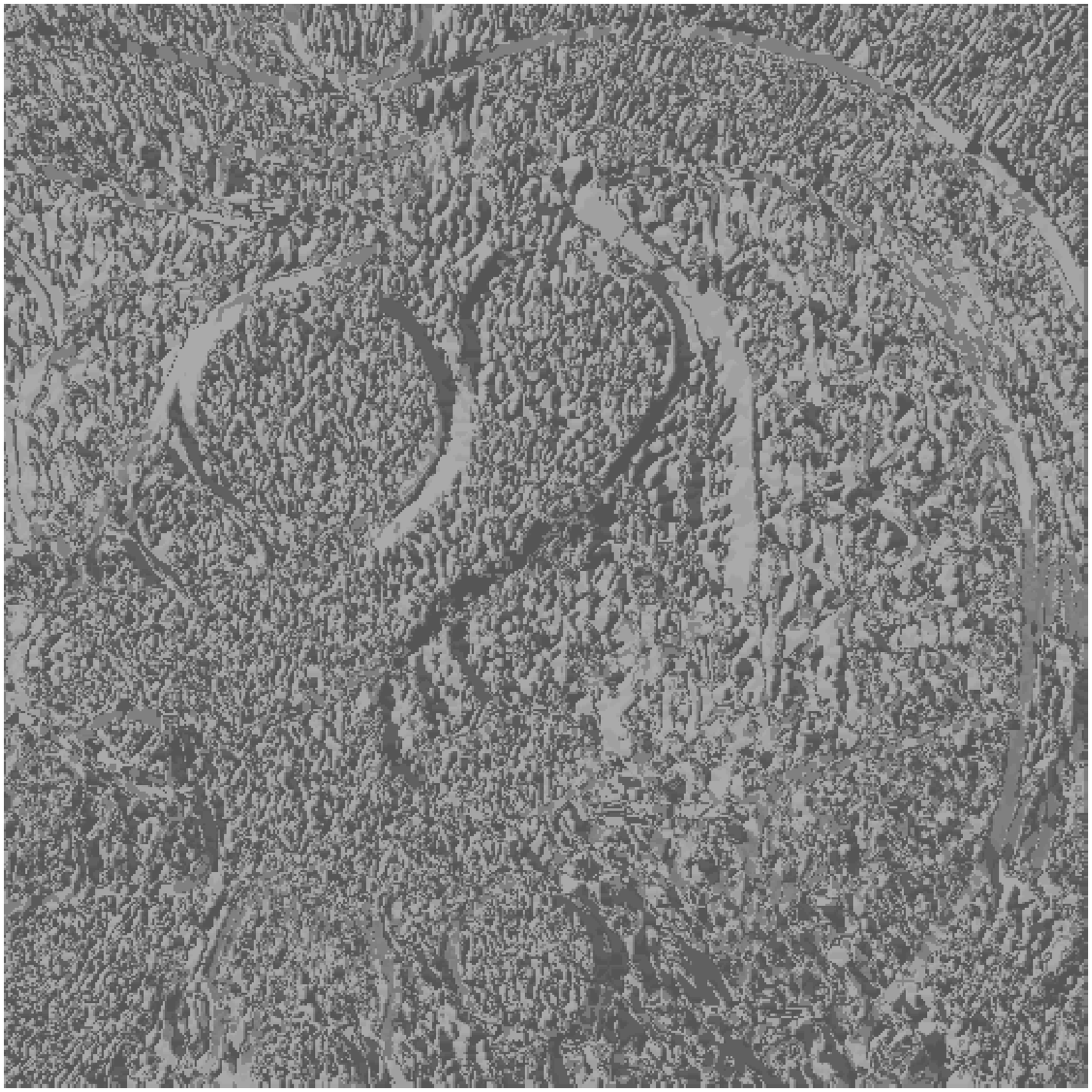}\label{fig:r1}}
		\subfloat[$\boldsymbol{M}$ for $r{=}3$]{\includegraphics[trim = 14cm 2cm 14cm 1cm,clip,width = 0.25\textwidth]{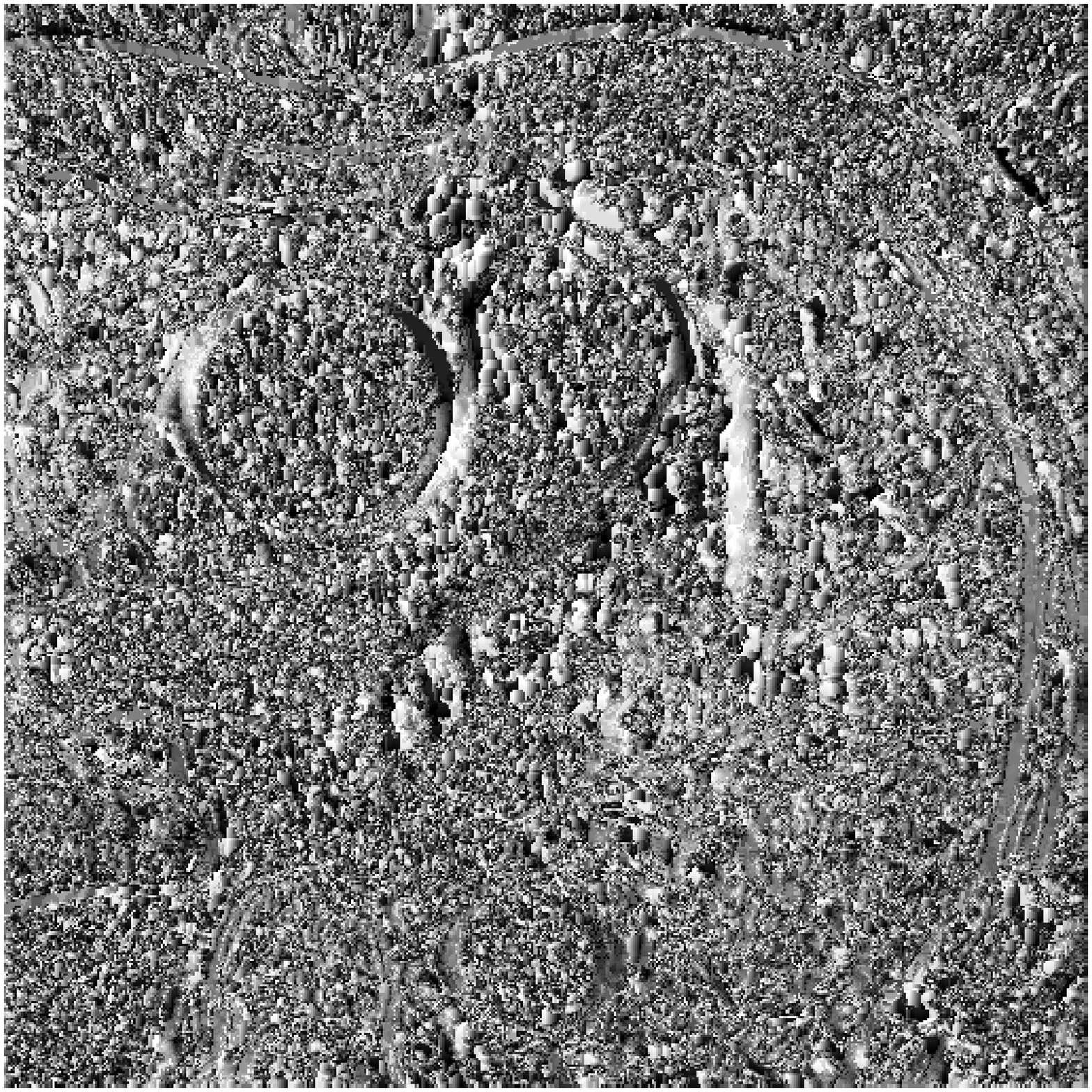}\label{fig:r3}}\par
		\subfloat[Absolute difference $|f_{2t-1}-f_{2t}|$]{\includegraphics[trim=14cm 2cm 14cm 1cm,clip,width=0.25\textwidth]{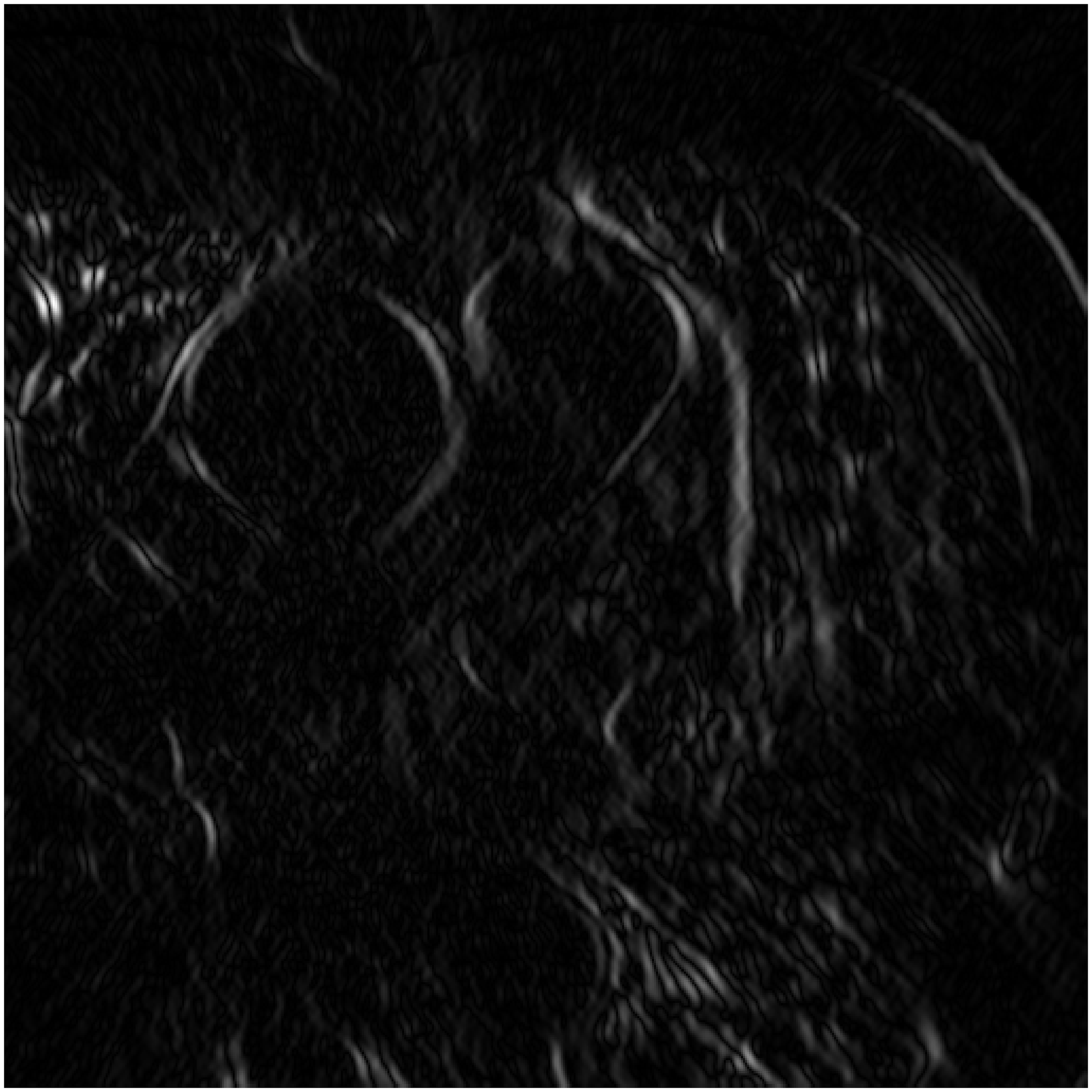}\label{fig:mse_mask}}
		\subfloat[$\boldsymbol{M}$ for $r_{\text{max}}{=}3$]{\includegraphics[trim = 14cm 2cm 14cm 1cm,clip,width = 0.25\textwidth]{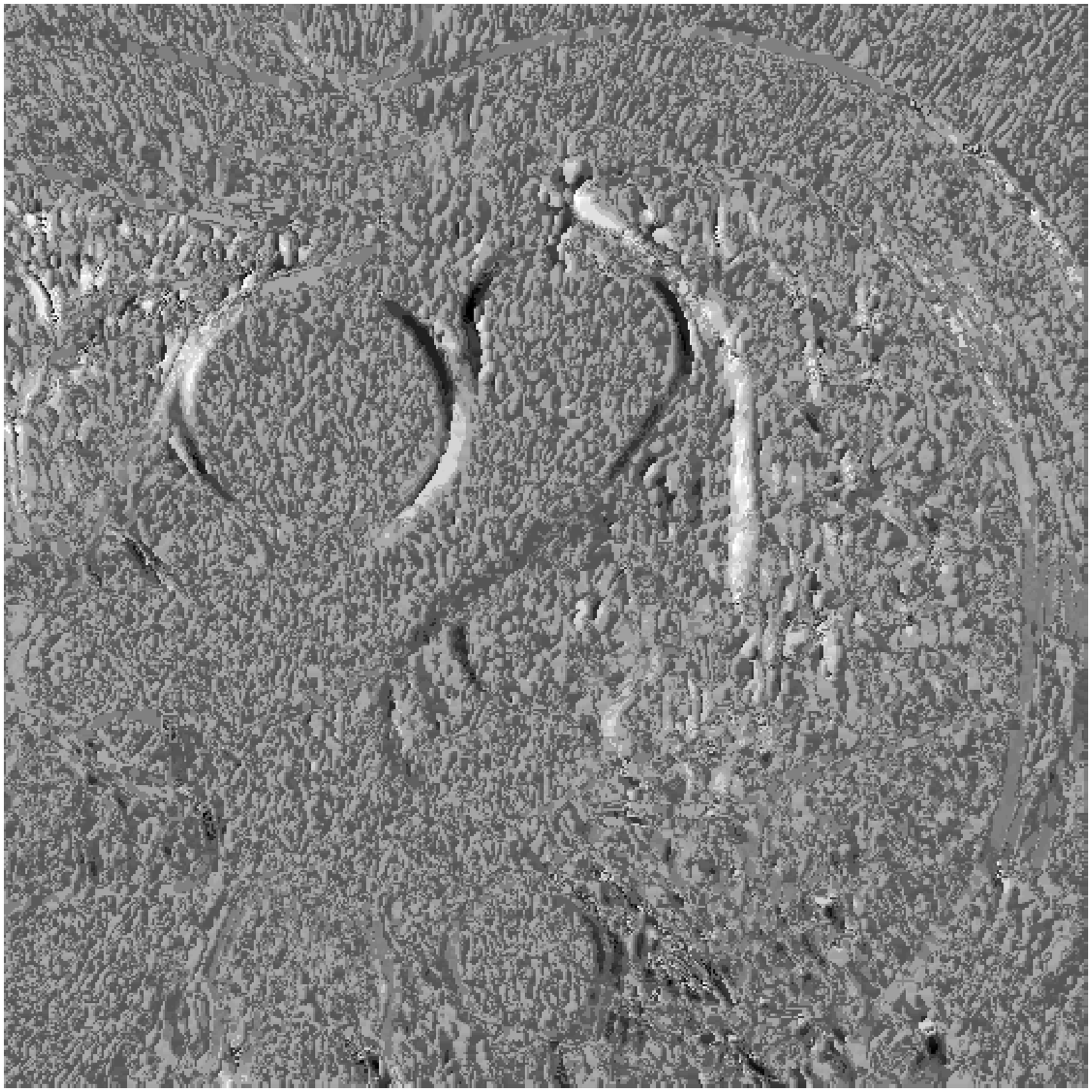}\label{fig:motion_map_rmax3}}
		\caption{Top: two subsequent frames of a CT data set showing a beating heart over time. Middle: corresponding motion maps $\boldsymbol{M}$ for different radii $r$. Bottom: Absolute difference \mbox{$|f_{2t-1}-f_{2t}|$} to determine the assignment of radii and the corresponding smoothed motion map for $r_{\text{max}}{=}3$.}
		\label{fig:motionMaps}
	\end{center}
	\vspace{-0.5cm}
\end{figure}

\begin{figure}[t]
	\centering	
	\psfragscanon
	\subfloat{
		\begin{tikzpicture}[scale=0.8, >=latex'] 
		\input{tikzlibrarydsp}
		\tikzstyle{box} = [draw]
		\draw[step=1,gray,thin,dotted] (-4.8,-2.8) grid (2.8,4.8);
		\coordinate (c0) at (3.5,1);
		\coordinate (c1) at (-4,4);
		\coordinate (c2) at (-4,3);
		\coordinate (c3) at (-4,2);
		\coordinate (c4) at (-4,1);
		\coordinate (c5) at (-4,0);
		\coordinate (c6) at (-4,-1);
		\coordinate (c7) at (-4,-2);
		\coordinate (c8) at (-3,4);
		\coordinate (c9) at (-3,3);
		\coordinate (c10) at (-3,2);
		\coordinate (c11) at (-3,1);
		\coordinate (c12) at (-3,0);
		\coordinate (c13) at (-3,-1);
		\coordinate (c14) at (-3,-2);
		\coordinate (c15) at (-2,4);
		\coordinate (c16) at (-2,3);
		\coordinate (c17) at (-2,2);
		\coordinate (c18) at (-2,1);
		\coordinate (c19) at (-2,0);
		\coordinate (c20) at (-2,-1);
		\coordinate (c21) at (-2,-2);
		\coordinate (c22) at (-1,4);
		\coordinate (c23) at (-1,3);
		\coordinate (c24) at (-1,2);
		\coordinate (c25) at (-1,1);
		\coordinate (c26) at (-1,0);
		\coordinate (c27) at (-1,-1);
		\coordinate (c28) at (-1,-2);
		\coordinate (c29) at (0,4);
		\coordinate (c30) at (0,3);
		\coordinate (c31) at (0,2);
		\coordinate (c32) at (0,1);
		\coordinate (c33) at (0,0);
		\coordinate (c34) at (0,-1);
		\coordinate (c35) at (0,-2);	
		\coordinate (c36) at (1,4);
		\coordinate (c37) at (1,3);
		\coordinate (c38) at (1,2);
		\coordinate (c39) at (1,1);
		\coordinate (c40) at (1,0);
		\coordinate (c41) at (1,-1);
		\coordinate (c42) at (1,-2);
		\coordinate (c43) at (2,4);
		\coordinate (c44) at (2,3);
		\coordinate (c45) at (2,2);
		\coordinate (c46) at (2,1);
		\coordinate (c47) at (2,0);
		\coordinate (c48) at (2,-1);
		\coordinate (c49) at (2,-2);
		\node[text width = 3cm, align = center] (t) at (-1,-3.5) {Possible positions for end nodes $j$};
		\draw[red,fill = white] (c1) circle (0.8em) node[black] {$1$};			
		\draw[red,fill = white] (c2) circle (0.8em) node[black] {$2$};
		\draw[red,fill = white] (c3) circle (0.8em) node[black] {$3$};	
		\draw[red,fill = white] (c4) circle (0.8em) node[black] {$4$};
		\draw[red,fill = white] (c5) circle (0.8em) node[black] {$5$};
		\draw[red,fill = white] (c6) circle (0.8em) node[black] {$6$};
		\draw[red,fill = white] (c7) circle (0.8em) node[black] {$7$};
		\draw[red,fill = white] (c8) circle (0.8em) node[black] {$8$};
		\draw[red,fill = white] (c9) circle (0.8em) node[black] {$9$};
		\draw[red,fill = white] (c10) circle (0.8em) node[black] {$10$};			
		\draw[red,fill = white] (c11) circle (0.8em) node[black] {$11$};
		\draw[red,fill = white] (c12) circle (0.8em) node[black] {$12$};	
		\draw[red,fill = white] (c13) circle (0.8em) node[black] {$13$};
		\draw[red,fill = white] (c14) circle (0.8em) node[black] {$14$};
		\draw[red,fill = white] (c15) circle (0.8em) node[black] {$15$};
		\draw[red,fill = white] (c16) circle (0.8em) node[black] {$16$};
		\draw[red,fill = white] (c17) circle (0.8em) node[black] {$17$};
		\draw[red,fill = white] (c18) circle (0.8em) node[black] {$18$};
		\draw[red,fill = white] (c19) circle (0.8em) node[black] {$19$};			
		\draw[red,fill = white] (c20) circle (0.8em) node[black] {$20$};
		\draw[red,fill = white] (c21) circle (0.8em) node[black] {$21$};	
		\draw[red,fill = white] (c22) circle (0.8em) node[black] {$22$};
		\draw[red,fill = white] (c23) circle (0.8em) node[black] {$23$};
		\draw[red,fill = white] (c24) circle (0.8em) node[black] {$24$};
		\draw[red,fill = white] (c25) circle (0.8em) node[black] {$25$};
		\draw[red,fill = white] (c26) circle (0.8em) node[black] {$26$};
		\draw[red,fill = white] (c27) circle (0.8em) node[black] {$27$};
		\draw[red,fill = white] (c28) circle (0.8em) node[black] {$28$};			
		\draw[red,fill = white] (c29) circle (0.8em) node[black] {$29$};
		\draw[red,fill = white] (c30) circle (0.8em) node[black] {$30$};	
		\draw[red,fill = white] (c31) circle (0.8em) node[black] {$31$};
		\draw[red,fill = white] (c32) circle (0.8em) node[black] {$32$};
		\draw[red,fill = white] (c33) circle (0.8em) node[black] {$33$};
		\draw[red,fill = white] (c34) circle (0.8em) node[black] {$34$};
		\draw[red,fill = white] (c35) circle (0.8em) node[black] {$35$};
		\draw[red,fill = white] (c36) circle (0.8em) node[black] {$36$};
		\draw[red,fill = white] (c37) circle (0.8em) node[black] {$37$};			
		\draw[red,fill = white] (c38) circle (0.8em) node[black] {$38$};
		\draw[red,fill = white] (c39) circle (0.8em) node[black] {$39$};	
		\draw[red,fill = white] (c40) circle (0.8em) node[black] {$40$};
		\draw[red,fill = white] (c41) circle (0.8em) node[black] {$41$};
		\draw[red,fill = white] (c42) circle (0.8em) node[black] {$42$};
		\draw[red,fill = white] (c43) circle (0.8em) node[black] {$43$};
		\draw[red,fill = white] (c44) circle (0.8em) node[black] {$44$};
		\draw[red,fill = white] (c45) circle (0.8em) node[black] {$45$};
		\draw[red,fill = white] (c46) circle (0.8em) node[black] {$46$};			
		\draw[red,fill = white] (c47) circle (0.8em) node[black] {$47$};
		\draw[red,fill = white] (c48) circle (0.8em) node[black] {$48$};	
		\draw[red,fill = white] (c49) circle (0.8em) node[black] {$49$};		
		\draw[black] (-4.35,-2.35) rectangle (2.35,4.35) node [black,above,xshift=-0.75em,yshift=-0.25em] {$r{=}3$};
		\draw[black] (-3.35,-1.35) rectangle (1.35,3.35) node [black,above,xshift=-0.75em,yshift=-0.25em] {$r{=}2$};
		\draw[black] (-2.35,-0.35) rectangle (0.35,2.35) node [black,above,xshift=-0.75em,yshift=-0.25em] {$r{=}1$};			
		\end{tikzpicture}	
	}\hspace{0.1cm}
	\subfloat{
		\raisebox{0.07\textwidth}{				
			\psfrag{5}{\textcolor{black}{$5$}}
			\psfrag{10}{\textcolor{black}{$10$}}
			\psfrag{15}{\textcolor{black}{$15$}}
			\psfrag{20}{\textcolor{black}{$20$}}
			\psfrag{25}{\textcolor{black}{$25$}}
			\psfrag{30}{\textcolor{black}{$30$}}
			\psfrag{35}{\textcolor{black}{$35$}}
			\psfrag{40}{\textcolor{black}{$40$}}
			\psfrag{45}{\textcolor{black}{$45$}}
			\includegraphics[width=0.0375\textwidth]{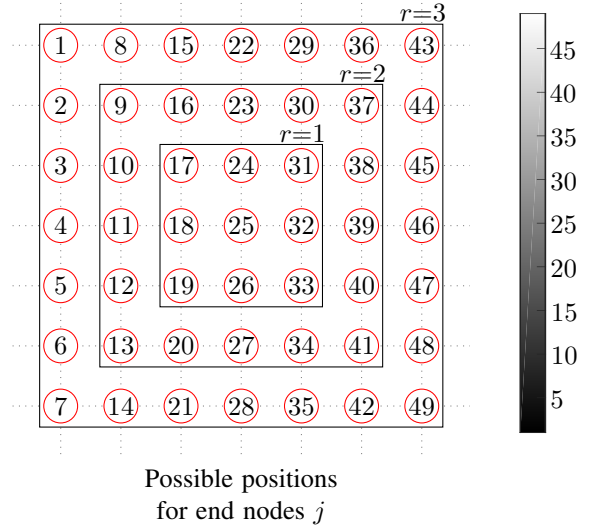}}
	}		
	\psfragscanoff
	\caption{Embedding of the indices for smaller radii into the indices of the maximum radius $r_\text{max}{=}3$. By assigning various radii to different regions, the variance of the motion maps can be smoothed. By allocating for every possible index a different shade of gray, as shown by the color bar on the right hand side, the motion maps in Fig.\,\ref{fig:motionMaps} are generated.}
	\label{fig:node_assignment}	
\end{figure}

\subsection{Reducing the Graph Description}
Assuming a predefined radius of $r{=}1$, so far, every start node $i$ is connected to nine end nodes $j$, as shown in Fig.\,\ref{fig:posA} for one single start node $i$.
In the corresponding adjacency matrix, given in Fig.\,\ref{fig:posB}, nine diagonals can be observed. The entries for the specific start node $i$ in Fig.\,\ref{fig:posA} are highlighted. Obviously, each of the nine possible positions corresponds to exactly one diagonal. The exact correspondences can be observed by the assigned numbers. Accordingly, for a fully connected neighborhood with radius \mbox{$r {=} 1$} for a frame of size $5\times 5$ pixels \mbox{$n_z {=} 169$} nonzero weights are distributed over $9$ diagonals. 

To reduce the graph description, the maximum prediction weight of every single start node $i$ is now set to $w_{ij,\text{max}} {=} 1$, while all other prediction weights related to start node $i$ are set to $0$. For the considered example, only \mbox{$n_z {=} 25$} nonzero weights remain, as can also be observed in Fig.\,\ref{fig:posC}.

\subsection{Construction of Motion Maps}
\label{subsec:construction}
After the reduction of the graph description as described above, every start node $i$ can now be represented by one specific \mbox{end node $j$} lying within a predefined neighborhood. Each node can also be assigned to one specific diagonal of the adjacency matrix. This allows for introducing a novel representation of the adjacency matrix, called motion map $\boldsymbol{M}$. Therefore, a matrix of size \mbox{$X{\times}Y$} is built, where every position $(x,y)$ corresponds to the spatial position of a start node $i$ in the even frame. Then, for every position $(x,y)$ the index of the position of the corresponding \mbox{end node $j$} with $w_{ij,\text{max}}{=}1$ is stored. In Fig.\,\ref{fig:ref} - Fig.\,\ref{fig:r3}, two subsequent frames of the data set, which was already used in Section~\ref{sec:introduction}, and their corresponding motion maps for a chosen neighborhood with radius $r{=}1$ and $r{=}3$ are shown. Since a radius of $r{=}3$ means more possible positions for end node $j$, the variance of the corresponding motion map is significantly higher. 
These motion maps can be compared to motion fields resulting from pixel-based approaches for MC. However, pixel-based motion models require a large overhead to transmit to the decoder.
 
For efficient encoding of these motion maps, several processing steps are performed. 
In a first step the motion maps are scanned in Peano-Hilbert order. Since Peano-Hilbert space filling curve fits only square images corresponding to a power of two, for different shaped data a space filling curve of size $2^{\lceil\log_2(\text{max}(X,Y))\rceil}$ is generated. Then, path coordinates which are outside of the required shape are skipped during scanning.
In the next step, the resulting symbol string $s$ is entropy coded using multiple-context adaptive arithmetic coding provided by the QccPack library~\cite{fowler2000qccpack}. 
This implementation allocates an arithmetic coding model which contains the probability models for the different contexts. We are using the previous symbol as the context of the current symbol and update the frequency-count information in the arithmetic model after encoding each symbol.

Further, we propose to enhance the coding efficiency by smoothing and masking the motion maps as described in the following.

\subsubsection{Smoothing of Motion Maps}
As described in~\cite{Lanz2016}, an increasing radius $r$ results in an increasing visual quality of the LP subband and a decreasing mean energy in the HP subband. However, the variance of the assigned end nodes rises with increasing radius $r$, which is adversely for encoding the resulting symbol stream.
Therefore, we propose to smooth the motion maps by assigning larger radii to areas with strong motion, to ensure good reconstruction results for these parts, whereas a smaller radius shall be sufficient, when almost no motion occurs. To determine the amount of motion between two subsequent frames, we calculate their absolute difference $|f_{2t-1}-f_{2t}|$ and map it into a range between $0$ and $1$. An example can be seen in Fig.\,\ref{fig:mse_mask}, where the normalized absolute difference of the reference and current frame of Fig.\,\ref{fig:ref} and~\ref{fig:cur} is shown. The spatial intensity results from a combination of spatial and temporal gradients. We use it as an indicator for the local amount of motion. 
Hence, for a maximum radius of $r_{\text{max}}{=}3$, radius $r{=}3$ is assigned to areas which offer a high value (white), radius $r{=}2$ is assigned to areas with medium values (gray), and radius $r{=}1$ is assigned to areas with values close to zero (black). An example can be seen in Fig.\,\ref{fig:motion_map_rmax3}. However, as can be seen in Fig.\,\ref{fig:node_assignment}, the indices of the positions for radius $r{=}1$ are no longer in a range of $1$ to $9$. The indices of the inner radii have to be adapted to the indexing of the maximum radius. By doing this, a unique mapping of the single start nodes to their related end nodes can be guaranteed. 

\subsubsection{Masking of Motion Maps}

\begin{figure}[t]
	\centering
	\psfragscanon		
	\includegraphics[trim = 14cm 2cm 14cm 1cm,clip,width=0.24\textwidth]{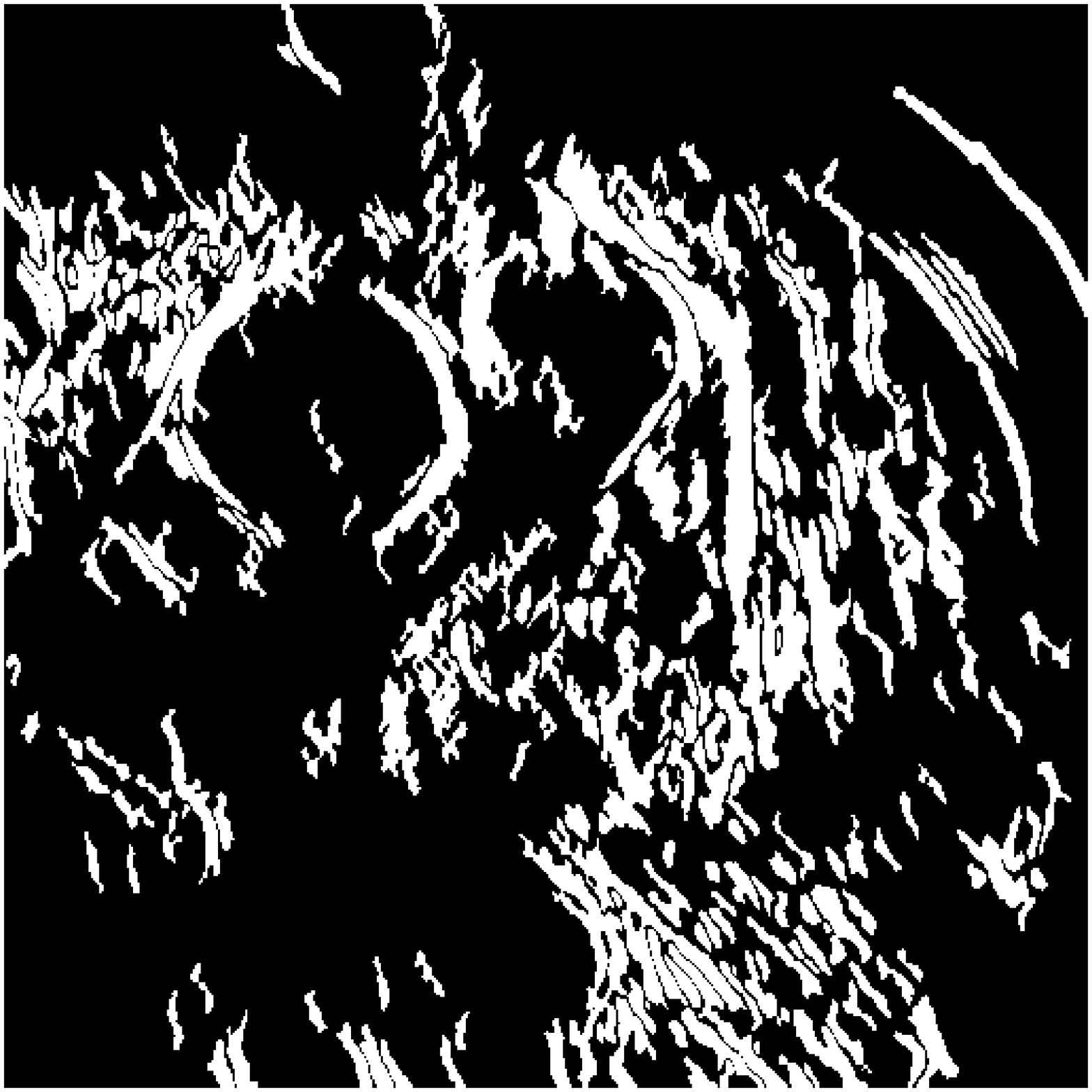}
	\psfragscanoff
	\caption{Binary mask $\boldsymbol{B}$ for a threshold $\tau(t)$ with \mbox{$\text{PSNR}_\text{target}{=}50$\,dB}.}
	\label{fig:binary_mask}
	\vspace{-8pt}
\end{figure}

\begin{figure}[t]
	\centering
	\psfragscanon		
	\psfrag{m}{Omitted pixel}
	\psfrag{n}{Considered pixel}
	\includegraphics[trim = 18cm 0cm 0cm 20cm,clip,width=0.36\textwidth]{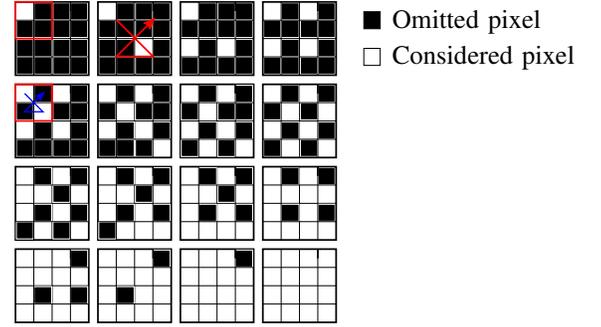}
	\psfragscanoff
	\caption{Construction scheme for sparse sampling masks with increasing densities.}
	\label{fig:masks_reg}
	\vspace{-9pt}
\end{figure}

\begin{figure*}[t]	
	\begin{tikzpicture}[scale=1.0, >=latex'] 
	\input{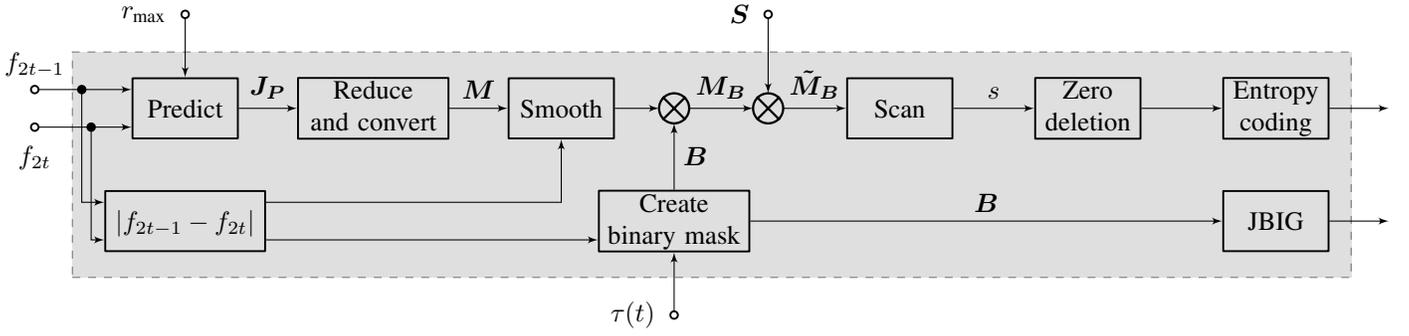}
	\tikzstyle{box} = [draw]
	\draw[dashed,fill=lightgray,opacity=0.5] (0.5,-0.25) rectangle (17.5,-3.25);
	\coordinate (c1) at (0,0);
	\coordinate (c2) at (18,0);	
	\node[dspnodeopen,dsp/label=above] (n1) at (0,-0.75) {$f_{2t-1}$}; 
	\node[dspnodefull] (n11) at (0.625,-0.75) {}; 
	\node[dspnodefull] (n21) at (0.75,-1.25) {};
	\node[dspnodeopen,dsp/label=left] (r) at (2,0.25) {$r_\text{max}$};
	\node[dspnodeopen,dsp/label=left] (S) at (9.75,0.25) {$\boldsymbol{S}$};
	\node[dspnodeopen,dsp/label=left] (t) at (8.5,-3.75) {$\tau(t)$};
	\node[dspnodeopen,dsp/label=below] (n2) at (0,-1.25) {$f_{2t}$}; 
	\node[dspfilter](pred) at (2,-1) {Predict};
	\node[dspfilter,minimum width = 2.125cm] (norm) at (2,-2.5) {$|f_{2t-1}-f_{2t}|$};
	\node[dspfilter,minimum width = 2cm,text height=2em](conv) at (4.5,-1) {Reduce\\ and convert};
	\node[dspfilter](smoo) at (7,-1) {Smooth};	
	\node[dspfilter,minimum width = 2cm,text height=2em](bin) at (8.5,-2.5) {Create\\ binary mask};
	\node[dspmixer] (mux1) at (8.5,-1) {};
	\node[dspmixer] (mux2) at (9.75,-1) {};
	\node[dspfilter](scan) at (11.5,-1) {Scan};
	\node[dspfilter,text height=2em](zdel) at (14,-1) {Zero\\ deletion};
	\node[dspfilter,text height=2em](cab1) at (16.5,-1) {Entropy\\ coding};
	\node[dspfilter](jbg1) at (16.5,-2.5) {JBIG};
	\draw[->] (n1) -- ($(pred.180) + (0,0.25)$);
	\draw[->] (n2) -- ($(pred.180) + (0,-0.25)$);
	\draw[->] (pred) -- node[above]{$\boldsymbol{J_P}$} (conv);	
	\draw[->] (conv) -- node[above]{$\boldsymbol{M}$} (smoo);	
	\draw[->] (smoo) -- (mux1);
	\draw[->] (mux1) -- node[above]{$\boldsymbol{M_B}$} (mux2);
	\draw[->] (mux2) -- node[above]{$\boldsymbol{\tilde{M}_B}$} (scan);		
	\draw[->] ($(norm.360)+(0,-0.25)$) -- ($(bin.180) + (0,-0.25)$);
	\draw[->] ($(norm.360)+(0,0.25)$) -| (smoo);
	\draw[->] (n11) |- ($(norm.180) + (0,0.25)$);
	\draw[->] (n21) |- ($(norm.180) + (0,-0.25)$);	
	\draw[->] (r) -- (pred);
	\draw[->] (S) -- (mux2);
	\draw[->] (bin) -- node[right]{$\boldsymbol{B}$} (mux1);
	\draw[->] (t) -- (bin);
	\draw[->] (scan) -- node[above]{$s$}(zdel);
	\draw[->] (zdel) -- (cab1);
	\draw[->] (bin) -- node[above]{$\boldsymbol{B}$} (jbg1);
	\draw[->] (cab1) -- (18,-1);
	\draw[->] (jbg1) -- (18,-2.5);
	\end{tikzpicture}
	\caption{Block diagram of the encoder for the proposed coding scheme. $r_\text{max}$ and $\boldsymbol{S}$ have to be known at the decoder, too.}
	\label{fig:block_diagram}
\end{figure*}
\begin{figure}[t]	
	\begin{tikzpicture}[scale=1.0, >=latex'] 
	\input{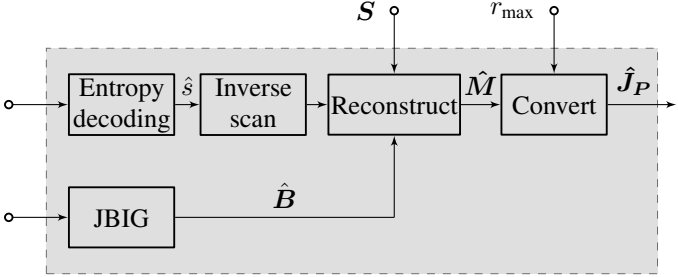}
	\tikzstyle{box} = [draw]
	\draw[dashed,fill=lightgray,opacity=0.5] (0.5,-0.25) rectangle (8.625,-3.25);
	\node[dspnodeopen,dsp/label=above] (s) at (0,-1) {};
	\node[dspnodeopen,dsp/label=below] (B) at (0,-2.5) {};
	\node[dspfilter](jbg2) at (1.5,-2.5) {JBIG};
	\node[dspfilter,text height=2em](cab2) at (1.5,-1) {Entropy\\ decoding};
	\coordinate (rec_node) at (5.125,-2.5) {};
	\node[dspfilter,text height=2em](iscan) at (3.25,-1) {Inverse\\scan};
	\node[dspfilter,minimum width = 1.75cm](rec) at (5.125,-1) {Reconstruct};
	\node[dspfilter](conv2) at (7.25,-1) {Convert};
	\node[dspnodeopen,dsp/label=left] (r2) at (7.25,0.25) {$r_\text{max}$};
	\node[dspnodeopen,dsp/label=left] (S2) at (5.125,0.25) {$\boldsymbol{S}$};
	\node (Jp) at (9,-1) {};
	\draw[->] (s) -- (cab2);
	\draw[->] (B) -- (jbg2);
	\draw[->] (cab2) -- node[above]{$\hat{s}$} (iscan);
	\draw[->] (iscan) -- (rec);
	\draw[->] (rec) -- node[above]{$\boldsymbol{\hat{M}}$}(conv2);
	\draw[-] (jbg2) -- node[above]{$\hat{\boldsymbol{B}}$} (rec_node);
	\draw[->] (rec_node) -- (rec);
	\draw[->] (r2) -- (conv2);
	\draw[->] (S2) -- (rec);
	\draw[->] (conv2) -- node[above,xshift=-3pt]{$\boldsymbol{\hat{J}_P}$}(Jp);	
	\end{tikzpicture}
	\caption{Block diagram of the decoder for the proposed coding scheme. $r_\text{max}$ and $\boldsymbol{S}$ have to be known.}
	\label{fig:block_diagram_dec}
	\vspace{-5pt}
\end{figure}

A further improvement regarding the coding efficiency can be reached by masking the motion maps by a binary mask $\boldsymbol{B}$. This binary mask equals $1$ in areas with strong motion and equals $0$ in areas with almost no motion. In case there is no motion assumed, the corresponding nodes are not transmitted. Instead, they are reconstructed at the decoder side by connecting them to their direct neighbors. The direct neighboring position corresponds to the main diagonal in the adjacency matrix $\boldsymbol{J_P}$. Therefore, the missing end nodes can easily be restored by filling the main diagonal with entries, where no other end node is not yet assigned.

To calculate the binary mask, the normalized absolute difference $|f_{2t-1}-f_{2t}|$ is binarized. Therefore, a certain threshold for every pair of frames is required. This threshold $\tau(t)$ is chosen with respect to the underlying mean squared error (MSE) of the two considered frames and a PSNR value which we want to guarantee for the reconstructed LP frame at the decoder side. By rearranging the general formula for calculating the PSNR value with a maximum possible amplitude $A_\text{max}$ 
\begin{equation}
	\text{PSNR}_\text{target} = 10\log_{10}\left(\frac{A_\text{max}^2}{\text{MSE}_\text{target}}\right)
\end{equation} to
\begin{equation}
\text{MSE}_\text{target} = \frac{A_{\text{max}}^2}{10^{\frac{\text{PSNR}_{\text{target}}}{10}}},
\end{equation} 
the corresponding value $\text{MSE}_\text{target}$ is achieved. Then, threshold $\tau(t)$ can be computed by relating $\text{MSE}_\text{target}$ to the actual $\text{MSE}$ of the two underlying frames:
\begin{equation}
\tau(t) = \frac{\text{MSE}_\text{target}}{\frac{1}{XY}\sum_{x=0}^{X-1}\sum_{y=0}^{Y-1}(f_{2t-1}(x,y)-f_{2t}(x,y))^2}.
\label{eq:tau}
\end{equation}
For example, taking $\text{PSNR}_\text{target} = 50$\,dB, the binary mask $\boldsymbol{B}$ for the two subsequent frames in Fig.\,\ref{fig:motionMaps} is shown in Fig.\,\ref{fig:binary_mask}. 

After multiplying $\boldsymbol{M}$ with the corresponding binary mask, the remaining masked motion map $\boldsymbol{{M}_B}$ has to be scanned and the resulting symbol stream $s$ has to be entropy coded.
To reduce the length of $s$, only the nonzero values shall be encoded and transmitted. Therefore, after scanning $\boldsymbol{M_B}$, all zero elements are deleted. To reconstruct the motion map at the decoder side without any loss, the exact positions of the nonzero elements have to be known. Since the shape of the binary masks depend on every pair of consecutive frames, all binary masks have to be transmitted as metadata to the decoder side.	
This is done by using JBIG compression standard~\cite{ITU1993} which was designed for bi-level image data such as scanned documents. 
By choosing higher values for $\text{PSNR}_\text{target}$, more information of the motion maps is available after masking them and therefore a higher quality of the reconstructed motion maps can be achieved. The coding costs for transmitting the masks will not increase significantly. However, the coding costs for transmitting the resulting symbol stream $s$ will strongly rise.

\subsection{Sparse Sampling of Motion Maps}

Further bit rate savings can be reached by applying sparse sampling to the masked motion maps $\boldsymbol{M_B}$. This processing step allows to meet different application scenarios with various channel capacities, e.g., wireless networks, for which a high sparsity would be preferable to save bandwidth.
Therefore, sparse sampling masks $\boldsymbol{S}$ with different density patterns are generated.  

Fig.\,\ref{fig:masks_reg} shows binary patches of size $4{\times}4$ which are copied periodically to fit the required size of the motion map. The upper left patch has the lowest density of considered pixels shown in white. By copying the red marked inner block of size $2{\times}2$ to the remaining quarters in the order the red arrow indicates, the density $d$ can be successively increased from $6.25$\% to $25$\%, as shown in the first row of Fig.\,\ref{fig:masks_reg}. For further increase, the remaining pixels of the inner block of size $2{\times}2$ are also considered one after another according to the blue arrow and are copied to the remaining quarters in the same order as before. Using this procedure, the density can successively be increased by $6.25$ percentage points, resulting in $16$ different masks.
The subsampled masked motion maps are further denoted as $\boldsymbol{\tilde{M}_B}$. 
Since the applied sparsity pattern depends on the available physical channel, it has to be determined in advance for the whole coding setup. Therefore, the applied sampling mask is known at the encoder and decoder side. An suitable selection of the sparsity pattern for a given channel capacity is not investigated so far.
Then, the missing positions of the symbol stream $s$ can be reconstructed by applying common interpolation methods. Therefore, the indexing of the possible end nodes in $\boldsymbol{\tilde{M}_B}$ requires a conversion to cartesian coordinates, resulting in two separate motion maps $\boldsymbol{\tilde{M}_B}_x$ and $\boldsymbol{\tilde{M}_B}_y$. For reconstruction, linear interpolation, nearest neighbor, and natural neighbor interpolation, which is based on Voronoi tessellation, are evaluated.

\begin{figure}[t]
	\captionsetup[subfigure]{justification=centering}
	\centering		
	\subfloat[Final motion map $\boldsymbol{\tilde{M}_B}$ for a sampling density of $d{=}100\%$. Coding costs for $s{=}22.62$~kB.]{\includegraphics[trim = 14cm 2cm 14cm 1cm,clip,width=0.24\textwidth]{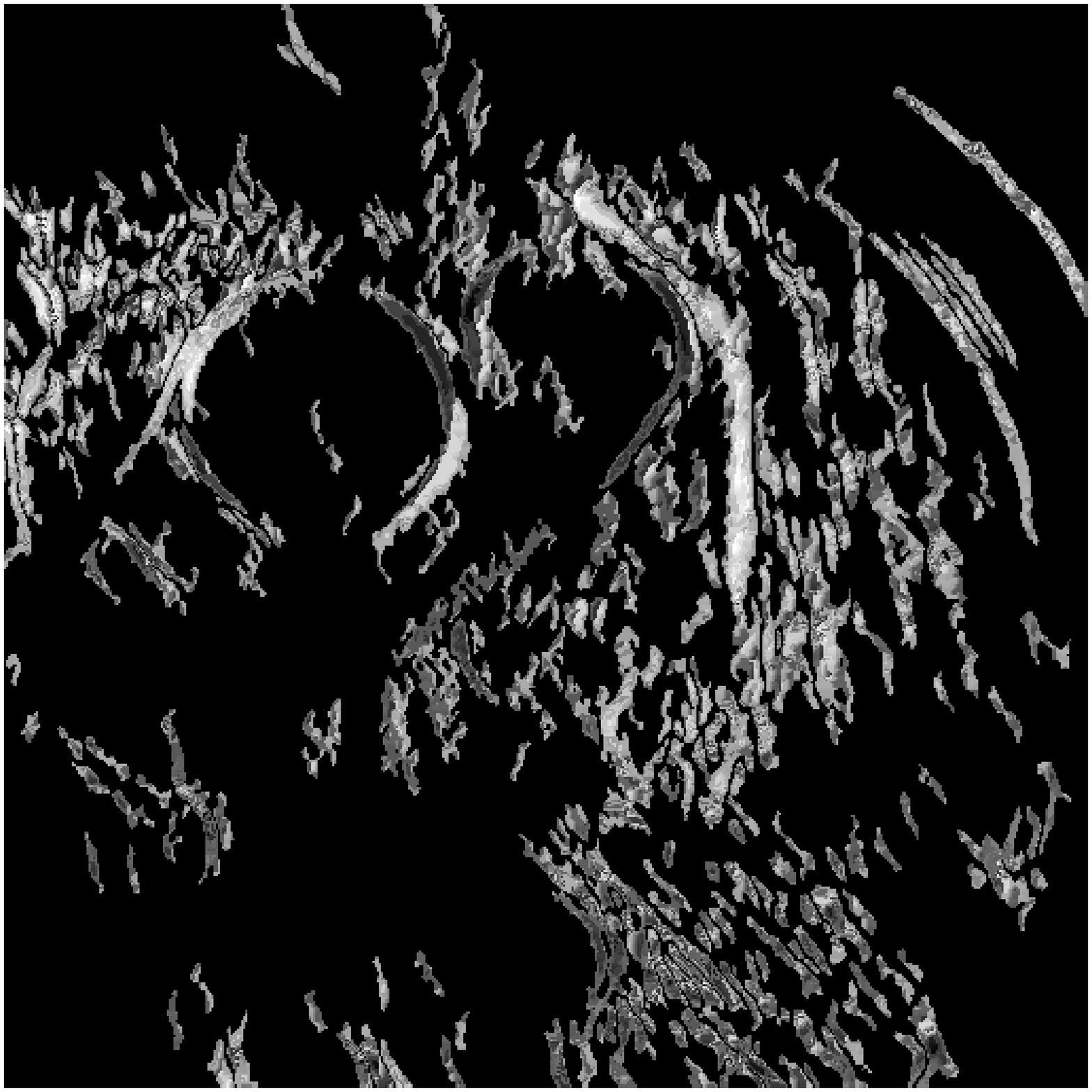}}
	\subfloat[Final motion map $\boldsymbol{\tilde{M}_B}$ for a sampling density of $d{=}50\%$.\hspace{1cm} Coding costs for $s{=}12.95$~kB.]{\includegraphics[trim = 14cm 2cm 14cm 1cm,clip,width=0.24\textwidth]{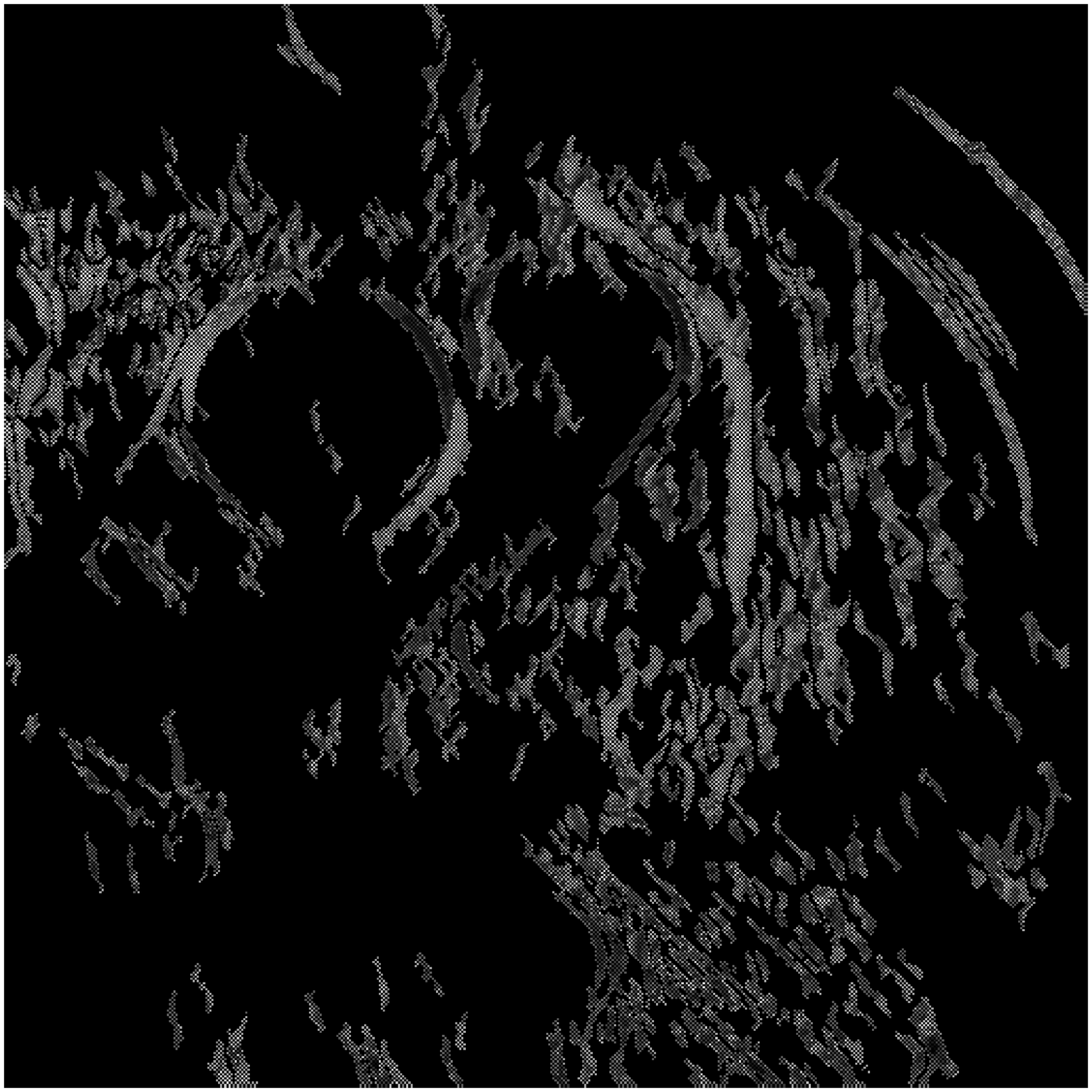}}
	\caption{Exemplary motion maps after all applied processing steps. The coding costs for $\boldsymbol{B}{=}4.66$~kB are the same for both cases.}
	\label{fig:maps}
	\vspace{-5pt}
\end{figure}
\bigskip

To summarize the entire encoding scheme, Fig.\,\ref{fig:block_diagram} shows a block diagram of the proposed method.
To predict $f_{2t}$ from $f_{2t-1}$, a graph with a certain radius is set up between two frames. Afterwards, the resulting adjacency matrix $\boldsymbol{J_P}$ is reduced and converted to a motion map $\boldsymbol{M}$. To save bit rate, the motion map is smoothed and masked by a binary mask $\boldsymbol{B}$. Both steps are based on the normalized absolute difference of the input frames $|f_{2t-1}-f_{2t}|$. The resulting masked motion map $\boldsymbol{M_B}$ is subsampled by applying a sparse mask $\boldsymbol{S}$, resulting in $\boldsymbol{\tilde{M}_B}$. 
After scanning the map in Peano-Hilbert order all zero entries in $s$ are deleted and the remaining symbol stream is encoded using multiple-context adaptive arithmetic coding. This requires to additionally encode $\boldsymbol{B}$, which is done by employing the JBIG algorithm. 
As an example, Fig.\,\ref{fig:maps} shows the resulting motion maps for two different densities $d{=}50\%$ and $d{=}100\%$ after all discussed processing steps. As input serves the same data as used before. 
The specific file sizes of the two output bit streams are given below. 
The structure of the decoder is shown in Fig.\,\ref{fig:block_diagram_dec}. To achieve the reconstructed motion map $\boldsymbol{\hat{M}}$, an inverse Peano-Hilbert scan followed by an interpolation of the missing values due to the sparse sampling mask is required. Afterwards, the reconstructed motion map is converted to an adjacency matrix again. It should be mentioned that neither the threshold $\tau(t)$ nor the chosen $\text{MSE}_\text{target}$ have to be transmitted, since this information is inherently stored in the binary mask $\boldsymbol{B}$.

\section{Simulation Results}
\label{sec:results}
The simulation setup comprises dynamic volumes from CT\footnote{The CT volume data set was kindly provided by Siemens Healthineers.} and MR devices, showing sequences of a beating heart. The dimensions for the spatial and temporal resolution are summarized in Table\,\ref{tab:data}. The bit depth for all sequences constitutes 12 bits per sample. 

In the following, we will at first analyze the impact of smoothing and masking the motion maps. Afterwards, we will evaluate the novel coding scheme for different sampling masks $\boldsymbol{S}$ and for various interpolation algorithms. Therefore, one Haar wavelet decomposition step in temporal direction is performed. For measuring the visual quality of the resulting LP subbands, we use $\text{PSNR}_{\text{LP}_t} [\text{dB}]$ as introduced in~\cite{Lanz2018}. This metric considers not only the similarity of the LP subband to the odd-indexed frames, but also the similarity to the even indexed frames, which is a very important aspect, if the LP subband is to be used as a downscaled representative for the original sequence. All results are summed up for each sequence and are then averaged over the whole data sets.
Additionally, we will compare the best performing approach of our proposed graph-based MCTF coder to some state-of-the-art methods including SL and EL coding schemes. 

\begin{table}[t]
	\centering
	\caption{For simulation a \mbox{3D+t} CT data set which comprises 127 temporal sequences, and 28 independent temporal sequences from a MR device are used.}
	\label{tab:data}
	\begin{tabular}{l|rrrr|c}
		\hline
		& \multicolumn{1}{c}{x} & \multicolumn{1}{c}{y} & \multicolumn{1}{c}{z} & \multicolumn{1}{c|}{t} & \multicolumn{1}{l}{$\#$ temporal sequences} \\ \hline
		CT  & 512                   & 512                   & 127                   & 10                     & 127                                             \\
		MR & 144                   & 192                   & 1                     & 25                     & 28                                              \\ \hline
	\end{tabular}
\end{table}
\begin{table}[t]
	\centering	
	\caption{PSNR in [dB] of the LP subband and file size in [kB] of the single subbands and the motion information $\boldsymbol{m_\text{tx}}$ for different radii and a maximums radius of $r_\text{max}{=}3$ as well as for the masked motion maps with a maximum radius of $r_\text{max}{=}3$.}
	\label{tab:radii}
	\setlength{\tabcolsep}{2pt}
	\resizebox{0.49\textwidth}{!}{%
		\begin{tabular}{c|l|r|r|r|r|r}
			& &\multicolumn{1}{c|}{\multirow{2}{*}{{$r{=}1$}}} &  \multicolumn{1}{c|}{\multirow{2}{*}{{$r{=}2$}}}  & \multicolumn{1}{c|}{\multirow{2}{*}{{$r{=}3$}}} & \multicolumn{1}{c|}{\multirow{2}{*}{{$r_\text{max}{=}3$}}} & \multicolumn{1}{c}{{$r_\text{max}{=}3$},}\\
			& & & & &  & \multicolumn{1}{c}{masked} \\\hline
			\multicolumn{1}{c|}{\multirow{6}{*}{\tabrotate{\textbf{CT}}}} & $\text{PSNR}_{\text{LP}_t}$   & 50.26  & 54.49  & 57.34  & 53.75 & 50.25  \\
			\multicolumn{1}{c|}{}                    					  & File size $\text{LP}_t    $   & 894.13 & 830.50 & 797.10 & 880.33 & 808.80\\
			\multicolumn{1}{c|}{}                    					  & File size $\text{HP}_t    $   & 817.12 & 727.54 & 643.45 & 817.00 & 802.26\\
			\multicolumn{1}{c|}{}                    					  & File size $\boldsymbol{m_\text{tx}}$   & 444.25 & 682.02 & 841.27 & 497.45 & 153.11\\
			\multicolumn{1}{c|}{}                    					  & File size $\sum           $   & 2155.49 & 2240.06 & 2281.81 & 2194.78 & 1764.17\\\hline
			\multicolumn{1}{c|}{\multirow{6}{*}{\tabrotate{\textbf{MR}}}} & $\text{PSNR}_{\text{LP}_t}$   & 67.85  & 71.63  & 73.97  & 70.00 & 66.50  \\
			\multicolumn{1}{c|}{}                    					  & File size $\text{LP}_t    $   & 195.63 & 192.66 & 191.33 & 194.18 & 191.14\\
			\multicolumn{1}{c|}{}                    					  & File size $\text{HP}_t    $   & 124.81 & 100.62 & 83.28 & 121.07 & 138.56\\
			\multicolumn{1}{c|}{}                     					  & File size $\boldsymbol{m_\text{tx}}$ 	  & 107.61 & 165.04 & 200.82 & 120.43 & 36.41\\
			\multicolumn{1}{c|}{}                    					  & File size $\sum           $   & 428.04 & 458.31 & 475.43 & 435.67 & 366.12
		\end{tabular}
	}
\end{table}

\subsection{Analysis of Smoothing and Masking the Motion Maps}

To demonstrate the enhancement of the coding efficiency due to the different processing steps of the novel coding scheme, Table\,\ref{tab:radii} shows the PSNR and rate results for increasing radii of the considered neighborhoods as well as the impact of smoothing and masking the motion maps. The required motion information, which has to be transmitted to the decoder side, is summarized by $\boldsymbol{m_\text{tx}}$. The resulting LP and HP subbands are compressed without any loss, using the wavelet-based volume coder JPEG\,2000. Therefore, the OpenJPEG~\cite{openjpeg} implementation with four spatial wavelet decomposition steps in $xy$-direction is used.
As already mentioned above, an increasing radius results in an increasing visual quality of the LP subband, but also in an increasing file size for the motion maps. To find an appropriate compromise between visual quality and file size, we smooth the motion maps as described in Section\,\ref{subsec:construction}.
In the second column from right, the results for a maximum radius \mbox{$r_\text{max}{=}3$} are given. The exact distribution of the possible radii considering $|f_{2t-1}-f_{2t}|$ is set to following intervals:
\begin{equation}
	\begin{aligned}
	\left[0;0.12\right[&\rightarrow r=1\\
	\left[0.12;0.29\right[&\rightarrow r=2\\
	\left[0.29;1\right]&\rightarrow r=3.
	\end{aligned} 
\end{equation}
These intervals are trained on the first temporal sequence of the CT data set. Although this is a very poor training, it gives satisfying results also for the MR data set, as the lower part of Table\,\ref{tab:radii} proves. A larger training set, which is different from the test data sets, will probably lead to even better results.  

The contribution of the smoothing process to the overall rate reduction compared to a fixed radius of $r{=}3$ constitutes approximately $4\%$ for both data sets.
By multiplying the motion maps with the binary masks, a further rate reduction can be reached. For calculating the threshold $\tau(t)$ according to equation~(\ref{eq:tau}), the covered range of the given bit depth has to be considered. Since the maximum amplitude values for CT data are much higher than for MR data, the mean squared error of CT data also covers larger ranges than for MR data. Therefore, a value of \mbox{$\text{PSNR}_\text{target}{=}50$\,dB} for the CT data set and \mbox{$\text{PSNR}_\text{target}{=}65$\,dB} for the MR data set is used. The impact to the file sizes of the single subbands and the overall file size by encoding the resulting symbol stream, as introduced in Section\,\ref{subsec:construction}, can be seen in the last column of Table\,\ref{tab:radii}. 
Accordingly, a further rate reduction of $19\%$ and $16\%$ can be achieved for the CT and MR data sets, respectively, while the visual quality of the LP subband is controlled by the threshold $\tau(t)$. 

\subsection{Analysis of Sparse Sampling and Reconstruction of the Motion Maps}
\label{subsect:opt}

\begin{figure}[t]	
	\begin{center}
		\advance\leftskip-0.4cm
		\setlength\figurewidth{0.4\textwidth}
%
%
\definecolor{mycolor1}{rgb}{0.00000,0.44700,0.74100}%
\definecolor{mycolor2}{rgb}{0.85000,0.32500,0.09800}%
\definecolor{mycolor3}{rgb}{0.92900,0.69400,0.12500}%
\definecolor{mycolor5}{rgb}{0.49400,0.18400,0.55600}%
\definecolor{mycolor4}{rgb}{0.46600,0.67400,0.18800}%
\begin{tikzpicture}

\begin{groupplot}[group style={group name = group,group size=1 by 2, horizontal sep = 1cm, vertical sep=0.7cm}]

\nextgroupplot[%
width=\figurewidth ,
height=0.5\figurewidth ,
scale only axis,
xmin=46.5,
xmax=50.5,
xmajorgrids,
ymin=30,
ymax=170,
ylabel={File size $\boldsymbol{m_\text{tx}}$ [kB]},
/pgf/number format/.cd,
use comma,
1000 sep={},
ymajorgrids,
axis background/.style={fill=white},
every axis title/.style={font=\bfseries,below right,at={(0,1)}},
title={CT 12 bit},
]

\addplot [color=mycolor4,solid,mark=*,mark options={scale=0.25, fill=mycolor4}]
table[row sep=crcr]{%
47.25	37.15\\
47.85	47.03\\
48.15	56.12\\
48.51	64.96\\
48.63	73.20\\
48.77	82.07\\
48.90	90.31\\
49.04	98.44\\
49.16	105.60\\
49.29	112.59\\
49.42	119.44\\
49.57	126.37\\
49.72	133.09\\
49.88	139.77\\
50.06	146.59\\
50.25	153.11\\
};
\label{reg_lin_yes};

\addplot [color=mycolor4,dashed,mark=*,mark options={scale=0.25, fill=mycolor4}]
table[row sep=crcr]{%
47.05	37.15\\
48.09	47.03\\
48.41	56.12\\
48.77	64.96\\
48.91	73.20\\
49.06	82.07\\
49.24	90.31\\
49.44	98.44\\
49.51	105.60\\
49.60	112.59\\
49.69	119.44\\
49.78	126.37\\
49.89	133.09\\
49.99	139.77\\
50.12	146.59\\
50.25	153.11\\
};
\label{reg_near_yes};

\addplot [color=mycolor4,dotted,mark=*,mark options={scale=0.25, fill=mycolor4}]
table[row sep=crcr]{%
47.25	37.15\\
47.89	47.03\\
48.18	56.12\\
48.50	64.96\\
48.63	73.20\\
48.78	82.07\\
48.93	90.31\\
49.08	98.44\\
49.20	105.60\\
49.33	112.59\\
49.46	119.44\\
49.61	126.37\\
49.75	133.09\\
49.91	139.77\\
50.07	146.59\\
50.25	153.11\\
};
\label{reg_nat_yes};

\nextgroupplot[%
width=\figurewidth ,
height=0.5\figurewidth ,
scale only axis,
xmin=58,
xmax=67,
xlabel={$\text{PSNR}_{\text{LP}_t}$ [dB]},
xmajorgrids,
ymin=8,
ymax=40,
ylabel={File size $\boldsymbol{m_\text{tx}}$ [kB]},
ymajorgrids,
axis background/.style={fill=white},
every axis title/.style={font=\bfseries,below right,at={(0,1)}},
title={MR 12 bit}
]

\addplot [color=mycolor4,solid,mark=*,mark options={scale=0.25, fill=mycolor4}]
table[row sep=crcr]{%
60.07	10.38\\
60.58	12.32\\
60.90	14.17\\
61.24	15.99\\
61.54	17.80\\
61.84	19.63\\
62.18	21.42\\
62.60	23.20\\
62.91	24.87\\
63.21	26.52\\
63.55	28.15\\
63.95	29.79\\
64.42	31.45\\
64.98	33.11\\
65.65	34.77\\
66.50	36.41\\
};

\addplot [color=mycolor4,dashed,mark=*,mark options={scale=0.25, fill=mycolor4}]
table[row sep=crcr]{%
58.80	10.38\\
59.75	12.32\\
60.29	14.17\\
60.84	15.99\\
61.15	17.80\\
61.47	19.63\\
61.79	21.42\\
62.16	23.20\\
62.49	24.86\\
62.84	26.52\\
63.23	28.15\\
63.65	29.79\\
64.15	31.45\\
64.75	33.11\\
65.47	34.77\\
66.50	36.41\\
};

\addplot [color=mycolor4,dotted,mark=*,mark options={scale=0.25, fill=mycolor4}]
table[row sep=crcr]{%
60.35	10.38\\
60.88	12.32\\
61.23	14.17\\
61.60	15.99\\
61.88	17.80\\
62.15	19.63\\
62.46	21.42\\
62.82	23.20\\
63.11	24.86\\
63.42	26.52\\
63.76	28.15\\
64.14	29.79\\
64.58	31.45\\
65.11	33.11\\
65.72	34.77\\
66.50	36.41\\
};

\end{groupplot}


\path (group c1r1.south west |-current bounding box.south west)--
coordinate(legendpos)
(group c1r2.south east |-current bounding box.south east);
\matrix[
matrix of nodes,
anchor=south,
inner sep=0.1em,
draw,
fill=white,
nodes={scale=0.9},
align=right,
]at([yshift=-4ex,xshift = 0ex]legendpos)
{	

\ref{reg_lin_yes}& Linear &[2pt] & \ref{reg_near_yes}& Nearest neighbor &[2pt] & \ref{reg_nat_yes}& Natural neighbor\\	};

\draw[->] (2,4)node[left]{$d:6.25\%$} -- node[right,xshift=2cm]{$100\%$}(6,4);

\end{tikzpicture}%
		\caption{$\text{PSNR}_{\text{LP}_t}$ results of the CT data (top) and MR data (bottom) compared against the file size in [kB] of the required motion information $\boldsymbol{m_\text{tx}}$ for our proposed coding scheme. Three different interpolation methods are evaluated, which are displayed over increasing densities $d$ of the applied sampling masks $\boldsymbol{S}$.}
		\label{fig:opt}
	\end{center}
	\vspace{-10pt}
\end{figure}
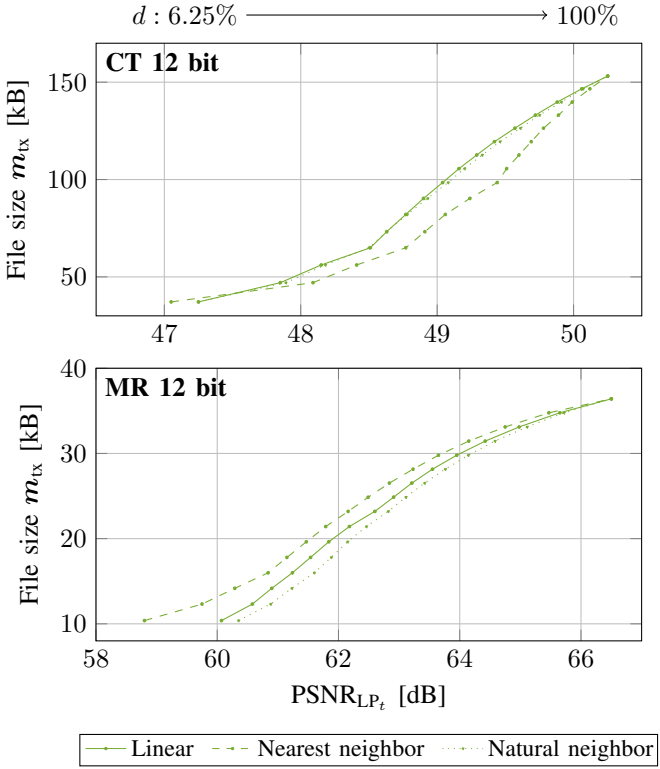
To analyze the performance of our proposed coding scheme regarding the sparse sampling masks, both data sets are evaluated for all possible sampling masks $\boldsymbol{S}$ and all considered interpolation methods.
The results are shown in Fig.\,\ref{fig:opt} for the CT as well as the MR data set.
Accordingly, it is possible to further reduce the required file size for encoding $\boldsymbol{m_\text{tx}}$, while the visual quality is degraded simultaneously. The optimal rate-distortion ratio depends on the capacity of the considered physical channel. Further, for both data sets, linear interpolation is never the best choice. Especially for the CT data set, the amplitude values of the motion maps have still a relatively high variance, even after the smoothing and masking steps. This leads to inferior results for linear interpolation. The MR data set covers a smaller range of possible amplitude values and therefore comprises smoother motion maps, which is why linear interpolation works better compared to the CT data set. 
In the following, for the CT data set nearest neighbor interpolation is chosen and for the MR data set natural neighbor interpolation is used. 

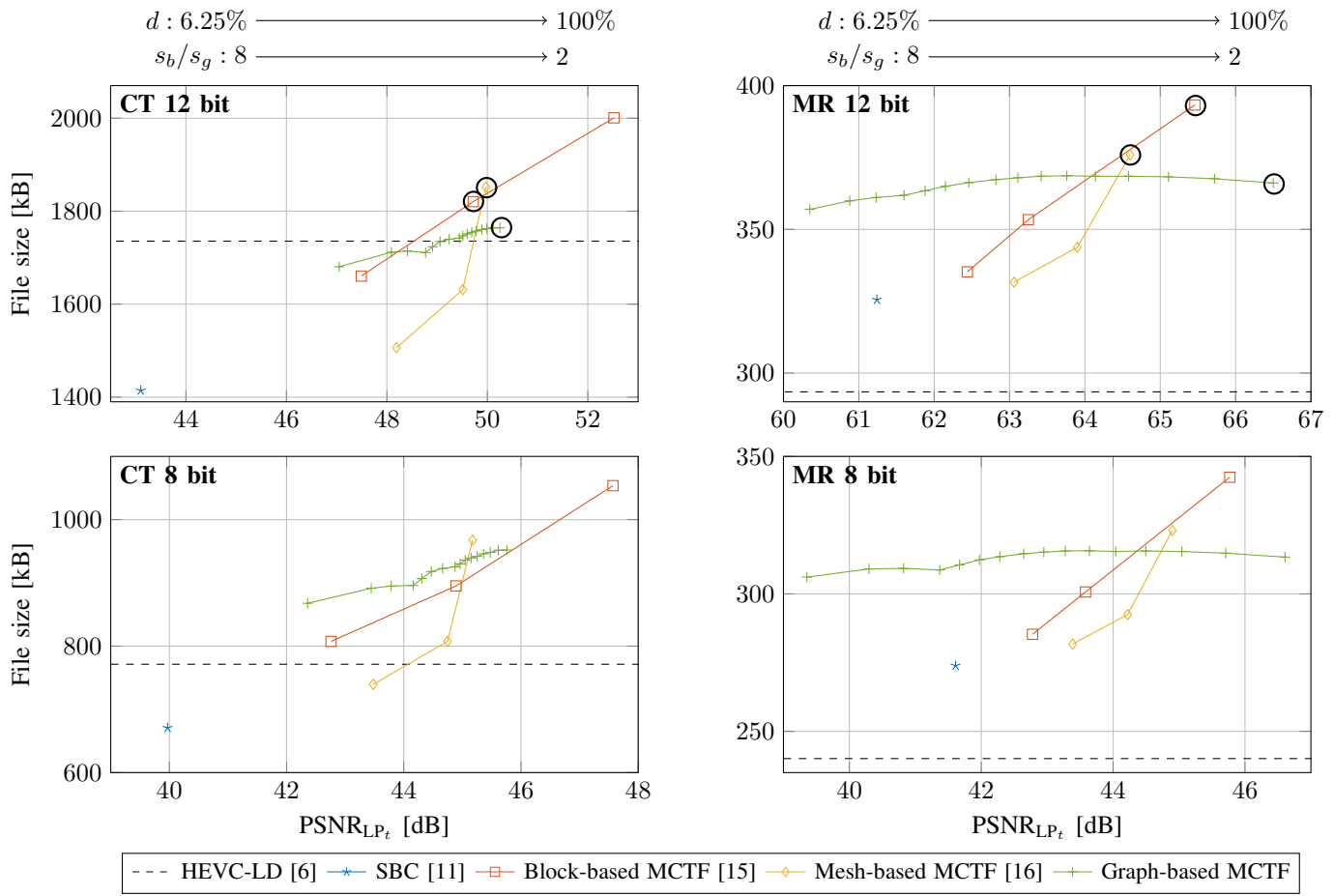
\begin{figure*}[t]
	\begin{center}
		\advance\leftskip-0.3cm
		\setlength\figurewidth{0.4\textwidth}
%
%
\definecolor{mycolor1}{rgb}{0.00000,0.44700,0.74100}%
\definecolor{mycolor2}{rgb}{0.85000,0.32500,0.09800}%
\definecolor{mycolor3}{rgb}{0.92900,0.69400,0.12500}%
\definecolor{mycolor4}{rgb}{0.49400,0.18400,0.55600}%
\definecolor{mycolor5}{rgb}{0.46600,0.67400,0.18800}%
\begin{tikzpicture}

\begin{groupplot}[group style={group name = group,group size=2 by 2, horizontal sep = 2cm, vertical sep=0.75cm}]

\nextgroupplot[%
width=\figurewidth ,
height=0.6\figurewidth ,
scale only axis,
xmin=42.5,
xmax=53,
xmajorgrids,
ymin=1390,
ymax=2070,
ylabel={File size [kB]},
/pgf/number format/.cd,
use comma,
1000 sep={},
ymajorgrids,
axis background/.style={fill=white},
every axis title/.style={font=\bfseries,below right,at={(0,1)}},
title={CT 12 bit}
]

\addplot [color=black,dotted]
table[row sep=crcr]{%
	42	4339\\
	54	4339\\
};\label{JPEG-LS}

\addplot [color=black,solid,mark = triangle]
table[row sep=crcr]{%
	35.5	1746\\
};\label{randomaccess}

\addplot [color=mycolor1,solid,mark = star]
table[row sep=crcr]{%
	43.10	1414.27\\
};\label{none}

\addplot [color=black,dashed]
table[row sep=crcr]{%
	42	1735.50\\
	54	1735.50\\	
};

\addplot [color=mycolor5,solid,mark=+,mark options={solid}]
table[row sep=crcr]{%
	47.05	1680.48\\
	48.09	1711.70\\
	48.41	1714.30\\
	48.77	1710.95\\
	48.91	1723.44\\
	49.06	1734.66\\
	49.24	1739.60\\
	49.44	1741.93\\
	49.51	1746.93\\
	49.60	1751.10\\
	49.69	1754.60\\
	49.78	1757.25\\
	49.89	1760.34\\
	49.99	1762.56\\
	50.12	1764.22\\
	50.25	1764.17\\	
};

\addplot [color=mycolor2,solid,mark = square]
table[row sep=crcr]{%
	47.49	1660.40\\
	49.72	1820.50\\
	52.52	2000.69\\	
};

\addplot [color=mycolor3,solid,mark = diamond]
table[row sep=crcr]{%
	48.19	1506.32\\
	49.51	1631.13\\
	49.97	1851.80\\	
};

\nextgroupplot[%
width=\figurewidth ,
height=0.6\figurewidth ,
scale only axis,
xmin=60,
xmax=67,
xmajorgrids,
ymin=290,
ymax=400,
ymajorgrids,
axis background/.style={fill=white},
every axis title/.style={font=\bfseries,below right,at={(0,1)}},
title={MR 12 bit}
]

\addplot [color=black,dotted]
table[row sep=crcr]{%
	42	1013\\
	54	1013\\
};

\addplot [color=black,solid,mark = triangle]
table[row sep=crcr]{%
	46.4	52.81\\
};

\addplot [color=mycolor1,solid,mark = star]
table[row sep=crcr]{%
	61.24	325.53\\
};

\addplot [color=black,dashed]
table[row sep=crcr]{%
	59	293.4695\\
	68	293.4695\\
};\label{HEVC}

\addplot [color=mycolor2,solid,mark = square]
table[row sep=crcr]{%
	62.44	335.21\\
	63.25	353.36\\
	65.46	393.28\\	
};\label{block}

\addplot [color=mycolor3,solid,mark = diamond]
table[row sep=crcr]{%
	63.06	331.64\\
	63.90	343.76\\
	64.60	375.92\\	
};\label{mesh}

\addplot [color=mycolor5,solid,mark=+,mark options={solid}]
table[row sep=crcr]{%
	60.35	356.96\\
	60.88	359.91\\
	61.23	361.09\\
	61.60	361.85\\
	61.88	363.43\\
	62.15	364.99\\
	62.46	366.25\\
	62.82	367.30\\
	63.11	367.94\\
	63.42	368.50\\
	63.76	368.64\\
	64.14	368.47\\
	64.58	368.48\\
	65.11	368.26\\
	65.72	367.61\\
	66.50	366.12\\		
};\label{graph_opt2}

\nextgroupplot[%
width=\figurewidth ,
height=0.6\figurewidth ,
scale only axis,
xmin=39,
xmax=48,
xlabel={$\text{PSNR}_{\text{LP}_t}$ [dB]},
xmajorgrids,
ymin=600,
ymax=1100,
ylabel={File size [kB]},
/pgf/number format/.cd,
use comma,
1000 sep={},
ymajorgrids,
axis background/.style={fill=white},
every axis title/.style={font=\bfseries,below right,at={(0,1)}},
title={CT 8 bit}
]

\addplot [color=mycolor1,solid,mark = star]
table[row sep=crcr]{%
	39.97	670.7\\
};

\addplot [color=black,solid,mark = triangle]
table[row sep=crcr]{%
	26.81	775.8\\
};

\addplot [color=black,dashed]
table[row sep=crcr]{%
	39	771.3288\\
	54	771.3288\\
};

\addplot [color=mycolor2,solid,mark=square]
table[row sep=crcr]{%
	47.5698070383686	1053.69381305364\\
	44.8936639824684	895.231522207185\\
	42.7591812285028	807.453194205217\\
};

\addplot [color=mycolor3,solid,mark=diamond]
table[row sep=crcr]{%
	45.1746267923912	967.687830647146\\
	44.7468179052128	807.554103100394\\
	43.4787888598344	739.473571296752\\
};

\addplot [color=mycolor5,solid,mark=+,mark options={solid}]
table[row sep=crcr]{%
	42.359577731537	867.721533587598\\
	43.4443904079546	891.539508489173\\
	43.7839090673826	894.948749692421\\
	44.1603508506294	895.932417261319\\
	44.3086426792841	906.939022514764\\
	44.4699191056542	917.865780327264\\
	44.6621938878937	922.790262057087\\
	44.8745945851856	925.468419352854\\
	44.9595464053385	930.036532664862\\
	45.0477189787685	935.956285371555\\
	45.1488671619491	939.334914800689\\
	45.2521380850864	941.913301242618\\
	45.3620673203094	945.926234928642\\
	45.4766179688994	948.216658464567\\
	45.6130474059797	951.692928764764\\
	45.75937406552	951.960752952756\\
};

\nextgroupplot[%
width=\figurewidth ,
height=0.6\figurewidth ,
scale only axis,
xmin=39,
xmax=47,
xlabel={$\text{PSNR}_{\text{LP}_t}$ [dB]},
xmajorgrids,
ymin=235,
ymax=350,
/pgf/number format/.cd,
use comma,
1000 sep={},
ymajorgrids,
axis background/.style={fill=white},
every axis title/.style={font=\bfseries,below right,at={(0,1)}},
title={MR 8 bit}
]

\addplot [color=mycolor1,solid,mark = star]
table[row sep=crcr]{%
	41.61	273.9\\
};

\addplot [color=black,solid,mark = triangle]
table[row sep=crcr]{%
	30.89	238.7\\
};

\addplot [color=black,dashed]
table[row sep=crcr]{%
	39	240.0974\\
	54	240.0974\\
};

\addplot [color=mycolor2,solid,mark=square]
table[row sep=crcr]{%
	45.7693731188832	342.426025390625\\
	43.5818705030999	300.690604073661\\
	42.777172685558	285.300606863839\\
};

\addplot [color=mycolor3,solid,mark=diamond]
table[row sep=crcr]{%
	44.8972234269165	323.085239955357\\
	44.2202528951324	292.467808314732\\
	43.385526333801	281.766008649554\\
};

\addplot [color=mycolor5,solid,mark=+,mark options={solid}]
table[row sep=crcr]{%
	39.3562058262673	306.102678571428\\
	40.2976361154826	309.084368024554\\
	40.821468557077	309.27587890625\\
	41.3712787821973	308.696602957589\\
	41.6707060806271	310.533796037947\\
	41.9723236224988	312.367885044643\\
	42.2819049594506	313.523786272321\\
	42.6444735788536	314.618198939732\\
	42.9493856311803	315.181675502232\\
	43.2753844753759	315.6142578125\\
	43.6412976055917	315.680733816964\\
	44.0388036437216	315.427699497768\\
	44.495870092932	315.614432198661\\
	45.042305879962	315.409109933036\\
	45.7088956443031	314.800223214286\\
	46.6072523417962	313.380057198661\\
};

\end{groupplot}


\path (group c1r1.south west |-current bounding box.south west)--
coordinate(legendpos)
(group c2r2.south east |-current bounding box.south east);
\matrix[
matrix of nodes,
anchor=south,
inner sep=0.2em,
draw,
nodes={scale=0.9},
]at([yshift=-4ex,xshift = 0ex]legendpos)
{	
	\node {\ref{HEVC}};  & \node {HEVC-LD~\cite{2016}}; & 
	\node {\ref{none}};  & \node {SBC~\cite{Karlsson1988}}; &
	\node {\ref{block}}; & \node {Block-based MCTF~\cite{lnt2012-40}}; & 
	\node {\ref{mesh}};  & \node {Mesh-based MCTF~\cite{8066388}}; &
	\node {\ref{graph_opt2}}; & \node {Graph-based MCTF}; \\
};

\draw[->] (2,5.25)node[left]{$d:6.25\%$} -- node[right,xshift=2cm]{$100\%$}(6,5.25);	
\draw[->] (2,4.75)node[left]{$s_b/s_g: 8$} -- node[right,xshift=2cm]{$2$}(6,4.75);

\draw[->] (11.25,5.25)node[left]{$d:6.25\%$} -- node[right,xshift=2cm]{$100\%$}(15.25,5.25);	
\draw[->] (11.25,4.75)node[left]{$s_b/s_g: 8$} -- node[right,xshift=2cm]{$2$}(15.25,4.75);

\draw[thick] (4.99,2.76) 	circle (3.8pt);
\draw[thick] (5.17,2.95) 	circle (3.8pt);
\draw[thick] (5.375,2.4) 	circle (3.8pt);

\draw[thick] (14.92,4.08) 	circle (3.8pt);
\draw[thick] (14.025,3.4) circle (3.8pt);
\draw[thick] (16,3) 	circle (3.8pt);

\end{tikzpicture}%
		\caption{$\text{PSNR}_{\text{LP}_t}$ results of the CT data (left) and MR data (right) compared against the overall file size in [kB]. The graph-based approach is evaluated for $r_\text{max}{=}3$. Our method is displayed for increasing densities, while the block- and mesh-based approaches are displayed over decreasing block and grid sizes, respectively. In addition, the rate results of lossless HEVC and SBC without any MC are shown. Top row: Original 12 bit data. Bottom row: Converted 8 bit data.}
		\label{fig:RD_CT}
	\end{center}
\end{figure*} 
\begin{table}[]
	\centering
	\caption{PSNR in [dB] and rate results in [kB] from encoding the data sets by HEVC-RA and JPEG-LS.}
	\begin{tabular}{l|l|l|r|r}
		&                                            &           & \multicolumn{1}{c|}{\textbf{12 bit}} & \multicolumn{1}{c}{\textbf{8 bit}} \\ \hline
		\multirow{4}{*}{{\tabrotate{\textbf{CT}}}} & \multicolumn{1}{l|}{\multirow{2}{*}{HEVC-RA}} & $\text{PSNR}_{\text{LP}_t}$ \hphantom{0-}{[dB]}     &  35.50           & 30.89 \\
		& \multicolumn{1}{l|}{}                      & File size $\Sigma$ [kB]& 1745.91 & 775.77 \\ \cline{2-5} 
		& \multicolumn{1}{l|}{JPEG-LS}               & File size $\Sigma$ [kB]& 4338.99 & 654.58 \\ \hhline{=|=|=|=|=} 
		\multirow{4}{*}{{\tabrotate{\textbf{MR}}}} & \multicolumn{1}{l|}{\multirow{2}{*}{HEVC-RA}} & $\text{PSNR}_{\text{LP}_t}$ \hphantom{0-}{[dB]}     &  46.40           & 26.81 \\
		& \multicolumn{1}{l|}{}                      & File size $\Sigma$ [kB]& 290.94 & 238.70 \\ \cline{2-5} 
		& \multicolumn{1}{l|}{JPEG-LS}               & File size $\Sigma$ [kB]& 1012.76 & 297.01    
	\end{tabular}
	\label{tab:jpeg}
\end{table}
\begin{table*}[t]
	\centering
	\caption{PSNR in [dB] and rate in [kB] of the single subbands for certain values of Fig.\,\ref{fig:RD_CT}. Absolute and relative differences against the proposed graph-based MCTF are also provided.}
	\label{tab:comp}
	\begin{tabular}{c|lr|c|c|c||r|r|r|r}
		& \multicolumn{2}{c|}{} & \multicolumn{1}{c|}{\textbf{Block-based}}  & \multicolumn{1}{c|}{\textbf{Mesh-based}} & \multicolumn{1}{c||}{\textbf{Graph-based}}& \multicolumn{2}{c|}{\textbf{$\Delta$ Proposed}}& \multicolumn{2}{c}{\textbf{$\Delta$ Proposed}}\\
		& \multicolumn{2}{c|}{} & \textbf{MCTF} & \textbf{MCTF} & \textbf{MCTF} & \multicolumn{2}{c|}{\textbf{to Block-based}} & \multicolumn{2}{c}{\textbf{to Mesh-based}}\\\hline
		\multirow{6}{*}{\tabrotate{\textbf{CT}}} & $\text{PSNR}_{\text{LP}_t}$  & [dB]& \hphantom{00}{49.72} & \hphantom{00}{49.97} & \hphantom{00}{50.25} & \multicolumn{2}{c|}{+\hphantom{0}{0.53}} &\multicolumn{2}{c}{+\hphantom{0}{0.28}} \\
		& File size $\text{LP}_t$ & [kB] & \hphantom{0}{878.70} & \hphantom{0}{822.44}& \hphantom{0}{808.80}& - 69.90 & - \hphantom{0}{8.64}\,$\%$ & - 13.64 & - \hphantom{0}{1.69}\,$\%$\\  
		& File size $\text{HP}_t$ 	  & [kB] & \hphantom{0}{867.81} & \hphantom{0}{827.73}& \hphantom{0}{802.26}& - 65.55 & - \hphantom{0}{8.17}\,$\%$ & - 25.47 & - \hphantom{0}{3.17}\,$\%$\\
		& File size $\boldsymbol{m_\text{tx}}$ & [kB] & \hphantom{00}{73.98}& \hphantom{0}{201.63} & \hphantom{0}{153.11} & + 79.13 & + 51.68\,$\%$ & - 48.52 & - 31.69\,$\%$\\ 
		& File size $\sum$     	  & [kB] & 1820.50				 &1851.80&1764.17& - 56.33 & - \hphantom{0}{3.19}\,$\%$ & - 87.63 & - \hphantom{0}{4.97}\,$\%$ \\\hline
		\multirow{6}{*}{\tabrotate{\textbf{MR}}} & $\text{PSNR}_{\text{LP}_t}$& [dB] & \hphantom{00}{65.46} & \hphantom{00}{64.60} & \hphantom{00}{66.50} & \multicolumn{2}{c|}{+\hphantom{0}{1.04}} & \multicolumn{2}{c}{+\hphantom{0}{1.90}} \\  
		& File size $\text{LP}_t$ & [kB] & \hphantom{0}{197.76} & \hphantom{0}{194.22}& \hphantom{0}{191.14}& - \hphantom{0}{6.62} & - \hphantom{0}{3.46}\,$\%$ & - \hphantom{0}{3.08} & - \hphantom{0}{1.61}\,$\%$ \\ 
		& File size $\text{HP}_t$ 	  & [kB] & \hphantom{0}{138.28} & \hphantom{0}{141.60}& \hphantom{0}{138.56}& + \hphantom{0}{0.28} & + \hphantom{0}{0.20}\,$\%$ & - \hphantom{0}{3.04} & - \hphantom{0}{2.19}\,$\%$ \\  
		& File size $\boldsymbol{m_\text{tx}}$ & [kB] & \hphantom{00}{57.23}& \hphantom{00}{40.11} &  \hphantom{00}{36.41} & - 20.82  & - 57.18\,$\%$ & - \hphantom{0}{3.70} & - 10.16\,$\%$ \\
		& File size $\sum$			  & [kB] & \hphantom{0}{393.28} & \hphantom{0}{375.92} &\hphantom{0}{366.12}& - 27.16 & - \hphantom{0}{4.62}\,$\%$ & - \hphantom{0}{9.80} & - \hphantom{0}{2.61}\,$\%$ \\
	\end{tabular}
\end{table*}

\begin{figure*}[h!]
	\captionsetup[subfigure]{position=top,labelformat=empty,justification=centering,farskip=2pt,captionskip=2pt} 
	\centering	
	\subfloat[Reference frame $f_1$]{	
		\tikz[remember picture] \node[inner sep=0] (ref1){	
			\hspace{2pt}				
			\begin{tikzpicture}[scale=0.72, >=latex'] 
			\node [inner sep=0pt] {	
				\includegraphics[trim = 14cm 2cm 14cm 1cm,clip,width = 0.18\textwidth]{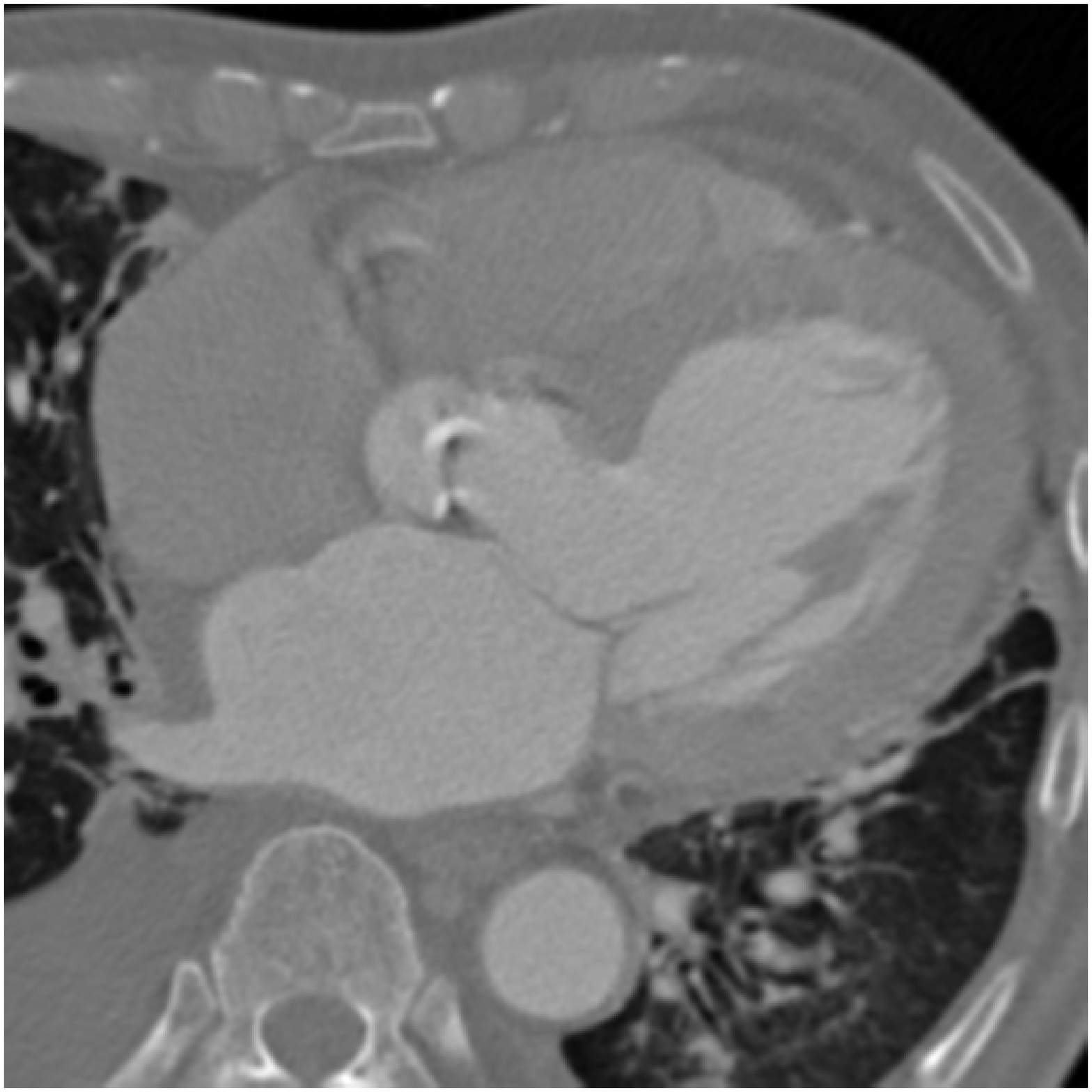}};
			\draw[draw=red,very  thick]  (-0.325,-0.9) rectangle (0.825,0.25);
			\end{tikzpicture}};
	}\quad
	\subfloat[Current frame $f_2$]{		
		\tikz[remember picture] \node[inner sep=0] (ref2){	
			\begin{tikzpicture}[scale=0.72, >=latex']
			\node [inner sep=0pt] {	
				\includegraphics[trim = 14cm 2cm 14cm 1cm,clip,width = 0.18\textwidth]{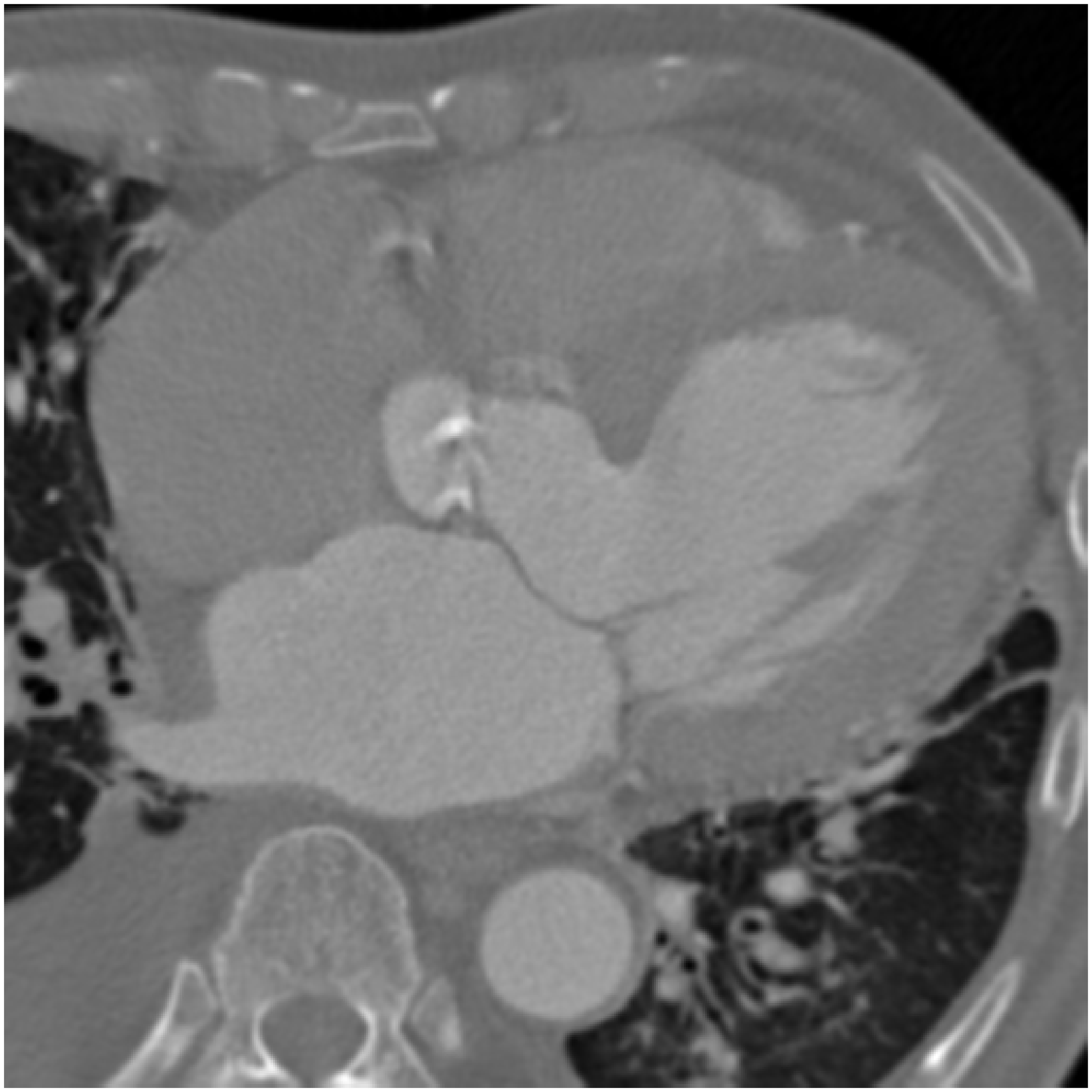}};
			\draw[draw=red,very  thick]  (-0.325,-0.9) rectangle (0.825,0.25);
			\end{tikzpicture}};
	}\\
	\captionsetup[subfigure]{position=bottom,labelformat=empty,justification=centering}
	\subfloat[Zoom into reference frame $f_1$]{	
		\tikz[remember picture] \node[inner sep=0] (zoom1){	
			\begin{tikzpicture}
			\node [inner sep=0pt] {	
				\includegraphics[trim = 17cm 6.25cm 17cm 3cm,clip,width = 0.24\textwidth]{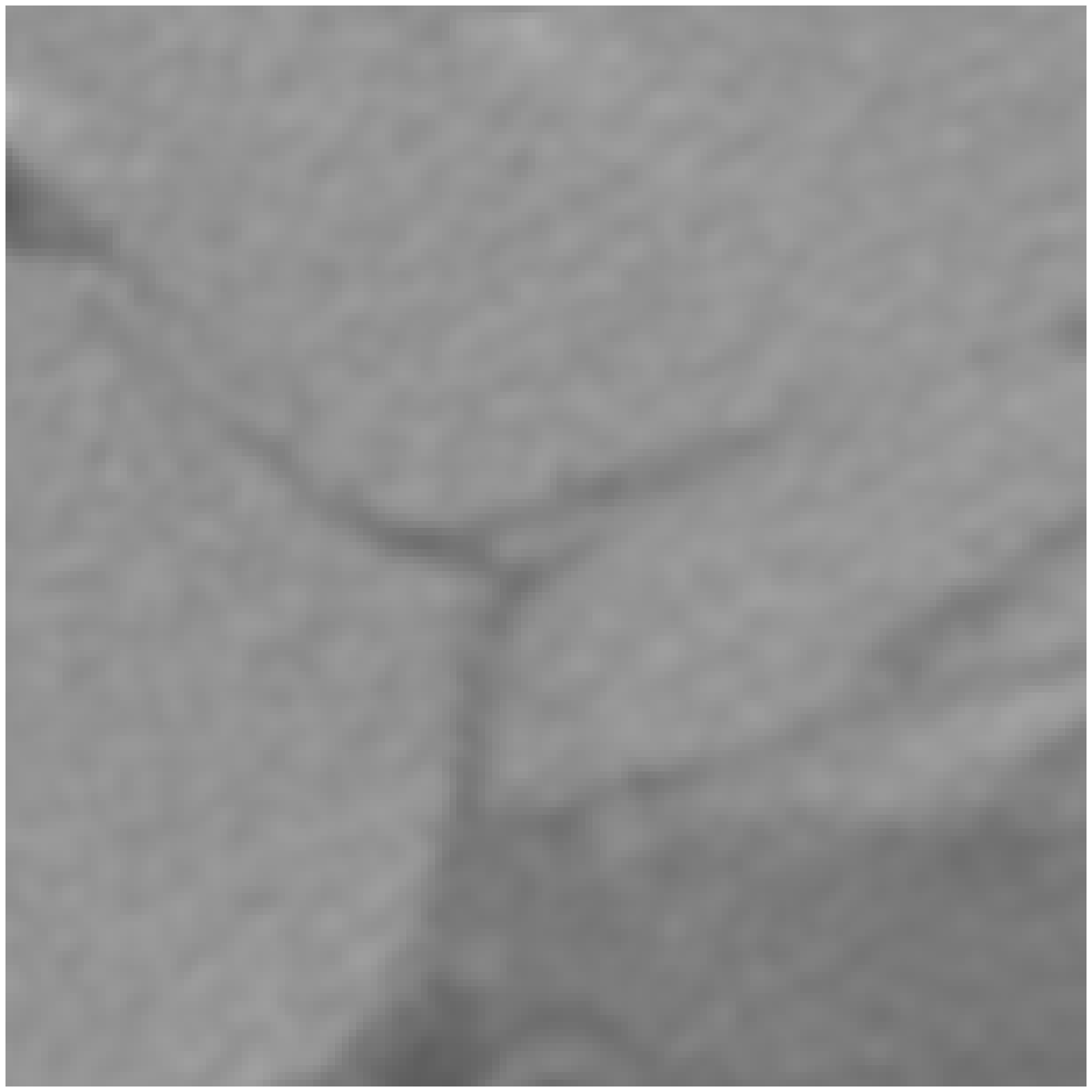}};		
			\end{tikzpicture}};
	}\hspace{4.5cm}
	\subfloat[Zoom into current frame $f_2$]{
		\tikz[remember picture] \node[inner sep=0] (zoom2){	
			\begin{tikzpicture}
			\node [inner sep=0pt] {	
				\includegraphics[trim = 17cm 6.25cm 17cm 3cm,clip,width = 0.24\textwidth]{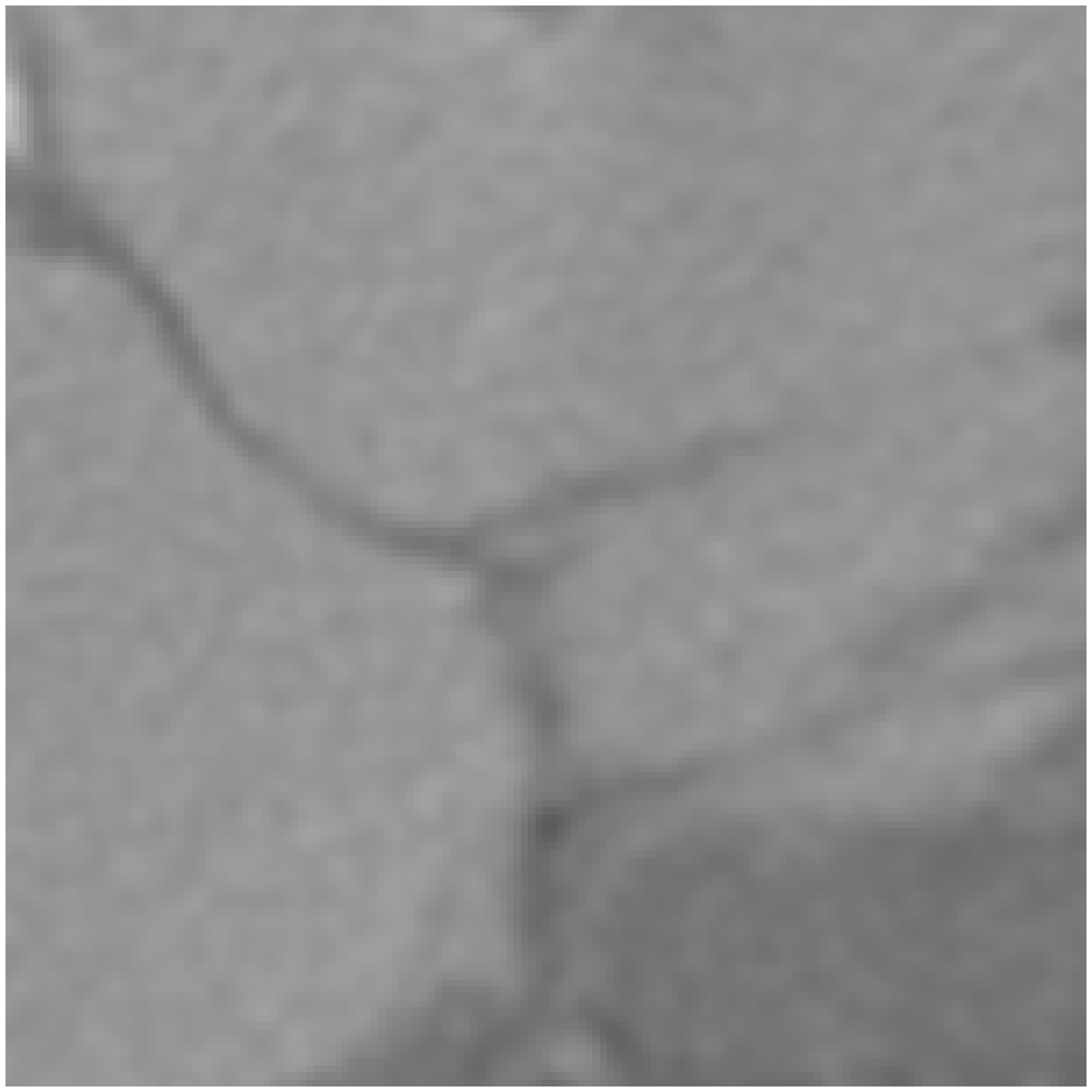}};
			\end{tikzpicture}};
	}\\	
	\subfloat[SBC: No compensation, $\text{LP}_1$]{
		\begin{tikzpicture}
		\node [inner sep=0pt] {	
			\includegraphics[trim = 17cm 6.25cm 17cm 3cm,clip,width = 0.24\textwidth]{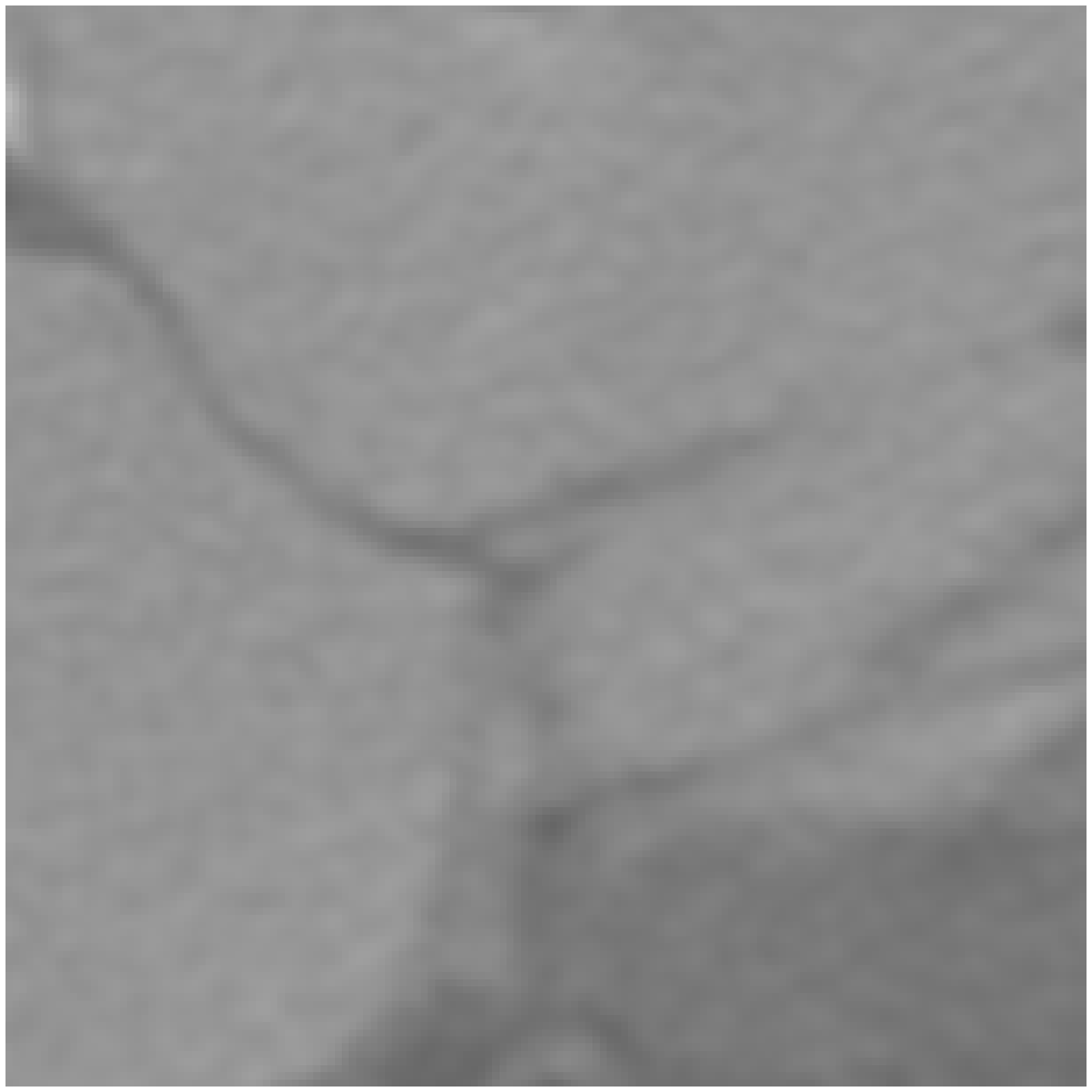}};
		\node[fill=white] (n1) at (1.3,1.9) {$42.15$\,dB};
		\draw[dashed,rotate around={85:(-0.1,-0.7)}, draw=red,very  thick]  (-0.1,-0.7) ellipse (1.2cm and 0.4cm);
		\draw[dashed,rotate around={140:(-1.2,0.6)}, draw=red,very  thick]  (-1.2,0.6) ellipse (0.8cm and 0.4cm);
		\end{tikzpicture}
	}
	\subfloat[MCTF: Block-based, $\text{LP}_1$]{
		\tikz[remember picture] \node[inner sep=0] (top){
			\begin{tikzpicture}
			\node [inner sep=0pt] {	
				\includegraphics[trim = 17cm 6.25cm 17cm 3cm,clip,width = 0.24\textwidth]{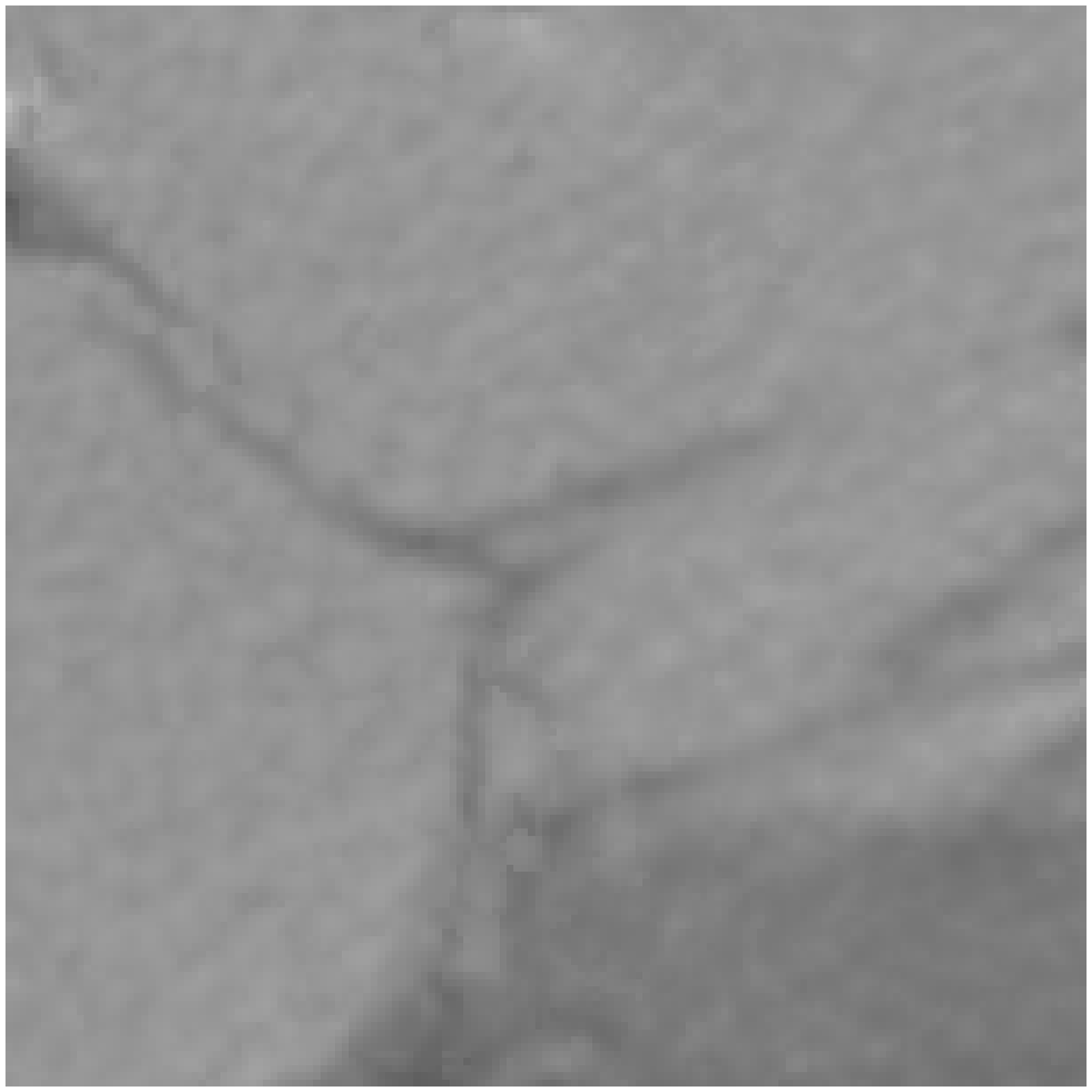}
			};
			\draw[white,fill=white] (0.55,1.65) rectangle (2.04,2.14);
			\node[fill=white] (n1) at (1.3,1.9) {$46.38$\,dB};
			\draw[rotate around={50:(-0.1,-1)}, draw=red,very  thick]  (-0.1,-1) ellipse (0.8cm and 0.6cm);
			\end{tikzpicture}};
	}
	\subfloat[SBC: No compensation, $\text{MC}(\text{LP}_1)$]{
		\begin{tikzpicture}
		\node [inner sep=0pt] {
			\includegraphics[trim = 17cm 6.25cm 17cm 3cm,clip,width = 0.24\textwidth]{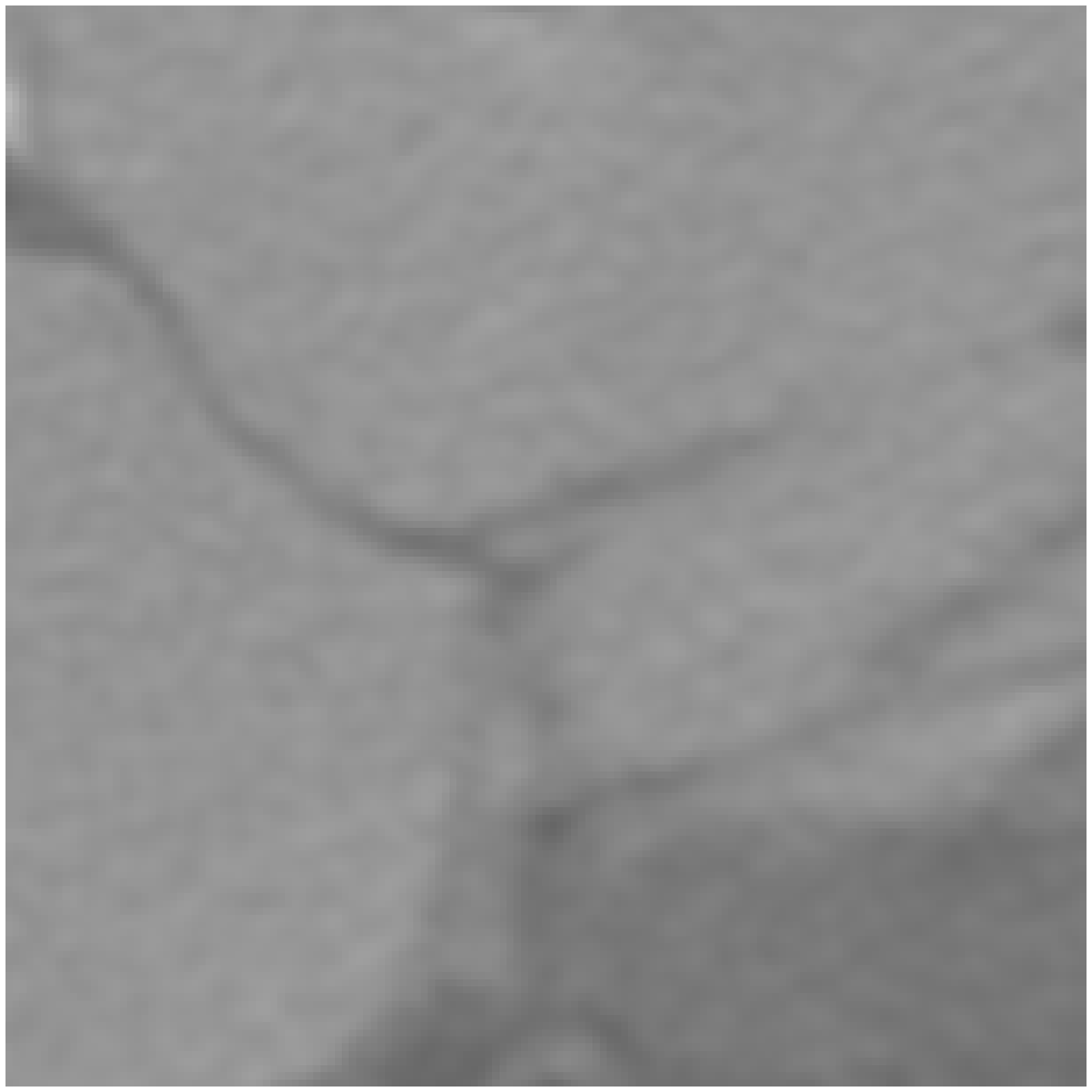}};
		\node[fill=white] (n1) at (1.3,1.9) {$42.25$\,dB};
		\draw[dashed,rotate around={85:(-0.1,-0.7)}, draw=red,very  thick]  (-0.1,-0.7) ellipse (1.2cm and 0.4cm);
		\draw[dashed,rotate around={140:(-1.2,0.6)}, draw=red,very  thick]  (-1.2,0.6) ellipse (0.8cm and 0.4cm);
		\end{tikzpicture}
	}
	\subfloat[MCTF: Block-based, $\text{MC}(\text{LP}_1)$]{
		\begin{tikzpicture}
		\node [inner sep=0pt] {	
			\includegraphics[trim = 17cm 6.25cm 17cm 3cm,clip,width = 0.24\textwidth]{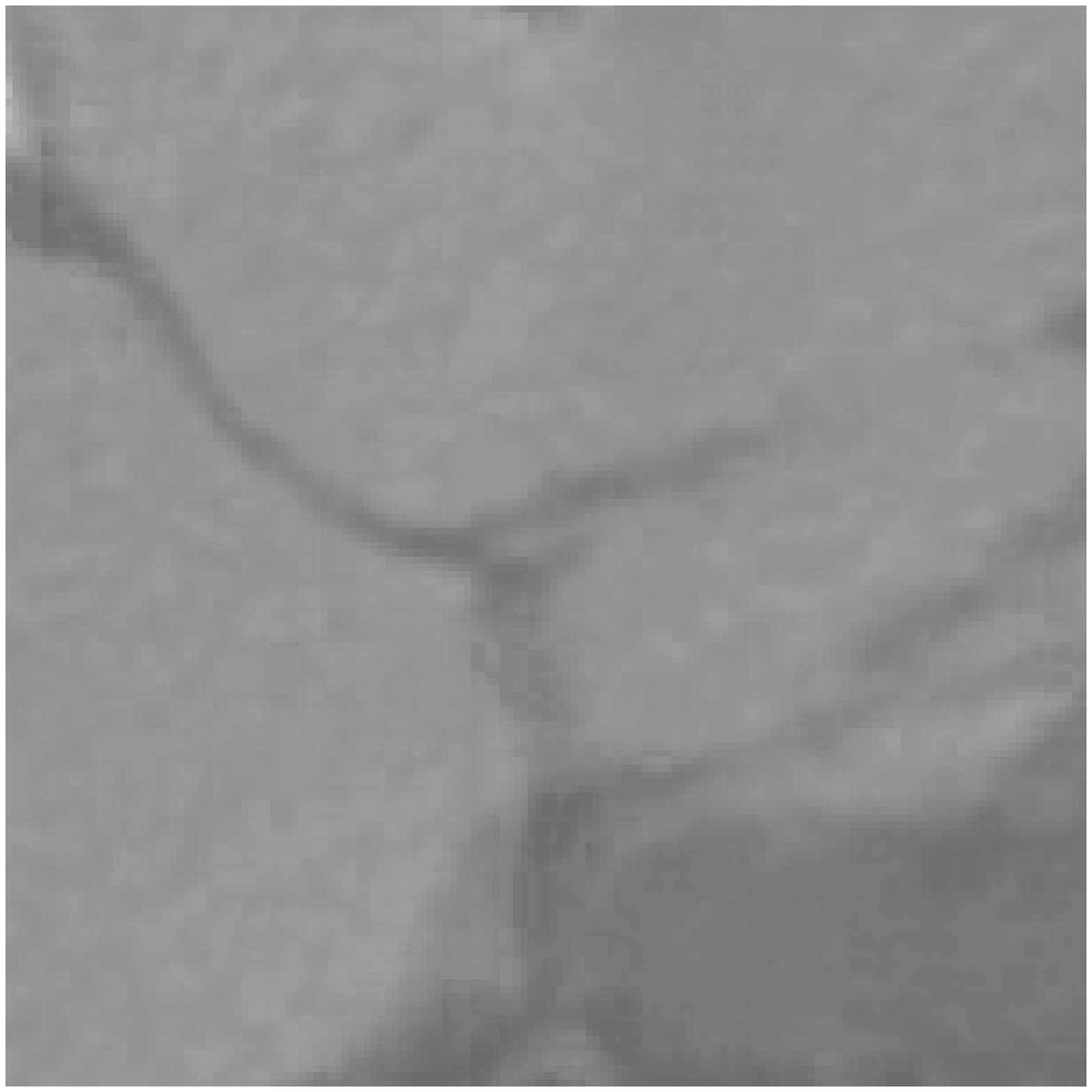}
		};		
		\node[fill=white] (n1) at (1.3,1.9) {$45.11$\,dB};
		\draw[dashed, rotate around={140:(-0.2,-0.2)}, draw=red,very  thick]  (-0.2,-0.2) ellipse (0.5cm and 0.5cm);
		\end{tikzpicture}
	}\\	
	\subfloat[MCTF: Mesh-based, $\text{LP}_1$]{
		\begin{tikzpicture}
		\node [inner sep=0pt] {	
			\includegraphics[trim = 17cm 6.25cm 17cm 3cm,clip,width = 0.24\textwidth]{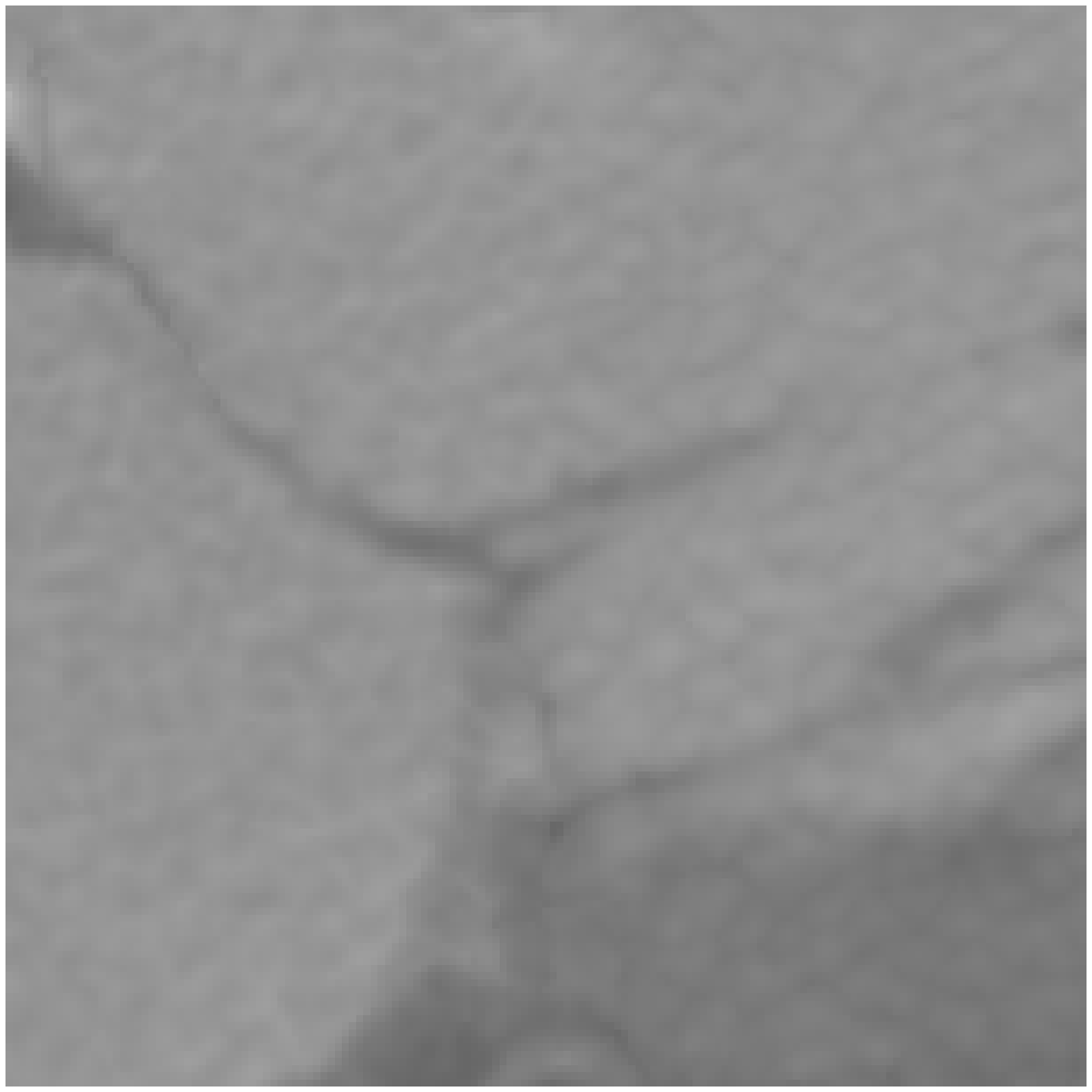}
		};
		\node[fill=white] (n1) at (1.3,1.9) {$45.84$\,dB};
		\draw [->,draw=red,very  thick] (-1,-0.4) -- (-0.2,-0.7);
		\draw[dashed,rotate around={140:(-1.3,0.7)}, draw=red,very  thick]  (-1.3,0.7) ellipse (0.6cm and 0.4cm);
		\end{tikzpicture}
	}	
	\subfloat[MCTF: Graph-based, $\text{LP}_1$]{	
		\tikz[remember picture] \node[inner sep=0] (bottom){
			\begin{tikzpicture}
			\node [inner sep=0pt] {		
				\includegraphics[trim = 17cm 6.25cm 17cm 3cm,clip,width = 0.24\textwidth]{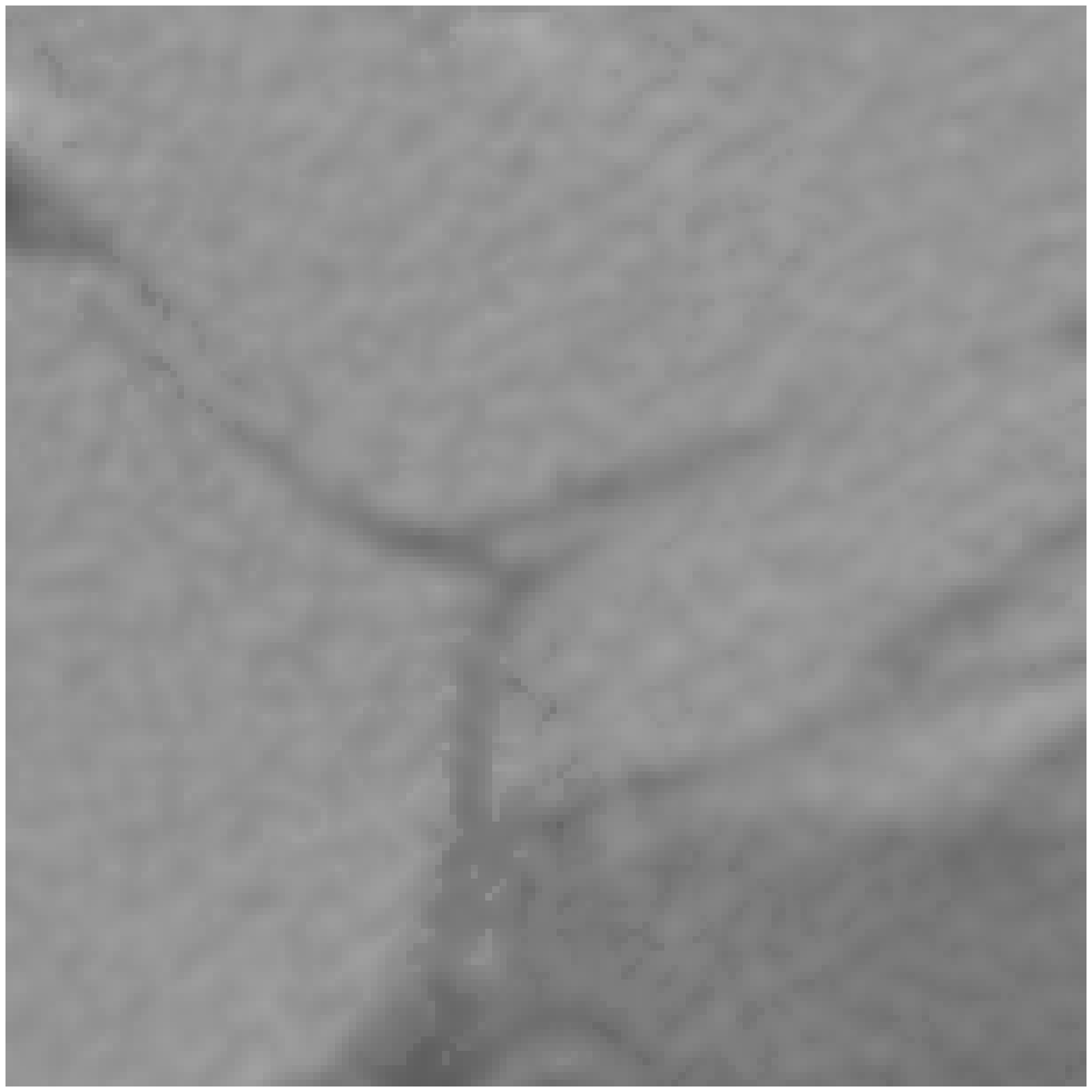}};
			\draw[white,fill=white] (0.55,1.65) rectangle (2.04,2.14);
			\node[fill=white] (n1) at (1.3,1.9) {$50.38$\,dB};
			\end{tikzpicture}};
	}	
	\subfloat[MCTF: Mesh-based, $\text{MC}(\text{LP}_1)$]{
		\begin{tikzpicture}
		\node [inner sep=0pt] {	
			\includegraphics[trim = 17cm 6.25cm 17cm 3cm,clip,width = 0.24\textwidth]{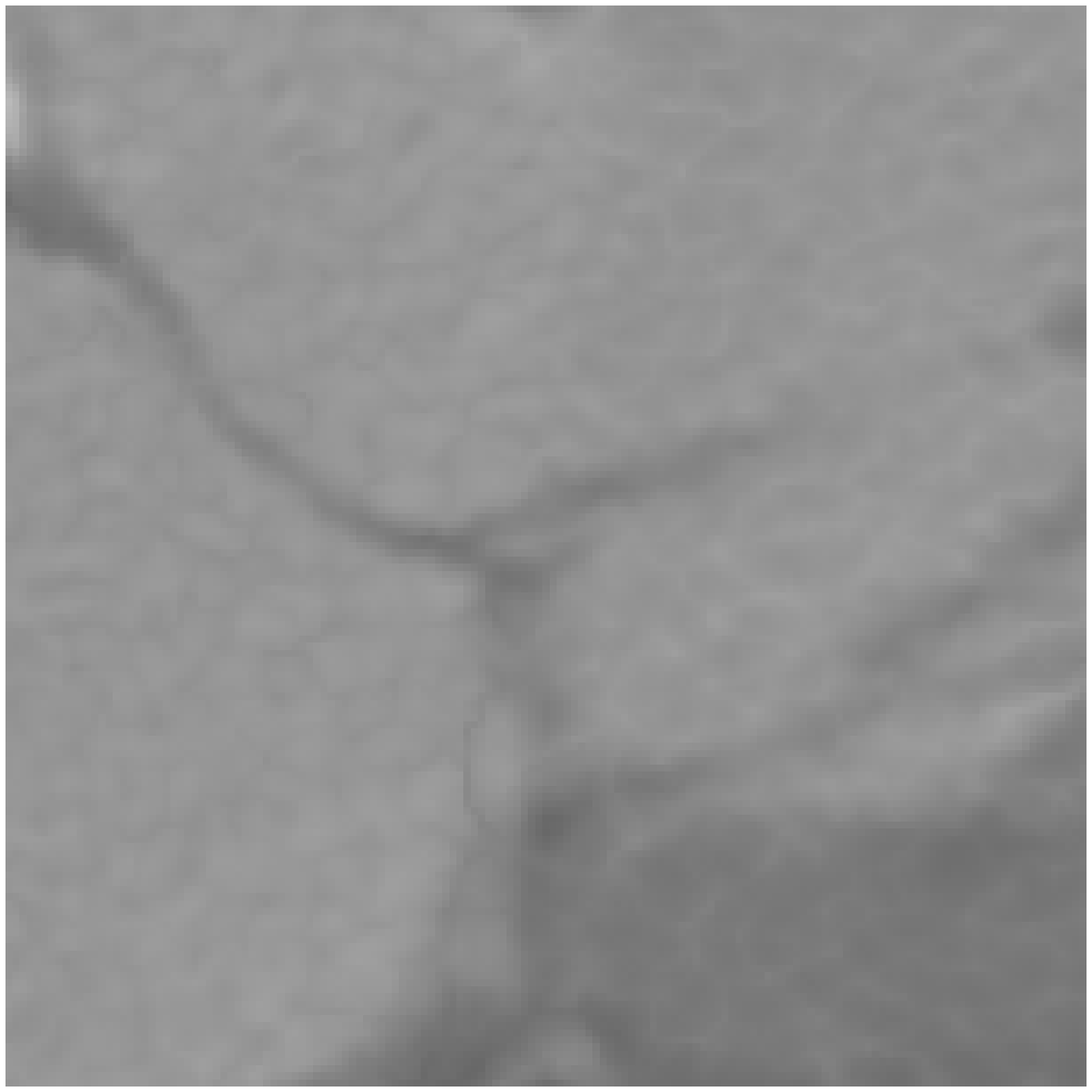}
		};
		\node[fill=white] (n1) at (1.3,1.9) {$44.53$\,dB};
		\draw [->,draw=red,very  thick] (-1.5,-1.5) -- (-0.50,-1.0);
		\draw[dashed, rotate around={140:(-0.2,-0.2)}, draw=red,very  thick]  (-0.2,-0.2) ellipse (0.5cm and 0.5cm);
		\end{tikzpicture}
	}	
	\subfloat[MCTF: Graph-based, $\text{MC}(\text{LP}_1)$]{		
		\begin{tikzpicture}
		\node [inner sep=0pt] {	
			\includegraphics[trim = 17cm 6.25cm 17cm 3cm,clip,width = 0.24\textwidth]{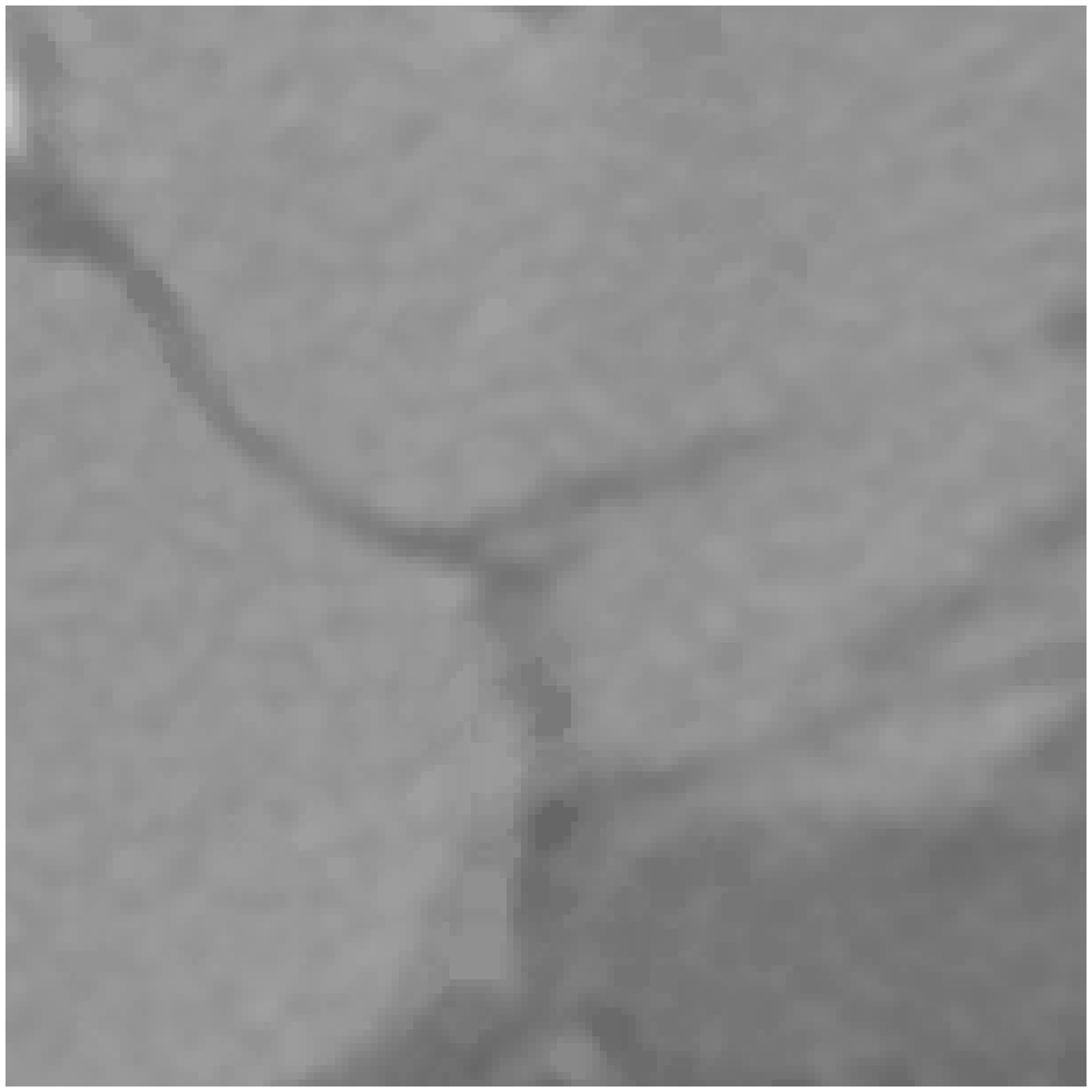}};
		\node[fill=white] (n1) at (1.3,1.9) {$49.14$\,dB};
		\end{tikzpicture}}
	\begin{tikzpicture}[overlay, remember picture]
	\draw[black, thick] ([shift={(1mm, 6.9cm)}]top.east)--([shift={(1mm, -20.5mm)}]bottom.east);
	\draw[red, thick, dashed] ([shift={(-0.134,0.18)}]ref1.center)--([shift={(1mm, 0mm)}]zoom1.north west);
	\draw[red, thick, dashed] ([shift={(0.694,0.18)}]ref1.center)--([shift={(-1.5mm, 0mm)}]zoom1.north east);
	\draw[red, thick, dashed] ([shift={(-0.234,0.18)}]ref2.center)--([shift={(1mm, 0mm)}]zoom2.north west);
	\draw[red, thick, dashed] ([shift={(0.594,0.18)}]ref2.center)--([shift={(-1.5mm, 0mm)}]zoom2.north east);
	\end{tikzpicture}
	\caption{Visual example for two successive frames at slice $z=46$ of the 12 bit CT volume. 
		Left: reference frame $f_1$ and details of the corresponding frames $\text{LP}_1$ using SBC without any motion compensation and MCTF with block-, mesh-, and graph-based motion compensation.
		Right: current frame $f_2$ and details of the corresponding frames $\text{MC}(\text{LP}_1)$ for all considered approaches.}
	\label{fig:examples}
\end{figure*}

\subsection{Evaluation of the Proposed Method}
\label{subsec:eval}

In the following, we will compare our proposed graph-based MCTF coder to state-of-the-art SL and EL coding schemes. As prominent representatives for SL coding, we employ the JPEG-LS as a conventional still image coder and the HEVC codec as a motion compensated predictive coder. By choosing the \textit{Lowdelay Main RExt} configuration (HEVC-LD) and lossless mode, it is possible to apply HEVC also to medical sequences. The test model HM-16.16 was used for simulation. 

For a fair comparison to state-of-the-art EL coding schemes, the degree of the three main types of scalability have to be considered. For our proposed graph-based MCTF coder, the frame rate of the BL is halved. Due to the encoding of the single subbands by JPEG\,2000, the spatial resolution is also decreased. The quality scalability is controlled by the density of the sparse sampling masks. 
SHVC and SELC also support all these scalability types. However, the temporal scalability of both codecs is achieved by subsampling the BL without any filtering process. Since MCTF employs subband filtering along the motion trajectories of the temporal axis, a fair comparison to SHVC and SELC is not possible. Additionally, SHVC is only implemented for $8$ and $10$ bit data. However, temporal scalability is inherently achieved by the \textit{Randomaccess Main RExt} configuration of the HEVC codec (HEVC-RA), which supports $12$ bit data. By decoding only the first temporal layer and omitting the spatial scalability aspect for compression, a visual comparison between the BL of HEVC and MCTF can be achieved.

Additionally, we compare our proposed graph-based MCTF to SBC without any MC as well as to block-based and mesh-based MCTF. These approaches are all based on the same system concept and allow for an evaluation with equal conditions.
Therefore, the search range of both compared MCTF methods is set to the maximum radius $r_\text{max}$ of the graph-based approach. Both compared MCTF methods are evaluated for block and grid sizes $s_b$ and $s_g$ of $2,4,$ and $8$, respectively.
Again, one Haar wavelet decomposition step in temporal direction is performed. As already mentioned, the resulting subbands are compressed without any loss, using the wavelet-based volume coder JPEG\,2000 implemented by OpenJPEG~\cite{openjpeg} with four spatial wavelet decomposition steps in $xy$-direction. The required motion vector fields of the block- and mesh-based approaches are encoded using the QccPack library~\cite{fowler2000qccpack}.
	
The performance of our proposed graph-based MCTF coder compared to SL coding with JPEG-LS and HEVC-LD as well as to HEVC-RA, SBC, block-based, and mesh-based MCTF with respect to the overall file size and the visual quality of the LP subband in terms of $\text{PSNR}_{\text{LP}_t}$, is graphically shown in Fig.\,\ref{fig:RD_CT}. 
Since the resulting file sizes from \mbox{JPEG-LS} on 12 bit data are significantly higher than from all other applied methods and therefore, cannot be displayed properly in Fig.\,\ref{fig:RD_CT}, we provide these results in Table~\ref{tab:jpeg}. For the same reasons regarding the visual quality of the BL from HEVC-RA, we provide the results for HEVC-RA also in this table.
To exclude that our results are only valid for 12 bit data, we perform the same experiments also on 8 bit data. 8 bit medical data sets are usually preconverted from 12 or 16 bits to 8 bits due to medical displays, which support very often only 8 bits per pixel. Therefore, we do the same for our used CT and MR data sets. 

Comparing the considered EL coding schemes, the compression efficiency is much higher for SBC than for any MCTF coder. According to~\cite{lnt2012-10}, this is caused by the correlated noisy structures that can be exploited by a traditional wavelet transform without MC. By applying MCTF, these structures can not be exploited with the same efficiency. Additionally, the corresponding motion information has to be coded and contributes to the overall file size. However, if the LP subband is to be used as a downscaled representative in telemedicine applications, the visual quality is of high relevance, which can significantly be increased by incorporating various compensation methods into the coding scheme.

Comparing our proposed method to the block- and mesh-based MCTF, the novel graph-based approach is less efficient for LP subbands with low quality. However, for high quality LP subbands the new method is able to outperform the other state-of-the-art methods. This is very advantageous for telemedicine applications, where a high visual quality of the LP subband is indispensable, if it is to be used as a downscaled representative for the original sequence. Block-based MCTF also achieves high quality LP subbands, but the required bit rate for transmitting the entire volume without any loss, is huge. This is caused by high frequency components arising from the block structures. Deformable motion models like mesh-based and graph-based MC can avoid these structures and result in lower bit rates. 

In general, our proposed method results in higher $\text{PSNR}_{\text{LP}_t}$ values and higher bit rates with an increasing sampling density. However, for the MR data the green curve is not monotonically increasing with increasing values of $d$. But if Fig.~\ref{fig:opt} is considered again, it can be observed that the file size for transmitting $m_\text{tx}$ is increasing as well as for the CT data. Consequently, the required bit rate for transmitting the resulting LP and HP frames decreases for sampling densities equal or higher than $62.5\%$. Since for higher sampling densities the corresponding LP and HP frames contain less artifacts, they can be compressed more efficiently. This leads to higher coding gains compared to the overhead, which the motion information generates.

For a closer examination, we choose some values in Fig.\,\ref{fig:RD_CT}, which operate at similar bit rates. All these values are marked with black circles in
Fig.\,\ref{fig:RD_CT}. The corresponding values for the quality in terms of $\text{PSNR}_{\text{LP}_t}$, the exact distribution of the rate into the single subbands, and the amount of bits needed to encode the motion information for both data sets can be found in Table\,\ref{tab:comp}. The differences against our proposed graph-based method are also provided. Accordingly, compared to the mesh-based approach, the visual quality of the LP subband can be increased by $0.28$\,dB and $1.90$\,dB for the CT and MR data set, respectively. Compared to the block-based approach, PSNR gains of $0.53$\,dB and $1.04$\,dB for the CT and MR data set can be achieved, respectively. We want to emphasize, that the required file size to transmit the LP subband and the overall file size to reconstruct the volumes in a lossless way, are always smaller than compared to the other approaches.

Comparing our proposed graph-based approach to SL coding with HEVC, an interesting fact can be observed. Contrary to the conventional assumption, that SL coding requires less bits than any scalable lossless coding scheme as introduced in Section~\ref{sec:introduction}, the opposite behavior occurs. For the 12 bit CT data set it is possible, to achieve smaller bit rates with wavelet-based EL coding than with SL coding. This may be caused by the noise characteristics of CT data. The noise in CT images is typically correlated to the acquisition process. Wavelet-based coding schemes seem to be able to exploit these correlated noisy structures better than DCT-based coding schemes, like HEVC. However, for the 8 bit CT data set, HEVC performs better than for 12 bit data. This leads to the assumption that 12 bit medical data can be compressed better with wavelet-based coding schemes. For telemedicine applications, where scalable lossless coding schemes are of high relevance, this observation should be focused in further research.

\subsection{Visual Example of the Proposed Method}

To demonstrate the visual performance of our proposed method, two original frames at $t{=}1$ and $t{=}2$ and slice $z{=}46$ from the 12 bit CT data set are shown in Fig.\,\ref{fig:examples} at the top. 
Additionally, in the left column a detail (position marked in red) of the corresponding LP frame $\text{LP}_1$ using SBC without any MC and MCTF with block-, mesh-, and graph-based MC is shown. To measure the visual quality of the LP frames, the similarity to $f_1$ and $f_2$ has to be considered. While the first evaluation can easily be done by evaluating $\text{LP}_1$ in terms of PSNR with respect to $f_1$, the second evaluation is more complex. To measure the similarity to $f_2$, the LP frames have to be warped to the corresponding time step. Then, these motion compensated LP frames $\text{MC}(\text{LP}_1)$ can also be evaluated in terms of PSNR with respect to $f_2$.
Accordingly, the right column of Fig.\,\ref{fig:examples} comprises the corresponding motion compensated LP frames $\text{MC}(\text{LP}_1)$.
As indicated by the low PSNR values of \mbox{$\text{LP}_1$} and \mbox{$\text{MC}(\text{LP}_1)$} for SBC, the LP frame is not suitable for being used in telemedicine applications, since fine structures, marked by the dashed ellipses, are blurred.

For a fair comparison, all methods used for MCTF are evaluated at approximately the same rate. Therefore, the block-based approach is evaluated at a block size of $s_b{=}4$, the mesh-based approach at a grid size of $s_g{=}2$ and the graph-based approach for a density of $d{=}100\%$. These configurations correspond to the values chosen in Table\,\ref{tab:comp}.
The block-based approach causes blocking artifacts in \mbox{$\text{LP}_1$}, which are marked by the red ellipse, and blurred structures in \mbox{$\text{MC}(\text{LP}_1)$}, which are marked by the dashed circle. In contrast, the mesh-based approach provides a smoother visual result. However, some erroneous structures appear, as the red arrows show. In addition, blurring artifacts occur, which are marked by the dashed circle and ellipse. The graph-based approach is capable to cope with all mentioned artifacts, which is also constituted by the high PSNR values. Additionally, our approach requires less file size than the block- and mesh-based approach with $s_b{=}4$ and $s_g{=}2$, respectively, as already shown in Table\,\ref{tab:comp}.

\section{Conclusion}
\label{conclusion}
Temporal scalability of dynamic volume data and a high visual quality of the corresponding representative is indispensable for telemedicine applications. Scalable lossless EL coding schemes based on MCTF are characterized by their high data fidelity due to subband filtering along the motion trajectories. In this context, graph-based motion compensation turned out to achieve superior results regarding the visual quality, but encoding of the motion information, which is stored in adjacency matrices, was not considered in current research. We proposed a method to uniquely convert these adjacency matrices into motion maps. After smoothing and subsampling these motion maps, they are encoded using multiple-context adaptive arithmetic coding. Missing values due to the subsampling step are reconstructed at decoder side. 
By applying this novel coding scheme, the visual quality of the lowpass subband can be improved by up to 0.53\,dB and 0.28\,dB for 12 bit CT data, as well as 1.04\,dB and 1.09\,dB for 12 bit MR data, respectively, compared to the block- and mesh-based approaches, while the file size can be reduced at the same time. 

At some points in the proposed processing chain we decided for some thresholds and methods empirically. 
Nevertheless, the final results are already very promising.
We expect further improvements by training the intervals for smoothing the motion maps on larger data sets, which include various medical devices. The same training data may be used to refine the choice of the $\text{PSNR}_\text{target}$.
Additionally, the choice of the sampling density as well as the selection of the reconstruction algorithm after sparse sampling can be improved by fitting these parameters automatically to the given requirements.
To achieve further coding gains binary arithmetic coding with several types of context information can be analyzed for encoding the binary masks more efficiently.
Beyond PSNR, a medical application-guided assessment, e.g., ROC metrics, may be considered in future works.

\section*{Acknowledgment}
We gratefully acknowledge that this work has been supported by the Deutsche Forschungsgemeinschaft (DFG) under contract number KA 926/4-3.

\ifCLASSOPTIONcaptionsoff
  \newpage
\fi


\bibliographystyle{IEEEtran}
\bibliography{Literatur}

%




\end{document}